\documentclass[a4paper,fleqn,usenatbib]{mnras}

\usepackage{newtxtext,newtxmath}

\usepackage[T1]{fontenc}
\usepackage{ae,aecompl}

\usepackage{graphicx}	
\usepackage{amsmath}	
\usepackage{amssymb}	

\usepackage{color}
\usepackage{BibDef}
\def\lsim{\mathrel{\rlap{\lower 3pt \hbox{$\sim$}} \raise 2.0pt \hbox{$<$}}}
\def\gsim{\mathrel{\rlap{\lower 3pt \hbox{$\sim$}} \raise 2.0pt \hbox{$>$}}}
\def\msun{\rm {M_\odot}}

\def\mach{\mathcal{M}}
\def\angstrom{\mathring{\mathrm{A}}}

\def\gmol{G\_mol}
\def\guv{G\_UV}
\def\grads{G\_RAD\_S}
\def\gradt{G\_RAD\_T}
\def\grt{G\_RT}

\title[The emergence of the SF--H$_2$ correlation] 
{The natural emergence of the correlation between H$_2$ and star formation rate surface densities in galaxy simulations}
\author[A. Lupi et al.]{Alessandro Lupi,$^{1}$\thanks{E-mail:
lupi@iap.fr} Stefano Bovino,$^2$ Pedro R. Capelo,$^3$\newauthor Marta Volonteri$^1$ and Joseph Silk$^{1,4,5,6}$\\
$^1$Sorbonne Universit\`{e}s, UPMC Univ Paris 6 et CNRS, UMR 7095, Institut d'Astrophysique de Paris,\\ 98 bis bd Arago, F-75014 Paris, France\\
$^2$ Hamburger Sternwarte, Universit\"{a}t Hamburg, Gojenbergsweg 112, DE-21029 Hamburg, Germany\\
$^3$ Center for Theoretical Astrophysics and Cosmology, Institute for Computational Science, University of Zurich, \\Winterthurerstrasse 190, CH-8057 Z\"{u}rich, Switzerland\\
$^4$AIM-Paris-Saclay, CEA/DSM/IRFU, CNRS, Univ Paris 7, F-91191 Gif-sur-Yvette, France\\
$^5$Department of Physics and Astronomy, The Johns Hopkins University, Baltimore, MD 21218, USA\\\
$^6$BIPAC, University of Oxford,1 Keble Road, Oxford OX1 3RH, UK}

\usepackage{etoolbox}
\makeatletter
\patchcmd\@combinedblfloats{\box\@outputbox}{\unvbox\@outputbox}{}{%
 \errmessage{\noexpand\@combinedblfloats could not be patched}%
}%
\makeatother

\begin{document}

\date{Draft \today}

\pagerange{\pageref{firstpage}--\pageref{lastpage}} \pubyear{2017}

\maketitle

\label{firstpage}

\begin{abstract}
In this study, we present a suite of high-resolution numerical simulations of an isolated galaxy to test a sub-grid framework to consistently follow the formation and dissociation of H$_2$ with non-equilibrium chemistry. The latter is solved via the package \textsc{krome}, coupled to the mesh-less hydrodynamic code \textsc{gizmo}. We include the effect of star formation (SF), modelled with a physically motivated prescription independent of H$_2$, supernova feedback and mass losses from low-mass stars, extragalactic and local stellar radiation, and dust and H$_2$ shielding, to investigate the emergence of the observed correlation between H$_2$ and SF rate surface densities. We present two different sub-grid models and compare them with on-the-fly radiative transfer (RT) calculations, to assess the main differences and limits of the different approaches. We also discuss a sub-grid clumping factor model to enhance the H$_2$ formation, consistent with our SF prescription, which is crucial, at the achieved resolution, to reproduce the correlation with H$_2$.
We find that both sub-grid models perform very well relative to the RT simulation, giving comparable results, with moderate differences, but at much lower computational cost. We also find that, while the Kennicutt--Schmidt relation for the total gas is not strongly affected by the different ingredients included in the simulations, the H$_2$-based counterpart is much more sensitive, because of the crucial role played by the dissociating radiative flux and the gas shielding.
\end{abstract}

\section{Introduction}
After the pioneering work by \citet{schmidt59}, several observations showed that the rapidity with which galaxies form stars correlates with the surface density of the total available gas. However, because of the time-consuming and challenging effort to measure this relation with very high resolution, only recently we have been able to collect large amounts of data to better constrain it. In particular, recent observations by \citet{bigiel10} have shown that this relation seems to deviate from the canonical single power law with slope 1.4 \citep{kennicutt98} for low surface densities.

Observations of nearby star-forming galaxies have also evidenced a linear correlation between the star formation rate (SFR) surface density and the molecular hydrogen (H$_2$) surface density on sub-kpc scales \citep{bigiel08,bigiel10}. On the other hand, the correlation with atomic hydrogen seems to be very weak. 

While many numerical simulations still use collisional ionisation equilibrium chemistry and do not follow the formation and evolution of molecular hydrogen, several groups have started to use molecular hydrogen based SF prescriptions, either following self-consistently the formation and dissociation of H$_2$ with non-equilibrium chemistry \citep{gnedin09,feldmann11,christensen12,tomassetti15} or via sub-grid prescriptions based on the assumption of chemical equilibrium \citep{pelupessy06,thompson14,hopkins14,somerville15,pallottini17,hopkins17,orr17}.

However, recent theoretical studies have revealed a lack of causal connection between H$_2$ and SF \citep{krumholz11,clark12}, and that the low temperatures of star-forming regions can be reached via molecular cooling as well as metal line cooling \citep[see][for a detailed study of equilibrium and non-equilibrium metal cooling]{capelo17}. They have also pointed out that dust-shielding from ultraviolet (UV) radiation is the main process driving the formation of H$_2$ and suggested that, more likely, the latter is controlled by SF, or, in general, by the gravitational collapse of atomic gas, not vice versa \citep{maclow16}.

Moreover, theoretical models of turbulent, magnetised clouds have tackled the problem of how SF should proceed inside giant molecular clouds (GMCs) \citep{krumholz05,padoan11,hennebelle11} without any assumption about the correlation with H$_2$, and found very good agreement with simulations and observations \citep{federrath12}.

Because of these uncertainties, a clear consensus is still missing, and the observational limits for measuring H$_2$, currently based on the abundance of CO, in particular for low-metallicity dwarf galaxies where CO appears to be under-abundant \citep{schruba12}, prevent us from fully disentangling causes and consequences. 

In view of this, a key role seems to be played by gas shielding from UV radiation. However, most of the simulations up to date do not consistently take into account local radiation from stellar sources and its effect on to the interstellar medium, but only consider a uniform extragalactic background 
\citep[e.g.][]{fiacconi15,kim16}. However, a correct inclusion of this effect requires very expensive full radiative transfer (RT) calculations. Several studies have started to incorporate it into numerical simulations, some with on-the-fly RT \citep{gnedin09}, and most of the others with sub-grid prescriptions \citep{christensen12,richings16,tomassetti15,hu16,hu17,forbes16}.

In this study, we present and test a model to self-consistently account for the formation and dissociation of H$_2$, via the chemistry package \textsc{krome} \citep{grassi14}, including SF, supernova (SN) feedback, and extragalactic and local stellar radiation. We consider the attenuation of the radiation via  dust and H$_2$ shielding (in \textsc{krome}), and shielding from gas via two different sub-grid models and {on-the-fly} RT calculations, to investigate the emergence of the correlation between H$_2$ and SF.

The paper is organised as follows: in Section~\ref{sec:setup}, we describe the model components and the simulation setup; in Section~\ref{sec:tests}, we describe the basic tests of the chemical network with and without photochemistry; in Section~\ref{sec:results}, we present the results of our simulations including all the aforementioned processes; in Section~\ref{sec:conclusions}, we discuss the limits of the model and we draw our conclusions.

\section{Simulation setup}
\label{sec:setup}
Here, we present and describe all the components of our model, which we implement in the code \textsc{gizmo} \citep{hopkins15}. Except for the mesh-less finite mass hydrodynamic scheme and the SN feedback prescription, which are the same as in \citet{lupi17}, all the other sub-grid prescriptions have been significantly revised with more physically motivated models. We also describe the initial conditions of our simulation suite, which are those of an isolated galaxy with structural parameters typical of $z=3$ galaxies.

\subsection{The hydrodynamic code \textsc{gizmo}}
\textsc{gizmo} \citep{hopkins15}, descendant from \textsc{Gadget3}, itself descendant from \textsc{Gadget2} \citep{springel05}, implements a new method to solve hydrodynamic equations, aimed at capturing the advantages of both usual techniques used so far, i.e. the Lagrangian nature of smoothed particle hydrodynamics codes, and the excellent shock-capturing properties of Eulerian mesh-based codes, and at avoiding their intrinsic limits.
The code uses a volume partition scheme to sample the volume, which is discretised amongst a set of tracer `particles' which correspond to unstructured cells. Unlike moving-mesh codes \citep[e.g. \textsc{arepo};][]{springel10}, the effective volume associated to each cell is not defined via a Voronoi tessellation, but is computed in a kernel-weighted fashion. Hydrodynamic equations are then solved across the `effective' faces amongst the cells using a Godunov-like method, like in standard mesh-based codes. In these simulations, we employ the mesh-less finite-mass method, where particle masses are preserved, and the standard cubic-spline kernel, setting the desired number of neighbours to 32.
Gravity is based on a Barnes-Hut tree, as in \textsc{Gadget3} and \textsc{Gadget2}. While the gravitational softening is kept fixed at 80 and 4~pc for dark matter and stars, respectively, for gas we use the fully adaptive gravitational softening, which ensures that both hydrodynamics and gravity are treated assuming the same mass distribution within the kernel, avoiding the main issues arising when two different resolutions are used. The minimum allowed softening, which sets the effective resolution of the simulation, is set to 1~pc, corresponding to an inter-particle spacing of $\sim 1.4$~pc, but this value is reached only in the very high density regions before a burst of SF occurs.

\subsection{Star formation}
\label{sec:sf}
SF is implemented using a stochastic prescription, where the SFR is defined as
\begin{equation}
\dot{\rho}_{\rm SF} = \varepsilon \frac{\rho_{\rm g}}{t_{\rm ff}},
\end{equation}
with $\varepsilon$ the SF efficiency parameter, $\rho_{\rm g}$ the local gas density, $t_{\rm ff}=\sqrt{3{\rm \pi}/(32 G \rho_{\rm g})}$ the free-fall time, and $G$ the gravitational constant. 
The commonly used prescription for SF assumes a constant value for $\varepsilon$, usually a few per cent, which is calibrated to reproduce the local Schmidt--Kennicutt law 
\citep[hereafter KS;][]{schmidt59,kennicutt98}.

However, such a simplistic assumption is unable to capture the clustered SF in GMCs, resulting in a more diffuse SF across the entire galaxy. 

Simulations of star-forming regions \citep{federrath09,federrath10} have demonstrated that the usual assumption of a constant SF efficiency used in numerical simulations does not correspond to what really occurs in GMCs, where turbulence can locally trigger very efficient SF while keeping the global GMC stable to collapse. \citet{federrath12} compared different theoretical models of SF in GMCs with both numerical simulations of turbulent clouds and observations and showed that the multi-free-fall version of the model by \citeauthor{padoan11} (\citeyear{padoan11}; hereafter PN) exhibits the best agreement with both observations and simulations. 
In this study, we implement the PN thermo-turbulent model in \textsc{gizmo}, assuming that our gas particles represent an entire GMC. The SF efficiency is defined by three parameters describing the degree of turbulence in a GMC: $\alpha_{\rm vir} = 2E_{\rm kin}/|E_{\rm grav}|$,\footnote{We use an approximate value for $\alpha_{\rm vir}$, which can be obtained under the assumption of a spherical homogeneous cloud with diameter $L$, i.e. $\alpha_{\rm vir} = 5\sigma_{\rm v}^2L/(6GM_{\rm cloud})$, where $\sigma_{\rm v} = L||{\nabla \otimes \mathbf{v}}||$ is the velocity dispersion in the cloud and $M_{\rm cloud}$ is the cloud mass. In the simulations, $L$ is computed as the grid-equivalent cell size, i.e. $L=\left[4\pi/(3N_{\rm ngb})\right]^{1/3}h \approx 0.5 h$ with $h$ the particle smoothing length and $N_{\rm ngb}=32$ the number of neighbours used in the kernel estimate, and $M_{\rm cloud}$ is  the particle mass.} where $E_{\rm kin}$ is the turbulent kinetic energy and $E_{\rm grav}$ the gravitational energy of the GMC; the Mach number $\mach$; and the turbulent parameter $b$ describing the ratio between compressive and solenoidal turbulence. For a statistical mixture of solenoidal and compressive turbulence, the typical value is $b=0.4$ \citep{federrath10}. The PN model assumes that the density distribution inside a GMC follows a log-normal probability density function (PDF) with mean density $\rho_0$ and variance $\sigma_s^2=\ln(1+b^2\mach^2)$ and that SF only occurs above a density $\rho_{\rm crit}/\rho_0=0.067\theta^{-2}\alpha_{\rm vir}\mach^2$, where $\theta=0.97 \pm 0.10$ corresponds to the thickness of the post-shock layer in the turbulent GMC in units of the cloud size. 
With these parameters, we can compute the mean SF efficiency of the cloud by integrating the actual SF efficiency over the entire PDF, obtaining
\begin{equation}
\varepsilon=\frac{\varepsilon_\star}{2\phi_{\rm t}}\exp\left({\frac{3}{8}\sigma_s^2}\right)\left[1+{\rm erf}\left({\frac{\sigma_s^2-s_{\rm crit}}{\sqrt{2\sigma_s^2}}}\right)\right],
\end{equation} where $\varepsilon_\star=0.5$ is the local SF efficiency to match observations \citep{heiderman10}, $1/\phi_{\rm t}=0.49 \pm 0.06$ is a fudge factor to take into account the uncertainty in the free-fall time-scale, and $s_{\rm crit}= \ln{(\rho_{\rm crit}/\rho_0)}$.

Compared to the standard prescription used, which needs to be calibrated, this model is more tied to the simulation, since all the information required to compute $\varepsilon$ can be obtained from the gas properties themselves, using the kernel-weighting scheme available in the code. However, it cannot be used at arbitrarily low resolution, but only when single GMCs can be resolved with at least a few resolution elements.

In this study, we allow SF when gas matches two criteria: i) $\rho_{\rm g}> \rho_{\rm SF,thr}$, where we assume that $\rho_{\rm g}=\rho_0$ in the SF model, and ii) $\mach>2$. The first criterion is not strictly necessary, but it is useful to prevent SF in very low density regions where the kernel weighting would extend to very large scales, totally uncorrelated with the typical sizes of GMCs. The second criterion, instead, is fundamental to limit SF in regions dominated by supersonic turbulence and avoid SF in trans-sonic and sub-sonic ones, where the assumption of a log-normal PDF is not valid anymore.
Note that, unlike other recipes, we do not have an explicit criterion on the gas temperature. Although the criterion on the Mach number, for a fixed velocity dispersion, gives a temperature criterion, we will show in Appendix~\ref{app:sfe} that a natural temperature barrier arises for SF, without the Mach number threshold playing any crucial role.

When the two criteria are matched, we stochastically spawn a new stellar particle with the same mass of the progenitor gas cell mass.

\subsection{Chemistry and radiative cooling}
Chemistry and radiative cooling are computed with the \textsc{krome} chemistry package \citep{grassi14}, solving the non-equilibrium rate equations for nine different species (H, H$^+$, He, He$^+$, He$^{++}$, H$^-$, H$_2$, H$_2^+$, and e$^-$). We use {\it model 1a} in \citet{bovino16},\footnote{We note that there are a few typos in the reaction rates of \citet{bovino16}, and we refer to \citet{capelo17} for details. We also note that we removed the temperature limit at 100 K in reaction 11 of \citet{bovino16}.} which includes photoheating, H$_2$ UV pumping, Compton cooling, photoelectric heating, atomic cooling, H$_2$ cooling, and chemical heating and cooling. Compared to the original network, we also include three-body reactions involving H$_2$ \citep{glover08,forrey13}, H$_2^+$ collisional dissociation by H, and H$^-$ collisional detachment by He \citep{glover09}. We note that these reactions are included for completeness.
In this study, we consider relatively high metallicity gas in the initial conditions, setting $Z = 0.5 \rm\, Z_\odot$.
In these conditions, H$_2$ is expected to mainly form on dust, modelled here assuming a fixed dust-to-gas ratio $D$ linearly scaling with $Z$, i.e. $D/D_\odot = Z/Z_\odot$, where $D_\odot = 0.00934$. 
The H$_2$ formation rate on dust is computed following \citet{jura75} as
\begin{equation}
R_f ({\rm H}_2) = 3 \times 10^{-17}n_{\rm H,tot} n_{\rm tot}Z/Z_\odot C_{\rho}\,\rm cm^{-3}s^{-1},
\label{eq:H2}
\end{equation}
where $n_{\rm H_{tot}}=n_{\rm H}+n_{\rm H^+}+2n_{\rm H_2}+2n_{\rm H_2^+}+n_{\rm H^-}$ is the total hydrogen nuclei number density, $n_{\rm tot} = \rho_{\rm g}/(m_{\rm H}\mu)$ is the total gas number density, with $m_{\rm H}$ the hydrogen mass and $\mu$ the mean molecular weight, and $C_\rho = \langle \rho_{\rm g}^2\rangle/\langle\rho_{\rm g}\rangle^2$ is the clumping factor, accounting for the enhanced H$_2$ formation in the high-density regions unresolved in the simulation. According to \citet{gnedin09}, the value of $C_\rho$ should be in the range $3-10$, although their simulations show that even a value of 100 could not be enough to recover the relation between SFR and molecular gas surface densities. Across the paper, we use two different clumping factor models, a constant $C_\rho=1$, and the case in which the unresolved gas above $\rho_{\rm SF,thr}$ follows the same log-normal PDF used for SF, giving 
\begin{equation}
C_\rho = \exp(\sigma_s^2) = 1+b^2\mach^2.
\label{eq:cfac}
\end{equation}
We also include H$^-$ photodetachment, though its effect should be irrelevant at high metallicity. This is not valid for very low metallicity and will be investigated in a future work.

The dissociation of H$_2$ occurs via two main mechanisms, the excitation to the vibrational continuum of an excited electronic state and the Solomon process \citep[see][]{bovino16}, when dissociating and ionising radiation are present.

To model the ionising radiation, we consider two different contributions: a uniform extragalactic background, following the model by \citet{haardt12}, and the local radiation from young massive stars. We consider eight energy bins, ranging from 0.75 to 1000 eV, to cover the characteristic energies for H$^-$ photodetachment, H$_2$ and H$_2^+$ dissociation, and the different hydrogen and helium ionisation states, following a similar approach to \citet{katz16}. We report the bins in Table~\ref{tab:bins}, with a description of the main processes occurring in every bin.

\begin{table}
\caption{Energy bins for photoheating processes included in the simulations. We use eight bins in the range 0.75--1000 eV to track the different ionisation states of the followed species. Columns 1--2 show the minimum and maximum bin energies, respectively, and column 3 shows the main process occurring in each bin. The second bin is actually split in three sub-bins to improve accuracy.}
\label{tab:bins}
\centering
\begin{tabular*}{\columnwidth}{@{\extracolsep{\stretch{1}}}*{3}{c}@{}}
\hline
$E_{\rm min}$ (eV) & $E_{\rm max}$ (eV) & Process\\
\hline
0.75 & 1.70 & H$^-$ photodetachment \\
1.70 & 11.20 & H$_2^+$ dissociation\\
11.20 & 13.60 & H$_2$ dissociation (Solomon)\\
13.60 & 15.20 & H ionisation\\
15.20 & 24.59 & H + H$_2$ ionisation \\
24.59 & 54.42 & H + He + H$_2$ ionisation\\
54.42 & 100.6 & H + He + He$^+$ + H$_2$ ionisation \\
100.6 & 1000.0 & H + He + He$^+$ + H$_2$ ionisation\\
\hline
\end{tabular*}
\end{table}

However, in reality, the extragalactic background is unable to reach the innermost parts of a galaxy, because of the shielding by the intervening gas. In order to account for this effect, we consider that the UV background is attenuated by a factor $\exp(-\tau)$, where $\tau=\sum_i\sigma_{i,\rm bin} N_i$ is the optical depth for every energy bin, $\sigma_{i,\rm bin}$ is the cross section in the bin for the $i$-th species (pre-computed by \textsc{krome}), and $N_i$ is the $i$-th species column density. We approximate $N_i$ assuming each gas cell as surrounded by gas with similar properties, which gives $N_i  \sim n_i \lambda$, where $n_i$ is the $i$-th species number density and $\lambda$ is the absorption length, which we define equal to the Jeans length $\lambda_{\rm J}$ of the gas cell. This choice is motivated by the results in \citet{safranekshrader17}, where the authors explicitly show that the Jeans length is a better approximation compared to the gradient-based shielding lengths. \footnote{Though better results are obtained when a temperature cap of 40 K to the Jeans length is used, the largest differences are observed in the CO distribution, which is not followed in this study.} In addition, we also consider local radiation from young massive stars, assuming that every stellar particle samples a full stellar population following a \citet{chabrier03} initial mass function (IMF) and using the updated stellar population synthesis models by \citet[][hereafter BC03; see Section~\ref{sec:feedback} for the model details]{bruzual03}. At high densities, gas self-shields from radiation, thus we need to include self-shielding by H$_2$ and dust to properly track the abundances in the high-density regions without overestimating the effective radiative flux. Both terms are implemented as a sub-grid recipe following \citet{richings14}.

Metal cooling rates, instead, are provided by look-up tables pre-computed with the photoionisation code \textsc{Cloudy} \citep{ferland13} by \citet{shen10,shen13}, as a function of density, temperature, and redshift, including the extragalactic UV background by \citet{haardt12}. Metal cooling rates are provided assuming solar abundances and then linearly rescaled with the total metallicity, which is followed as a passive scalar.

Cooling is not allowed below 10~K, to avoid inconsistencies in the cooling rates, which are not well defined below this temperature.

\subsection{Supernova and wind feedback}
\label{sec:feedback}
Since we cannot resolve single stars in our simulations, we assume that our stellar particles correspond to an entire stellar population, following a \citet{chabrier03} IMF. According to stellar evolution, we consider three different processes: type II supernovae (SNe), type Ia SNe, and low-velocity winds by stars in the range $1-8\rm\, \msun$.
\begin{itemize}
\item After $\sim 4$~Myr, the most massive stars start to explode as type II SNe. For stars with mass between $8$ and $40\rm\, \msun$, we release $E_{\rm SN}=10^{51} \rm erg/SN$ via thermal injection on to the gas cells enclosed in the stellar particle kernel sphere,\footnote{For the stellar particles, we define the kernel size using 64 neighbours, in order to guarantee that the volume around the star is fully covered.} together with mass and metals. Due to the limited resolution, the energy-conserving phase of the bubble expansion cannot be followed in the simulation and the additional energy released would be rapidly lost because of radiative cooling, resulting in an ineffective SN feedback. In order to avoid gas from rapidly getting rid of this additional energy, we implement a delayed-cooling prescription, where gas cooling is inhibited for the time needed to reach the momentum-conserving phase \citep{stinson06}. In particular, we shut off cooling for the survival time of the blast-wave $t_{\rm max} = 10^{6.85}E_{51}^{0.32} n_{\rm H_{tot}}^{0.34} \tilde{P}_{04}^{-0.70}$ yr, where $E_{51}$ is the released energy in units of $10^{51}$ erg and $\tilde{P}_{04}=10^{-4}P_0k_{\rm B}^{-1}$, with $P_0$ the ambient pressure and $k_{\rm B}$ the Boltzmann constant. This model has been proven to better reproduce the KS relation and the outflow mass-loading factor in isolated galaxy simulations compared to other more physically motivated models, as stated by \citet{rosdahl17}. The only limitation is that it produces `unphysical' high temperatures in a region of the density--temperature diagram where cooling is expected to be effective. We assume that type II SNe return all their mass but $1.4\rm\, \msun$ and we follow metal production (via iron and oxygen yields whose production rates are thought to be metallicity independent) using the tabulated results by \citet{woosley07} fitted by \citet{kim14}. The total metal mass can then be defined as:
\begin{equation}
M_{\rm Z} = 2.09M_{\rm oxygen} + 1.06M_{\rm iron},
\end{equation}
where $M_{\rm oxygen}$ and $M_{\rm iron}$ are oxygen and iron mass, respectively, computed convolving the yield function with the IMF.
For stars above $40\rm\, \msun$, we assume a black hole (BH) would form via direct collapse (although we do not simulate sink formation), thus we release neither mass nor metals.
\item Type Ia SNe occur in evolved binary systems, when one of the stars has become a white dwarf and has accreted enough mass to exceed the Chandrasekhar mass limit \citep{chandrasekhar31}. Type Ia SNe explode according to a distribution of delay times, as described by \citet{maoz12}, scaling as $t^{-1}$ between 0.1 and 1 Gyr after a burst of SF. Type Ia SNe leave no remnant and release to the environment $1.4\rm\, \msun$, $M_{\rm iron}=0.63\rm\, \msun$ and $ M_{\rm oxygen}=0.14\rm\, \msun$. Since type Ia SNe occur a long time after the burst of SF, these events are usually located far away from the progenitor molecular cloud, and they are not clustered as type II SNe. In this case, we do not shut off cooling, as described in \citet{stinson06}.  
\item Because of the low mass, stars between 1 and 8 $\rm \msun$ evolve on longer time-scales compared to their massive counterparts and do not explode as SNe. During their evolution, however, they release part of their mass as stellar winds. We model stellar winds by injecting only mass and metals on to the neighbouring cells of the stellar particle. Assuming the initial-final mass relation for white dwarfs by \citet{kalirai08}, the mass loss for these stars can be computed as $w_{m}=0.394 + 0.109m\rm\, \msun$, where $m$ is the mass of the star. Also in this case, we convolve $w_m$ with the assumed IMF to compute the net mass loss as a function of stellar age. We assume that stellar winds only carry the progenitor star metallicity, hence they do not enrich the interstellar medium.
\end{itemize}

By taking into account these processes, the stellar feedback in our simulations is able to return $42$ per cent of the stellar particle mass in a Hubble time, prolonging star formation even in cases of no fresh gas inflows \citep[e.g.][]{ciotti91,leitner11,voit11}.

\subsection{Stellar radiative feedback}
\label{sec:radiation}
In addition to the previously described processes, we also include radiative feedback from the stellar population. This form of feedback is particularly important for young stars, because of the strong ionising and dissociating flux produced, which affects the abundance of molecular gas. We assume that every stellar particle spawned in the simulation corresponds to an instantaneous burst of SF and we model the evolution of the stellar luminosity $L_\star$ with age and metallicity using the BC03 models.  
We consider here two different sub-grid models accounting for this contribution, to compare them with on-the-fly RT calculations and identify the most accurate to be used in cosmological simulations, where RT is still prohibitive.

In model \textit{(a)}, we loop over all the active gas cells, collecting all the stars within a maximum distance $R_{\rm max}$ from the cell, defined as the critical radius at which the surrounding medium would become optically thick, i.e. the optical depth $\tau \sim 1$. In order to compute this radius, we consider the local cell properties and integrate the gas density $\rho_{\rm g}(R)$ outwards along the direction $R$ of maximally decreasing density, obtaining
\begin{equation}
\label{eq:tau}
\tau = \frac{\sigma_{\rm eff}}{m_{\rm H}} (\rho_{\rm g}R_{\rm max} - |\nabla\rho_{\rm g}|R_{\rm max}^2/2),
\end{equation}
where $\sigma_{\rm eff}$ is the minimum cross section for the photoionisation processes considered and $\nabla\rho_{\rm g}$ is the density gradient. Although $\sigma_{\rm eff}$ is different for every process considered, we approximate it using a unique value $\sigma_{\rm eff}=10^{-20}\,\rm cm^2$, small enough to avoid an unphysical underestimation of $R_{\rm max}$, and at the same time large enough to avoid looking for all the stars in the galaxy. Among the two solutions of Eq.~\eqref{eq:tau}, we take the smallest $R_{\rm max}$. When a solution to the equation cannot be found, we assume $R_{\rm max}={\rm max}(5h,\rho_{\rm g}/|\nabla\rho_{\rm g}|)$, where $h$ is the kernel size of the gas cell and $\nabla\rho_{\rm g}$ is the gas density gradient around it.\footnote{We choose the factor 5 because it represented the best compromise between limiting the computational overhead and collecting radiation from large enough distances.} After collecting the stellar particles within $R_{\rm max}$, we compute the total radiative flux in each bin which reaches the target gas cell as 
\begin{equation}
F = \sum_i \frac{L_{\star,i}}{4\pi d_{i}^2} \exp{(-\tau_{i})},
\end{equation}
where $L_{\star,i}$ and $d_{i}$ are the luminosity and the distance of the $i$-th stellar particle, respectively, and $\tau_i$ is the optical depth of the intervening gas between the cell and the source. Since we cannot estimate the real optical depth between the cell and the source, we approximate $\tau_i$ using the average species abundances around the stellar particle and an absorption length defined as the minimum between $d_i$ and $l_{\rm Sob} = \rho_\star/|\nabla\rho_\star|$, where $\rho_\star$ and $\nabla\rho_\star$ are the gas density and the gas density gradient around the star, respectively. This model can be considered an improved version of that in \citet{christensen12}, where shielding was not considered. 

In model \textit{(b)}, instead, we benefit from the tree structure used for gravity computation, storing the total luminosity and the luminosity-weighted centre of the stellar sources in every node, as in \citet{hopkins14,hopkins17}. The stellar luminosity is attenuated at the source using the Sobolev approximation, giving $L_{\rm eff,\star} = L_\star \exp{(-\tau)}$, where $\tau$ is defined as in model \textit{(a)}. Then, during the gravity force calculation, we compute the total flux $f$ reaching a gas cell collecting the contribution of all the stellar particles in the simulated volume. Finally, we attenuate the resulting flux at the absorption point using, as for the extragalactic UV background, the Jeans length of the cell.\footnote{As \citet{hartwig15} pointed out, the Jeans length tends to overestimate the H$_2$ column densities ($N_{\rm H_2}$) by up to an order of magnitude for $N_{\rm H_2}\lesssim 10^{16}\rm\, cm^{-2}$ and to underestimate it for larger $N_{\rm H_2}$. Hence, we expect our H$_2$ densities to be slightly higher than in reality. Nevertheless, since a simple correction factor is not available, it represents a simple and reasonable approximation that avoids time-consuming calculations.} 

\subsection{Radiative transfer with \textsc{gizmo}}
For the RT simulations, we employ the momentum-based RT method with the M1 closure scheme \citep{levermore84}. The implementation directly follows that of \citet{rosdahl13}, suitably modified for a mesh-less structure.\footnote{A detailed description of how to implement radiative transfer methods in mesh-less codes can be found at \url{https://physastro.pomona.edu/wp-content/uploads/2016/07/David-Khatami-Final-Thesis.pdf}.} In particular, the Riemann problem is solved using the Harten, Lax and van Leer (HLL) solver \citep{harten83}, which better models beams and shadows compared to the simpler and more diffusive Global Lax Friedrichs \citep{lax54} function (obtained by setting the HLL eigenvalues to $\pm c$, where $c$ is the speed of light in vacuum), but increases the asymmetry of the ionisation front around isotropic sources.
The RT calculations are performed using an operator-splitting approach in three steps: injection, transport, and thermochemistry.
During the injection step, the radiative energy emitted by the stellar source during the current time-step is distributed among the cells within the kernel volume of the stellar particle, as done for SN feedback.
During the transport step, photon-advection is performed by solving the Riemann problem among the interacting cells, as done for the hydrodynamics, and finally, for the thermochemistry step, we couple the code with \textsc{krome}, computing the radiative flux in a cell (for each energy bin) as
\begin{equation}
F= \frac{c N_\gamma}{h^3},
\end{equation}
where $N_\gamma$ is the number of photons per bin in the cell.
The energy bins (see Table~\ref{tab:bins}) and stellar spectra used in the RT simulation are the same used for the sub-grid models {\it (a)} and {\it (b)}.

The time-step for the RT calculations is constrained by the Courant condition
\begin{equation}
\Delta t_{\rm RT} \leq C_{\rm CFL} \frac{h}{c},
\end{equation}
where $C_{\rm CFL}=0.2$ is the Courant factor, and $h$ the particle/cell size.
Compared to the common hydrodynamic case (where $v_{\rm max} \sim 1000\rm\, km s^{-1}$), the time-step condition for RT can be hundred--to--thousand times more severe, making the simulations much more computationally expensive.
The simplest solution to this issue is the {\it reduced speed of light approximation}, introduced by \citet{gnedin01}, where the Courant condition is relaxed by replacing c with $\tilde{c} = f_{\rm c} c$. The exact value of $f_{\rm c}$ depends on the boundary conditions of the problem and the associated typical time-scales and can be expressed as
\begin{equation}
f_{\rm c} = \min(1;\sim 10 t_{\rm cross}/\tau_{\rm sim}),
\end{equation}
where $t_{\rm cross} = r_{\rm S}/c$, with $r_{\rm S}=[3\dot{N}/(4\pi\alpha^B n_{\rm H}^2)]^{1/3}$ the Str\"{o}mgren (\citeyear{stromgren39}) sphere radius, $\dot{N}$ the ionising photons injection rate, and $\alpha^B\sim 2.6\times 10^{-13}\rm\, cm^3s^{-1}$ the case-B recombination rate at $T\sim 10^4$~K. $\tau_{\rm sim}$  is the shortest relevant time-scale of the simulation.
In our simulations, we assume that the typical radiation source is a young massive star ($\dot{N} \sim 1-5 \times 10^{48}\rm\, s^{-1}$), surrounded by high-density gas ($n_{\rm H} \gsim 10\rm\, cm^{-3}$), and we require $\tau_{\rm sim}\sim 1$~Myr to resolve the stellar evolution of these massive stars. Putting everything together, we obtain a minimum value of $f_{\rm c} \sim 3\times 10^{-4}$. However, in our simulations, we assume $f_{\rm c} = 10^{-3}$ as a good compromise between higher accuracy and larger computational cost.

For an exhaustive description of the radiative transfer implementation in \textsc{gizmo} and numerical tests, we refer to Hopkins et al. (in preparation).

\begin{figure*}
\centering
\includegraphics[width=0.4\textwidth]{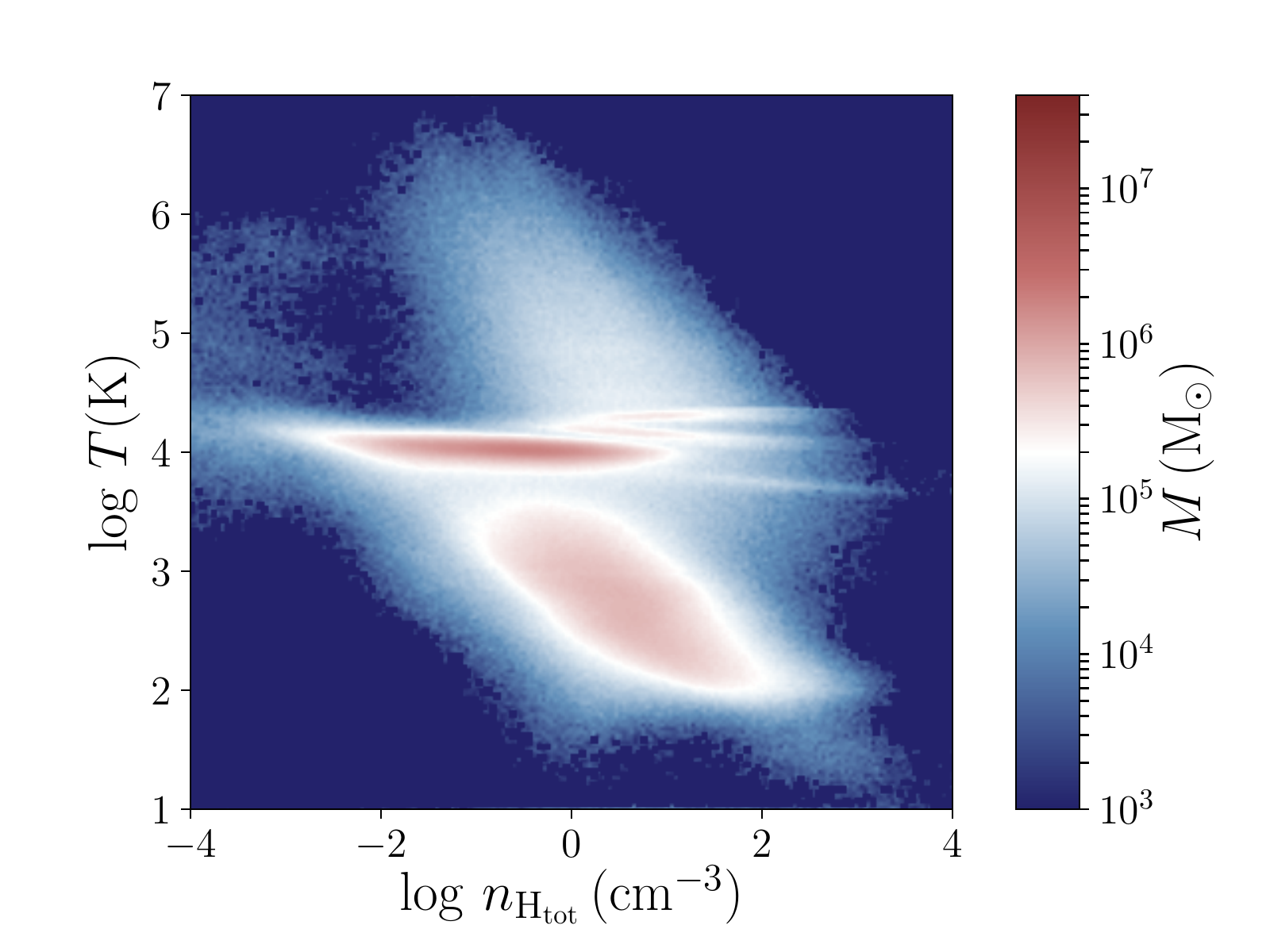}
\includegraphics[width=0.4\textwidth]{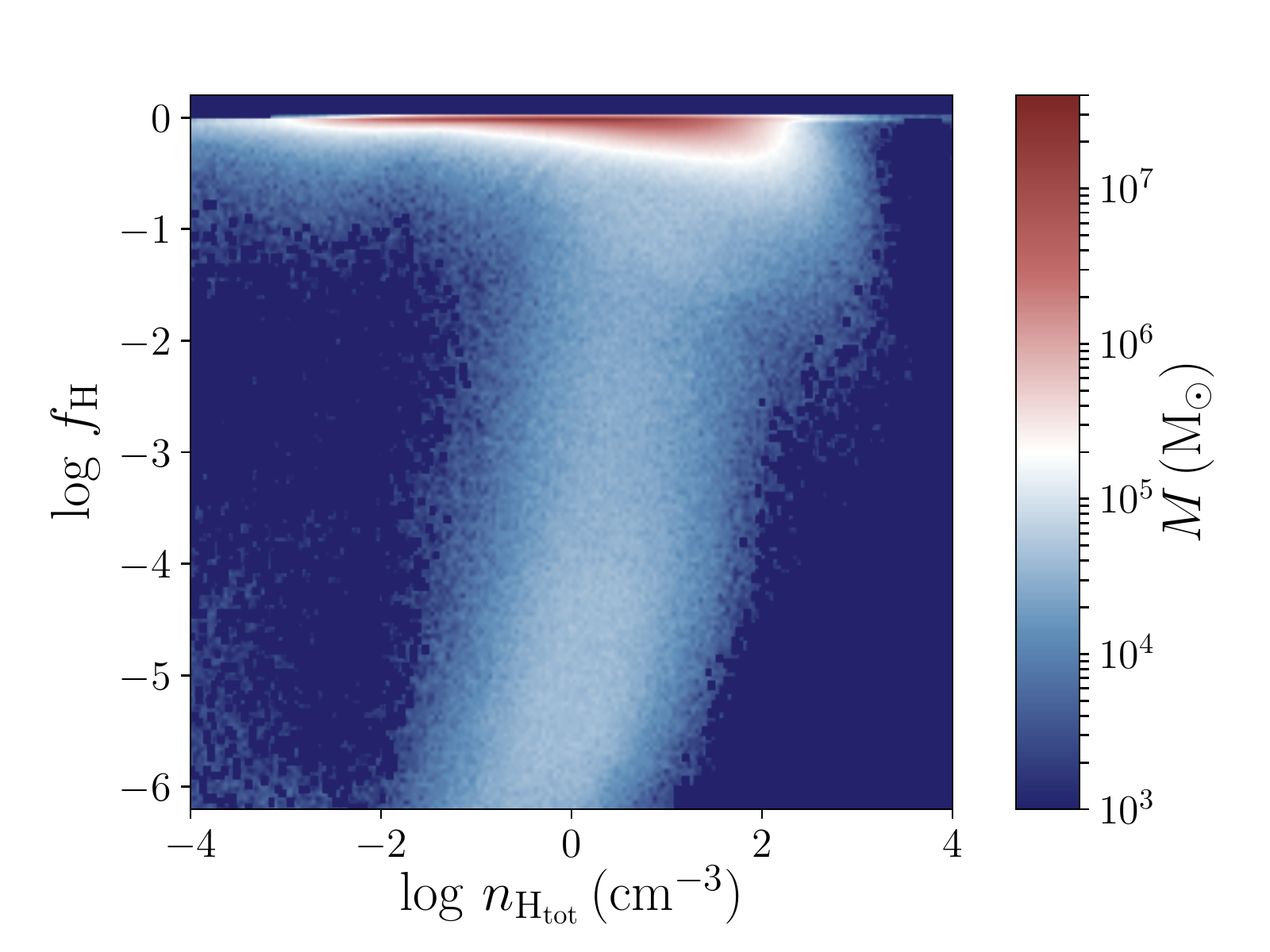}\\
\includegraphics[width=0.4\textwidth]{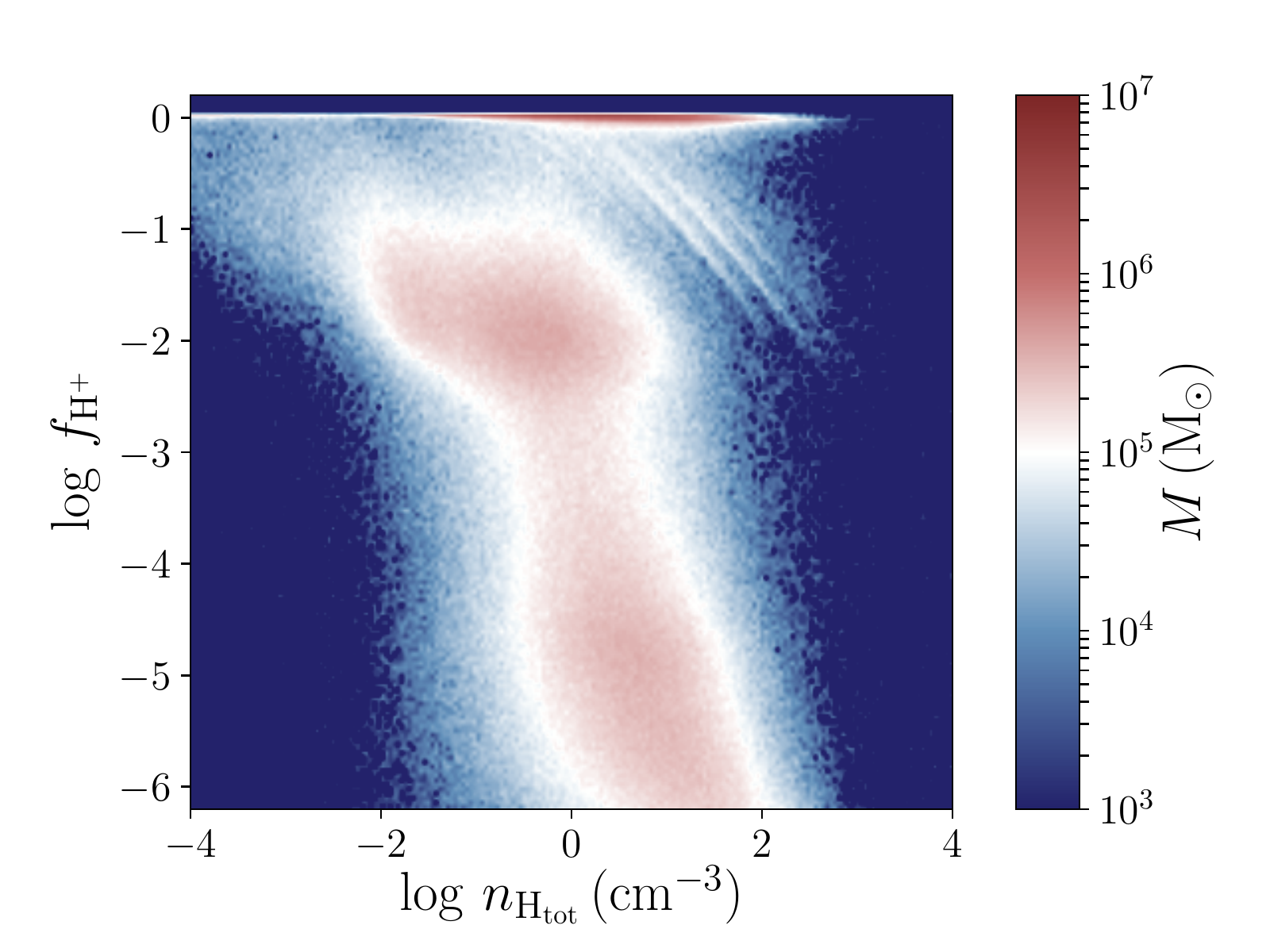}
\includegraphics[width=0.4\textwidth]{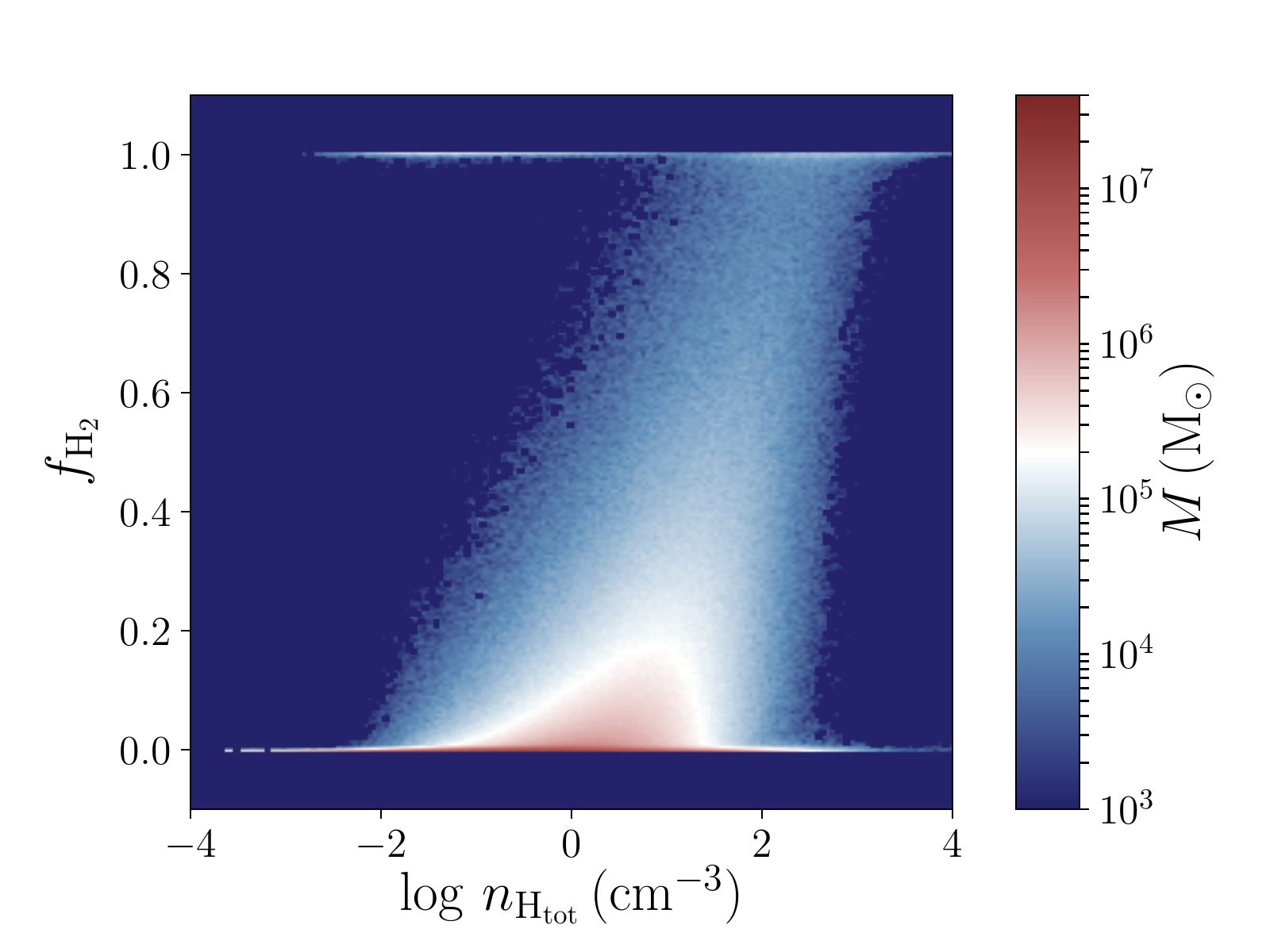}
\caption{Average gas properties for the \gmol{} run around $t=$ 350~Myr. The top-left panel shows the gas distribution in the density--temperature plane. The other three plots, instead, report the neutral ($f_{\rm H}$), ionised ($f_{\rm H^+}$), and molecular hydrogen ($f_{\rm H_2}$) mass fractions as a function of the total hydrogen density $n_{\rm H_{tot}}$. Although most of gas above 100~cm$^{-3}$ is fully molecular as expected, high values of $f_{\rm H_2}$ are also visible at lower densities, because of the missing dissociating radiation.}
\label{fig:noUVp}
\end{figure*}

\subsection{Initial conditions}
In this study, we follow the evolution of an isolated galaxy at $z = 3$. The galaxy is a composite system of a dark matter (DM) halo, a gaseous and stellar disc, and a stellar bulge. The DM halo is described by a spherical Navarro--Frenk--White \citep[NFW;][]{navarro96} density profile within the virial radius, $R_{\rm vir}$, and by an exponentially decaying NFW profile outside $R_{\rm vir}$, as described in \citet{springel99}. The stellar and gaseous disc is described by an exponential surface density profile and by an isothermal sheet \citep{spitzer42}, whereas the stellar bulge is described by a spherical \citet{hernquist90} profile. The galaxy has a virial mass $M_{\rm vir} = 2 \times 10^{11}$~M$_{\odot}$ \citep{adelberger05} and a virial radius $R_{\rm vir} = 45$~kpc, with 2 and 0.4 per cent of its virial mass in the disc and bulge, respectively. The mass fraction of gas in the disc is 60 per cent, typical of high-redshift galaxies \citep{tacconi10}. The modelled system is similar in structure to the main galaxy in the gas-rich galaxy merger simulated in \citet{capelo15}. This same galaxy model has also been used for a study on the effects of non-equilibrium metal (C, O, and Si) cooling \citep{capelo17}.

\begin{table}
\caption{Description of the full simulation suite. We show in the first column the name of the run and in the second the presence or not of the extragalactic UV background. The third column reports the model used (if any) for stellar radiative feedback (see Section~\ref{sec:radiation}) and the last one the clumping factor $C_{\rho}$ used in the run.}
\label{tab:suite}
\begin{tabular*}{\columnwidth}{@{\extracolsep{\stretch{1}}}*{1}{l}*{3}{c}@{}}
\hline
Run & UV background & Rad. feedback & $C_\rho$\\
\hline
\gmol & no & / & 1\\
\vspace{0.2cm}
\guv & yes & / & 1\\

\vspace{0.2cm}
\grt\_1 & yes & RT & 1\\

\grads & yes & model (a) & Variable\\
\gradt & yes & model (b) & Variable\\
\grt & yes & RT & Variable\\
\hline
\end{tabular*}
\end{table}
The mass resolution is $1594\,\msun$ for gas and stellar particles, and $1.97\times 10^5\,\msun$ for DM particles, corresponding to a total number of particles in the initial conditions of 1.5M in gas, 1.5M in DM, 1M in the stellar disc, and 500k in the stellar bulge. 

At the beginning of the simulations, the instantaneous cooling of the gas in the central regions of the galaxy would result in the vertical collapse of the disc and into an initial burst of SF. To prevent this spurious effect, we shut off SF for the first 10~Myr and let pre-existing stars explode as SNe, assuming their stellar ages are uniformly distributed in the range [0--2] Gyr. The SF density threshold is set to a total hydrogen density $n_{\rm H_{tot}}=1 \rm\, cm^{-3}$ to avoid SF in very low density regions where the kernel size would extend to very large scales uncorrelated with typical GMC sizes, as stated in Section~\ref{sec:sf}.
The galaxy is evolved for 400~Myr, a time long enough to follow the evolution after the initial relaxation, but still sufficiently short to allow us to neglect environmental effects like accretion flows and mergers.
We run a suite of six simulations where we include different sub-grid prescriptions which affect the evolution of molecular gas, to assess the relative contribution of the different processes and to assess the best sub-grid model for stellar radiative feedback which better matches RT calculations. The full simulation suite is reported in Table~\ref{tab:suite}.

\begin{figure*}
\centering
\includegraphics[width=0.66\textwidth,trim={0cm 1cm 0cm 2cm},clip]{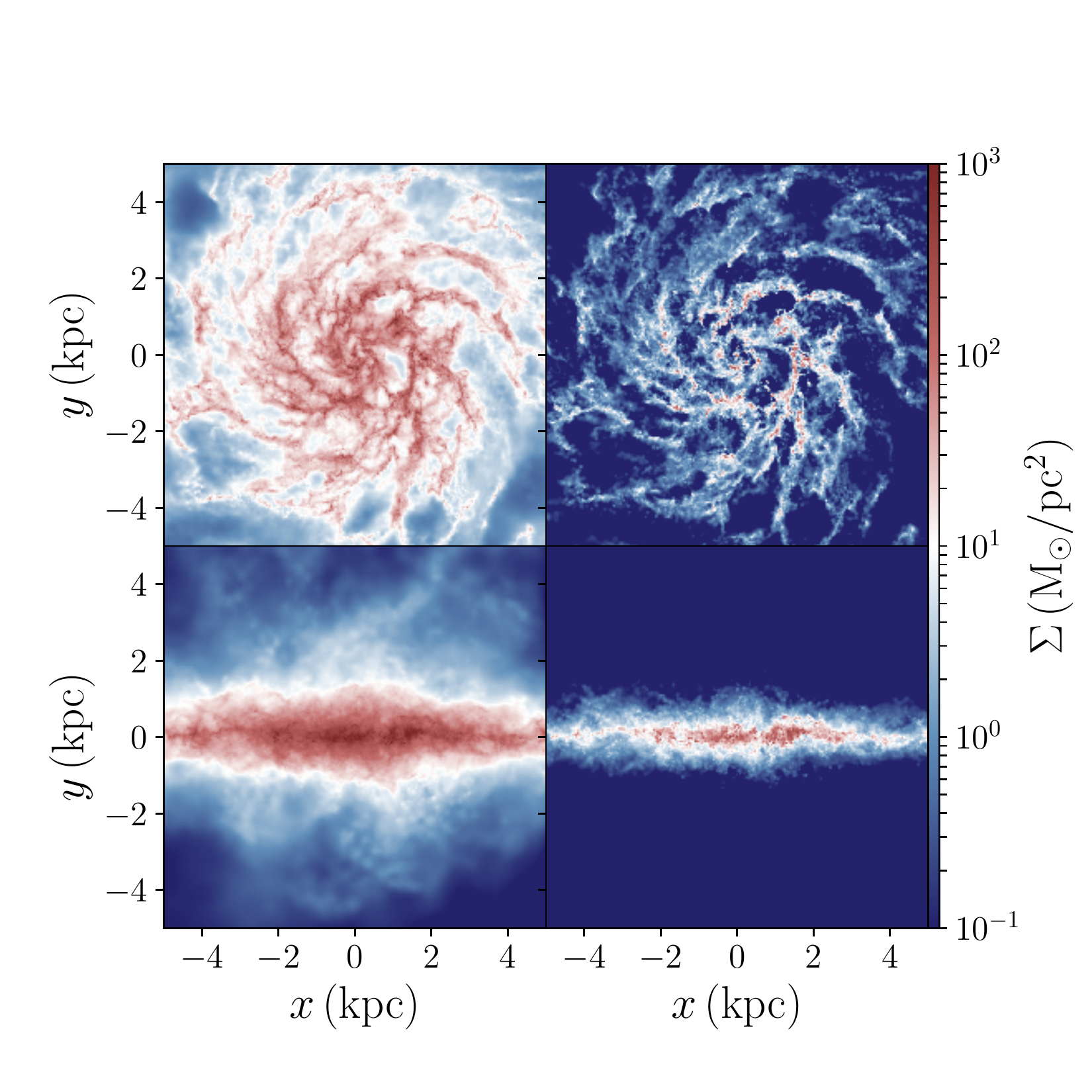}
\caption{Projected gas density maps for our \gmol{} run at $t=350$~Myr, face-on (top panels) and edge-on (bottom panels). The left-hand panels show the total hydrogen density and the right-hand panels the  H$_2$ component only. Molecular gas forms almost everywhere in the galaxy, where the gas is cold enough. The absence of the UV background allows H$_2$ to form also at low densities, unless the gas is heated up by SNe.}
\label{fig:noUVmap}
\end{figure*}

\section{Tests of the chemistry network}
\label{sec:tests}
We discuss here the results of the runs \gmol{} and \guv, aimed at testing the coupling between \textsc{gizmo} and the chemistry package \textsc{krome}.

\subsection{Test 1: non-equilibrium chemistry in \textsc{gizmo} and molecular gas evolution}
\label{sec:noUV}
We describe here the results of \gmol, where we follow the evolution of molecular gas when only collisional dissociation/ionisation and recombination processes are considered. In this case, we expect the cold gas to always be fully molecular, even at low densities. On the other hand, the only available sources for gas heating are SNe and adiabatic compression. All the analyses at a fixed time are carried out at $t=350$~Myr and not at the final time of the simulation, to show the galaxy properties at the median snapshot used for the SF law, obtained by stacking over a period of 100~Myr ($300\leq t\leq 400$~Myr) around $t=350$~Myr.

In Fig.~\ref{fig:noUVp}, we report the gas properties averaged over 100~Myr around $t=350$~Myr, using logarithmic bins 0.04 decade wide on both axes, except for the bottom-right plot, where we use linear bins of width 0.05 on the y-axes.
In the top-left panel, showing the density--temperature diagram of the gas, we observe that the gas clusters in two distinct regions, one around $T\sim 10^4$~K, extending from $n_{\rm H_{tot}} \sim 0.01\rm\, cm^{-3}$ up to $n_{\rm H_{tot}} \sim 10\rm\, cm^{-3}$, and the other in the range 100~$\lesssim T\lesssim$~5000~K, for $n_{\rm H_{tot}} \gsim 1\rm\, cm^{-3}$. The first one corresponds to the low to intermediate density gas which was heated up by SNe and has cooled down via Ly$\alpha$ and metal cooling to the critical temperature where molecular gas cooling and metal cooling start to dominate the cooling rate. The second, instead, is the region where gas efficiently cools down to the very low temperatures where SF occurs. Finally, we also observe a high-temperature tail ($T\gsim3\times10^4$~K)  for $0.01 \lesssim n_{\rm H_{tot}} \lesssim 100\rm\, cm^{-3}$, corresponding to the gas heated up by SNe and for which cooling is still disabled. 

The other three panels in Fig.~\ref{fig:noUVp}, instead, report the hydrogen species abundances, i.e. the neutral hydrogen $f_{\rm H}$, ionised hydrogen $f_{\rm H^+}$, and molecular hydrogen $f_{\rm H_2}$ mass fractions (normalised to the total hydrogen nuclei abundance) as a function of the total hydrogen density $n_{\rm H_{tot}}$. We clearly see that most of the gas is neutral/ionised, and only a small fraction becomes molecular. However, the apparent necessary condition to become fully molecular is the low temperature, independent of the density, as clearly visible in the bottom-right panel. We also find that a molecular mass fraction of 10 per cent can already be obtained at fairly low densities around few cm$^{-3}$, although the transition to fully molecular gas is observed at $n_{\rm H_{tot}}\sim 100\rm\, cm^{-3}$.

\begin{figure*}
\centering
\includegraphics[width=1\textwidth,trim={0cm 0.3cm 0cm 1cm},clip]{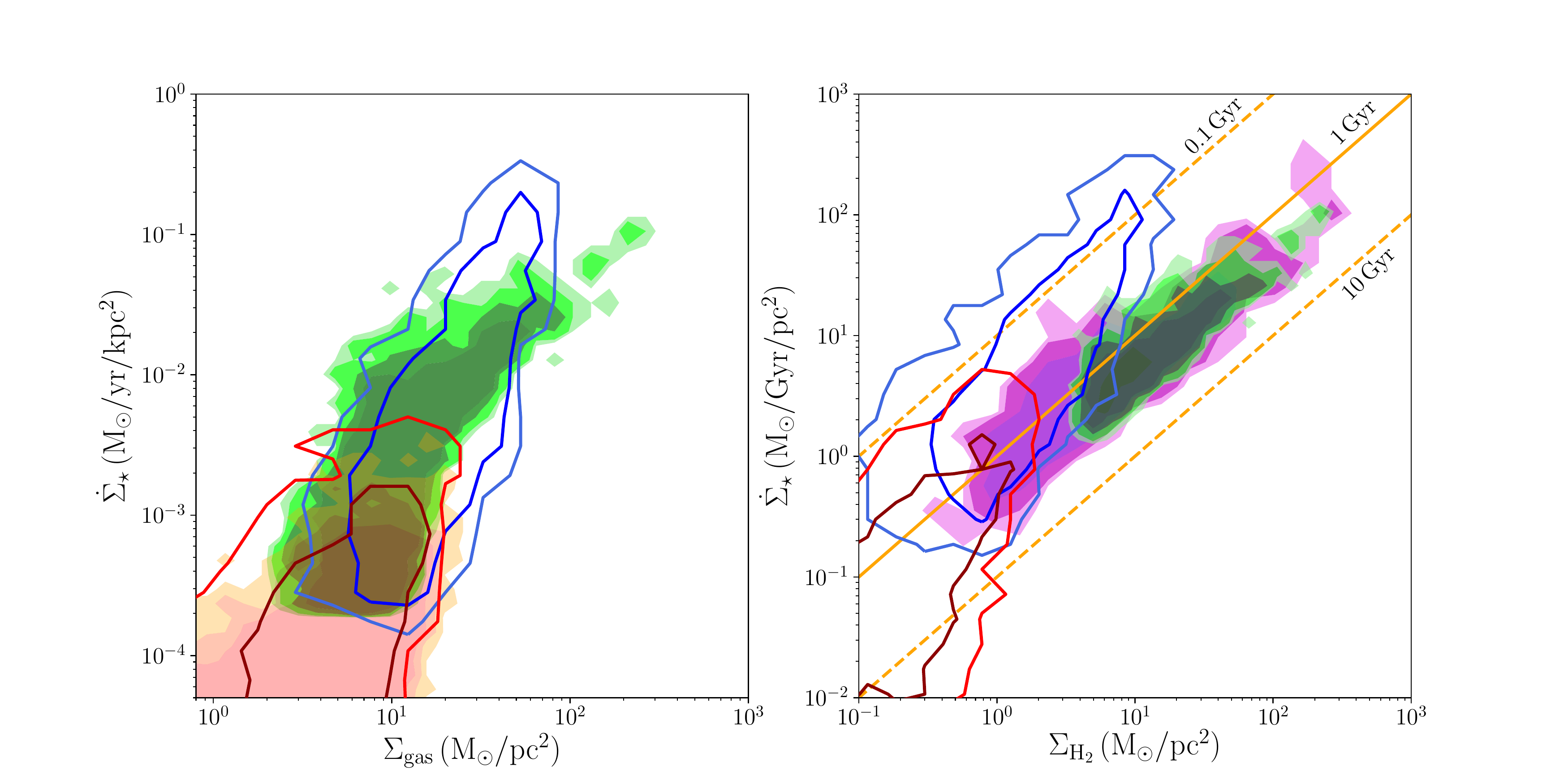}
\caption{SF law for our \gmol{} run in the last 100~Myr of the simulation, plotted as blue and red line contours. We show the relation for the total (H+H$_2$) gas in the left-hand panel and that for H$_2$ only in the right-hand panel and compare them with the observational data by \citet[green and red]{bigiel10} and \citet[purple]{schruba11}. The orange lines in the right-hand panel correspond to 0.1, 1, and 10~Gyr depletion times (defined as $\dot\Sigma_{\star}/\Sigma_{\rm H_2}$). Our data well match the observational data in total gas, whereas in H$_2$ the slope is steeper, with the contours extending to depletion times of 0.1~Gyr also at moderately low densities of about 1~$\rm \msun/pc^2$.}
\label{fig:noUVks}
\end{figure*}

In Fig.~\ref{fig:noUVmap}, we show the projected gas density maps of the galaxy in the face-on (top panels) and edge-on (bottom panels) views at $t=350$~Myr, where the left-hand panels correspond to the total hydrogen gas and the right-hand ones to the H$_2$ component only. The molecular gas density is higher in the higher density spiral arms of the galaxy, but a not negligible fraction is also observed in the low-density medium, in agreement with the results in Fig.~\ref{fig:noUVp}.

\begin{figure*}
\centering
\includegraphics[width=0.4\textwidth]{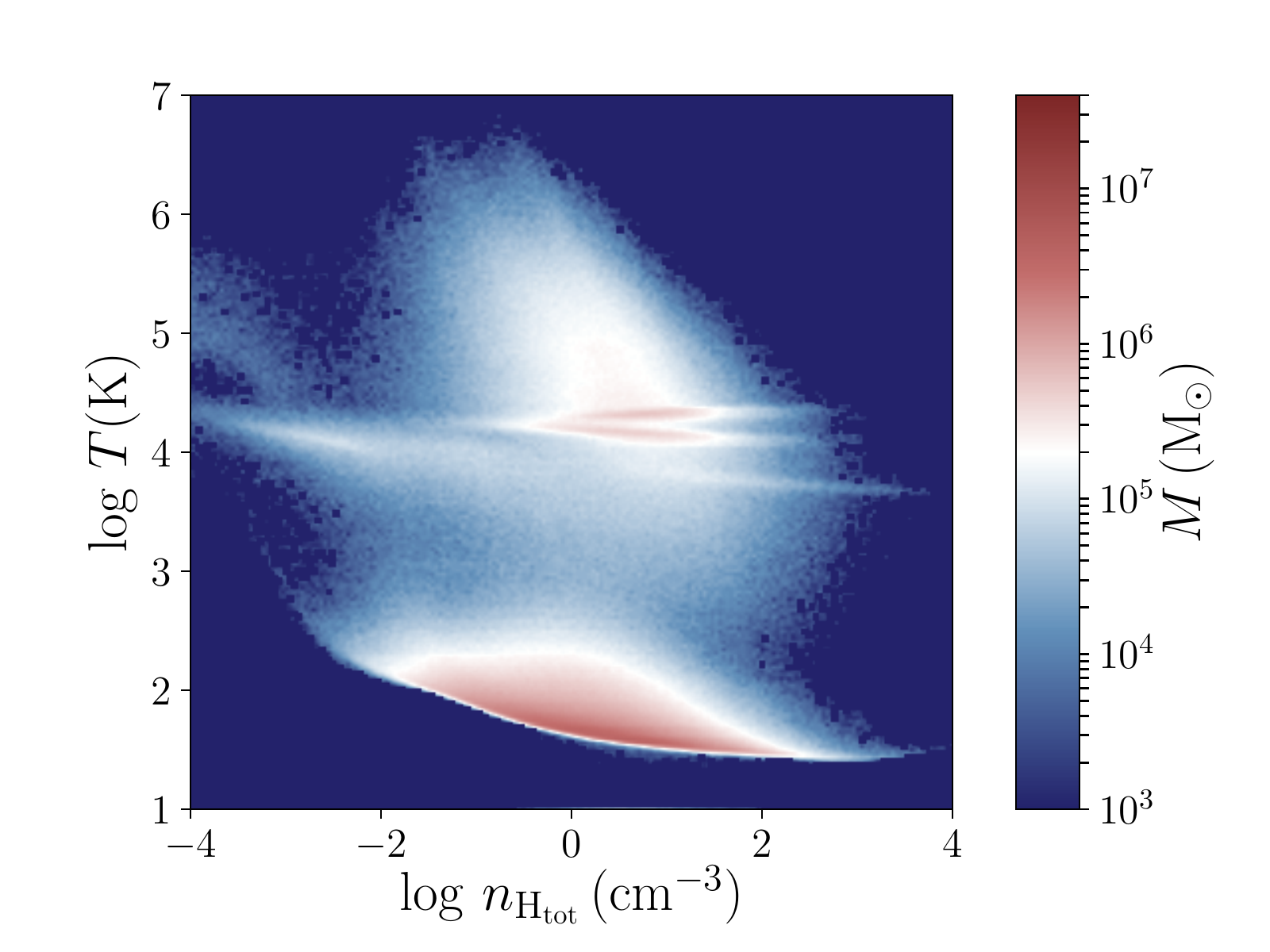}
\includegraphics[width=0.4\textwidth]{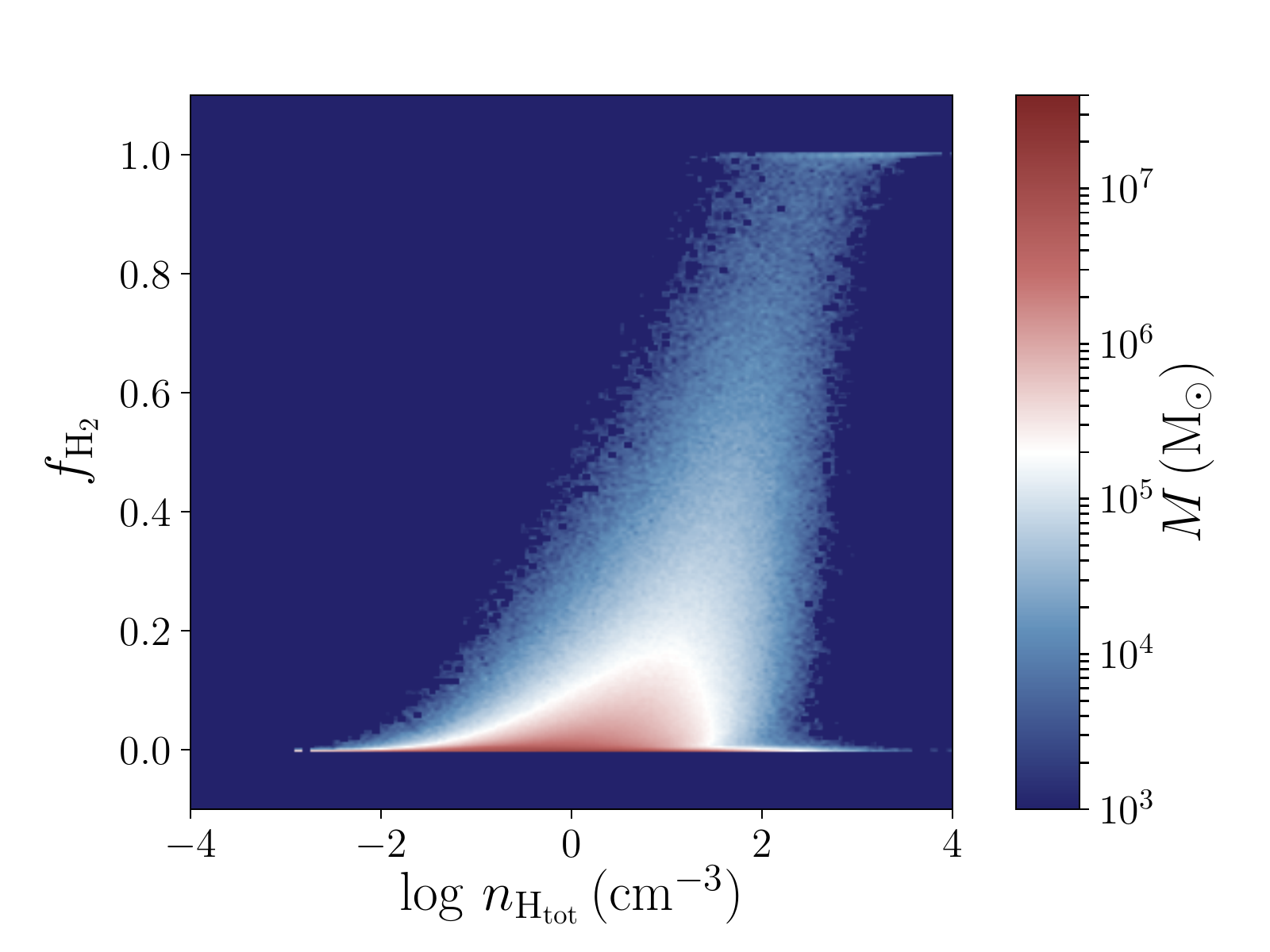}
\caption{Gas distribution in the density--temperature plane (left-hand panel) and H$_2$ mass fraction as a function of $n_{\rm H_{tot}}$ (right-hand panel) for the \guv{} run averaged over 100~Myr around $t=350$~Myr. Compared to \gmol{}, cold gas is observed down to $n_{\rm H_{tot}}\sim 10^{-2}\rm\, cm^{-3}$. In addition, the low-density tail of fully molecular gas observed in \gmol{} is not present anymore in \guv{} (right-hand panel), because of the presence of dissociating radiation.}
\label{fig:UVnolocp}
\end{figure*}

Thanks to the detailed chemistry model used, which allows us to follow the molecular gas abundance, we can also compare the relation between the SFR and the gas density in the galaxy, both in total gas (H+H$_2$) and H$_2$ only. To better compare our simulations with observations, we measure both SFR and gas densities in patches of 750~pc side length, matching the resolution in \citet{bigiel10}. In addition, instead of using the instantaneous SFR computed from the gas properties, we compute the stellar luminosity of our stellar particles in the GALEX FUV band (1250--1750~$\angstrom$), using the BC03 models, and convert it into a SFR using the relation by \citet{salim07}, to be consistent with the approach used by \citet{bigiel10}.\footnote{To test the possible bias in the measure using this approach, we compared the resulting SFR with that obtained by averaging the stellar mass formed over 10, 50, and 100~Myr, respectively, and with the instantaneous SFR estimated from the gas density, finding almost perfect agreement, with small differences only at low SFRs ($\Sigma_\star \lesssim 5\times 10^{-3}\rm\, \msun/yr/kpc^2$). Still, this was expected, because of the quiet SFR history of the galaxy.}  
Since the SFR in the galaxy does not vary significantly during the run (see left-hand panel of Fig.~\ref{fig:compsfr}), we could stack the data from 10 different snapshots in the last 100~Myr of the run ($300 \leq t\leq400\rm\, Myr$), every 10~Myr, to increase the number of data points.
The results are reported in Fig.~\ref{fig:noUVks}, where we show the total gas (H+H$_2$) KS relation in the left-hand panel and the H$_2$ SF law in the right-hand panel, and compare our data with the observations by \citet{bigiel10,schruba11}. These are plotted in filled contours of more than 1, 2, 5, and 10 points per bins, binning in both gas and SFR density with 0.1 decade-wide bins. The green contours correspond to the \citet{bigiel10} data inside $r_{25}$, the red ones to the \citet{bigiel10} data outside $r_{25}$, and the 
purple ones to \citet{schruba11}. 
Our data points, instead, are plotted as line contours of more than 2 and 10 points per bin, using 0.2 decade-wide bins in both gas and SFR density bins.
 
To highlight the SFR dependence on the radial distance from the galaxy centre, we split our point distribution in two data sets, corresponding to the patches inside (blue contours)/outside (red contours) 4~kpc (corresponding to the radius where $\sim 98$ per cent of SF occurs). The simulated data points match well the observational data, except for the highest densities (corresponding to the central 2~kpc of the galaxy), where SFR is higher. 
On the other hand, in the right-hand panel our points lay above the observed relation, and the slope is also different. While the different slope is consistent with a too large molecular gas fraction at low density, which moves the data points to the right and steepens the relation, at intermediate to high densities the molecular gas fraction is too low for the measured SFR, probably because the resolution is too low and underestimates the real H$_2$ formation rate in the high-density regions. As we will show also for the other runs, this is a common behaviour, which can be solved using a sub-grid clumping factor, as we will describe in Section~\ref{sec:clumping}.

\begin{figure*}
\centering
\includegraphics[width=\textwidth,trim={0cm 0.3cm 0cm 1cm},clip]{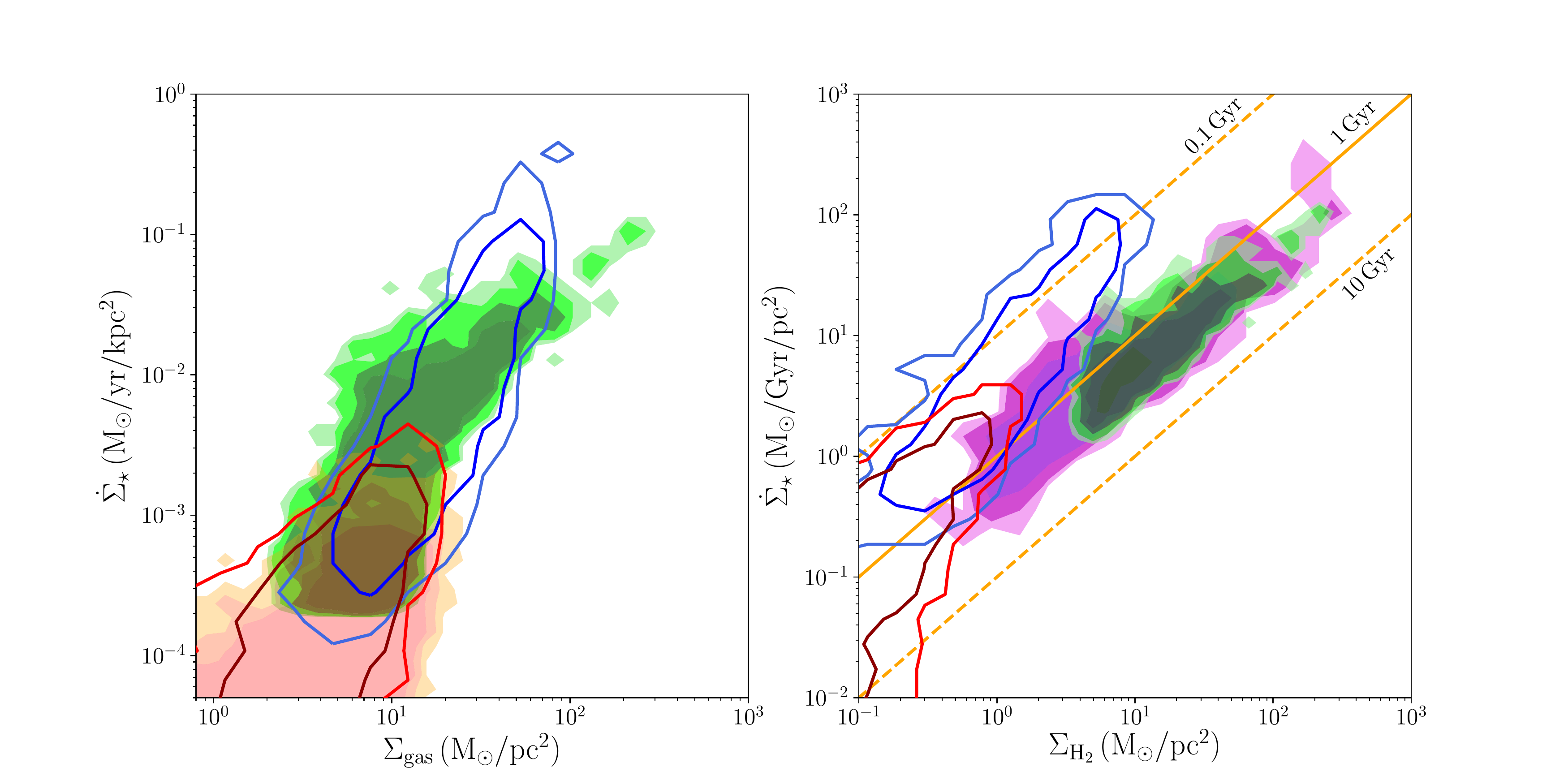}
\caption{Same as Fig.~\ref{fig:noUVks} for the \guv{} run. In the left-hand panel, the agreement with observational data is still good, with narrower contours overlapping with the \citet{bigiel10} data within $r_{25}$. In the right-hand panel, instead, the slope has slightly improved, but the normalization is offset by a factor of a few.}
\label{fig:UVnolocks}
\end{figure*}

\subsection{Test 2: non-equilibrium chemistry and photochemistry under a uniform extragalactic background}
\label{sec:UVnoloc}
We now describe the results of the \guv{} run, where we also include photochemistry. In this case, the additional heating source for gas comes from a uniform extragalactic UV background \citep{haardt12} at $z=3$. 
Because of the gas shielding prescription implemented, we expect only the gas at low density to be ionised, and almost no difference compared to the \gmol{} run at intermediate and high densities. In Fig.~\ref{fig:UVnolocp}, we show the gas distribution in the density--temperature plane and the H$_2$ mass fraction. We observe in the left-hand panel that the gas with $n_{\rm H_{tot}}\lesssim 10^{-2} \rm\, cm^{-3}$ is heated up above $10^4$~K, like in \gmol{}, except that here the heating is mainly due to the UV background instead of the SNe. On the other hand, at $n_{\rm H_{tot}}\gsim 10^{-2} \rm\, cm^{-3}$, we observe a larger mass of gas at low temperature. The reason for this difference is that the external UV background heats the gas, slowing down gas collapse and fragmentation, reducing the SFR and the subsequent SN heating. The absence of the UV background in \gmol{} allows gas to fragment and form stars more easily also in the galaxy outskirts. As a consequence, SNe explosions in a very low density medium become very effective at sweeping away gas, heating it and reducing the density even more.
In the right-hand panel, we clearly see that the gas becomes fully molecular above $n_{\rm H_{tot}}= 100\rm\, cm^{-3}$, but no molecular gas is observed at low density, unlike in the \gmol{} run, because of the enhanced dissociation rate due to the external UV background.

Fig.~\ref{fig:UVnolocks} shows the SF law for the \guv{} run, obtained with the same procedure used for \gmol. In the left-hand panel, we clearly see that the inclusion of the UV background decreases the spread in the relation at low to intermediate gas densities, without changing the slope of the relation. Also in this case, our data points lie on the observed relation, both for the inner and outer part of the galaxy. Again, at high densities we still do not observe a bending, similar to the \gmol{} run. In the right-hand panel, instead, the trend is significantly different, with a narrower and less steep distribution compared to the observed one, especially in the central 4~kpc of the galaxy.

\section{The effect of stellar radiative feedback}
\label{sec:results}
We describe here the results of our fiducial runs \grads, \gradt, and \grt, where we include radiative feedback from stars and a variable clumping factor, comparing the same properties discussed in the previous section. Unlike in previous studies, where the clumping factor was constant for all the gas, here we self-consistently compute $C_\rho$ assuming the same PDF used for SF, but only for gas above $\rho_{\rm SF,thr}$.

Despite the ever increasing computational power, the resolution achieved in our simulations is still prohibitive for cosmological simulations. Appendix~\ref{app:lowres} presents analogues of the simulations described here, but at low resolution, showing that, provided that the gas particle mass does not exceed the typical mass of a GMC, results are consistent with those discussed in this section.

\begin{figure*}
\centering
\includegraphics[width=0.9\textwidth,trim={0cm 2cm 0cm 2cm},clip]{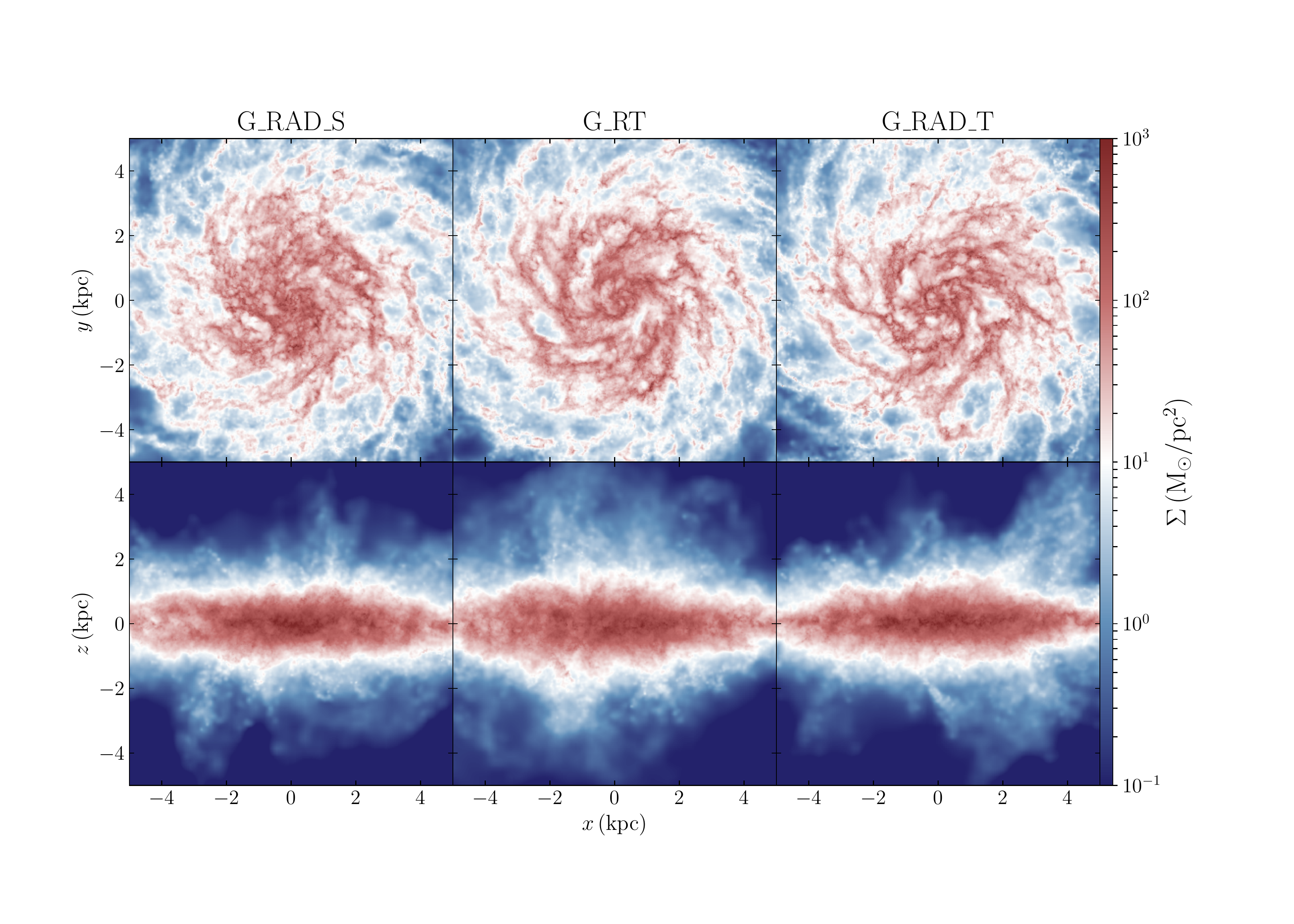}
\caption{Projected gas density maps of total hydrogen in the \grads, \gradt, and \grt{} runs at $t=350$~Myr, face-on (top panels) and edge-on (bottom panels). The gas distribution is smoother in \grads{} compared to \gradt{} and \grt{}, with less defined spiral arms. Though not shown here, this difference persisted in other snapshots. In the edge-on view, instead, the slightly thicker disc in \grt{} is simply due to the feedback resulting from a burst of SF occurred around $t=300\rm\, Myr$.}
\label{fig:compmaps}
\end{figure*}

\begin{figure*}
\centering
\includegraphics[width=0.9\textwidth,trim={0cm 2cm 0cm 2cm},clip]{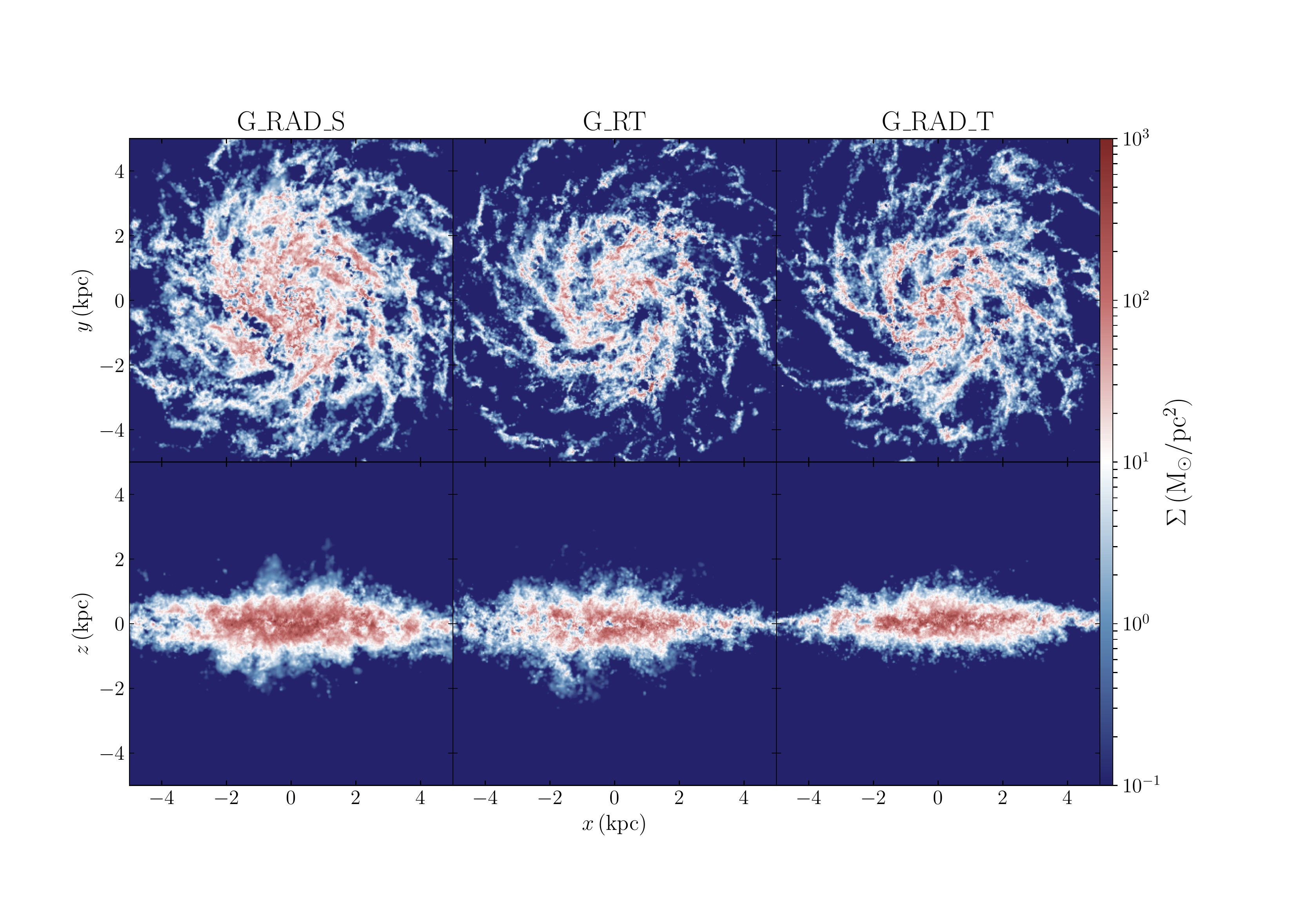}
\caption{Projected gas density maps of the H$_2$ component for our \grads, \gradt, and \grt{} runs at $t=350$~Myr, face-on (top panels) and edge-on (bottom panels). In \grads{}, H$_2$ can also form at slightly lower densities compared to the other two runs, due to an underestimation of the radiative flux in this regime. \gradt{}, instead, is closer to \grt{}, thanks to the gravity tree-based model which accurately accounts for all stellar sources in the galaxy, although there is still more H$_2$ than in \grt{}.}
\label{fig:compmapsH2}
\end{figure*}

\subsection{Molecular gas distribution in the galaxy}
We compare here the molecular gas distribution in the galaxy at $t=350$~Myr. We first discuss the qualitative behaviour from the projected gas density maps in Figs~\ref{fig:compmaps}~and~\ref{fig:compmapsH2}. The small differences in the radiative feedback models can significantly affect the SF history of the galaxy, producing visible differences in the galaxy maps, which however do not have a strong impact on the conclusions.

\begin{figure*}
\centering
\includegraphics[width=0.8\textwidth]{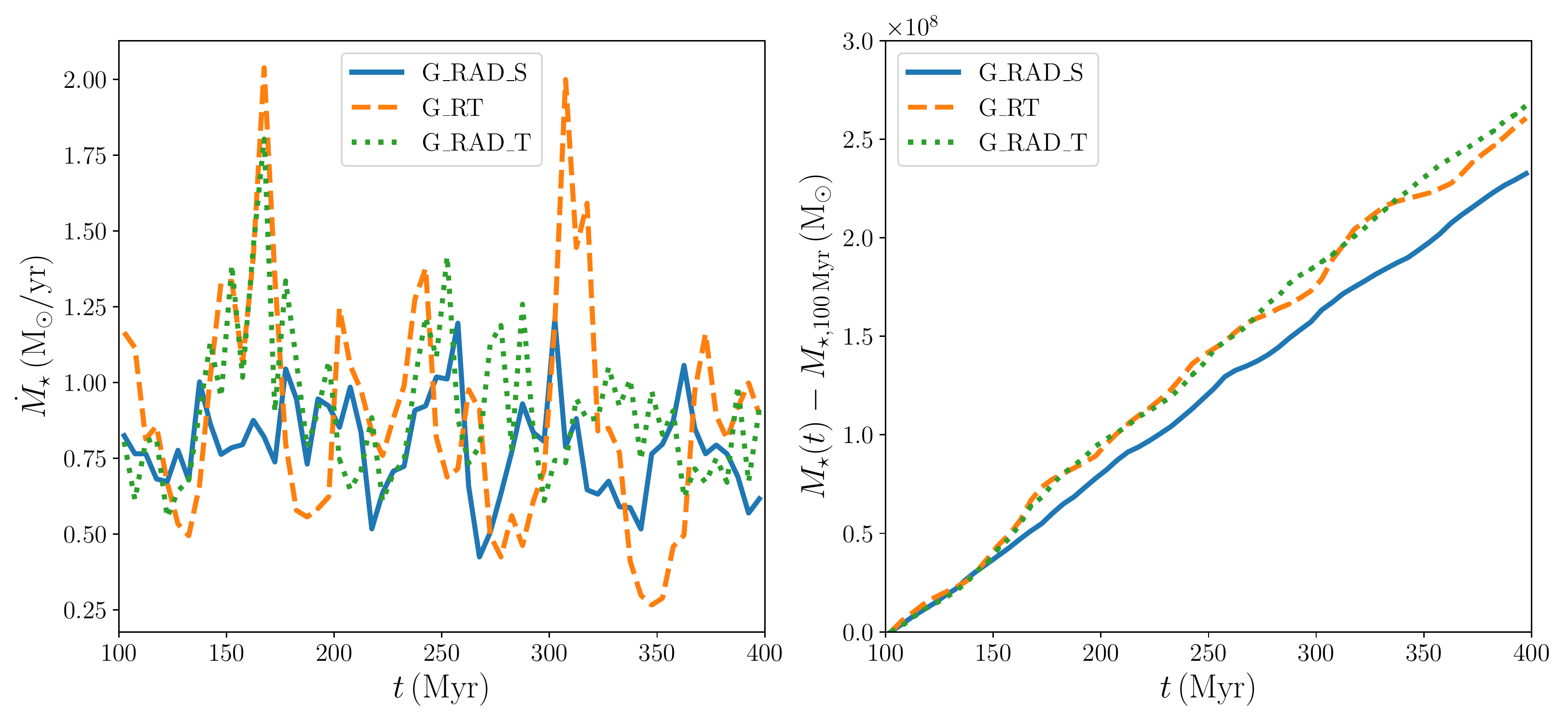}
\caption{SF in our \grads, \grt, and \gradt{} runs in the latest 300~Myr of the simulations. We show the SFR in the left-hand panel and the total stellar mass formed in the right-hand panel. The green dotted line corresponds to \gradt, the orange dashed line to \grt, and the blue solid line to \grads. Though the SFR shows bursts of SF of moderate intensity in all the runs, the SF history is quiet ($\sim 1\,\msun\rm /yr$) during the entire simulations. A slightly lower SFR can be noticed in \grads{}, probably due to a stronger radiative flux close to the stellar sources compared to the other runs. As a consequence, the total stellar mass formed (right-hand panel) in \grads{} is a bit lower than in \grt{} and \gradt{}.}
\label{fig:compsfr}
\end{figure*}

In the face-on view (top panels of Fig.~\ref{fig:compmaps}), \grads{} exhibits a slightly smoother gas distribution compared to \gradt{} and \grt, with a less defined spiral pattern. In the bottom panels (edge-on view), the differences are smaller. \grads{} shows lower densities at higher distances from the galaxy plane, compared to the other runs, although the difference is negligible in terms of mass. \grt, instead, displays a slightly thicker disc amongst the three runs, compatible with the stronger feedback following a burst of SF which occurred at $t\sim300\rm\, Myr$ (see Fig.~\ref{fig:compsfr}).

\begin{figure*}
\begin{flushleft}
\large \hspace{0.14\textwidth} G\_RAD\_S \hspace{0.225\textwidth} G\_RT \hspace{0.22\textwidth} G\_RAD\_T 
\end{flushleft}
\centering
\includegraphics[width=0.3\textwidth,trim={0cm 0cm 0cm 0.5cm},clip]{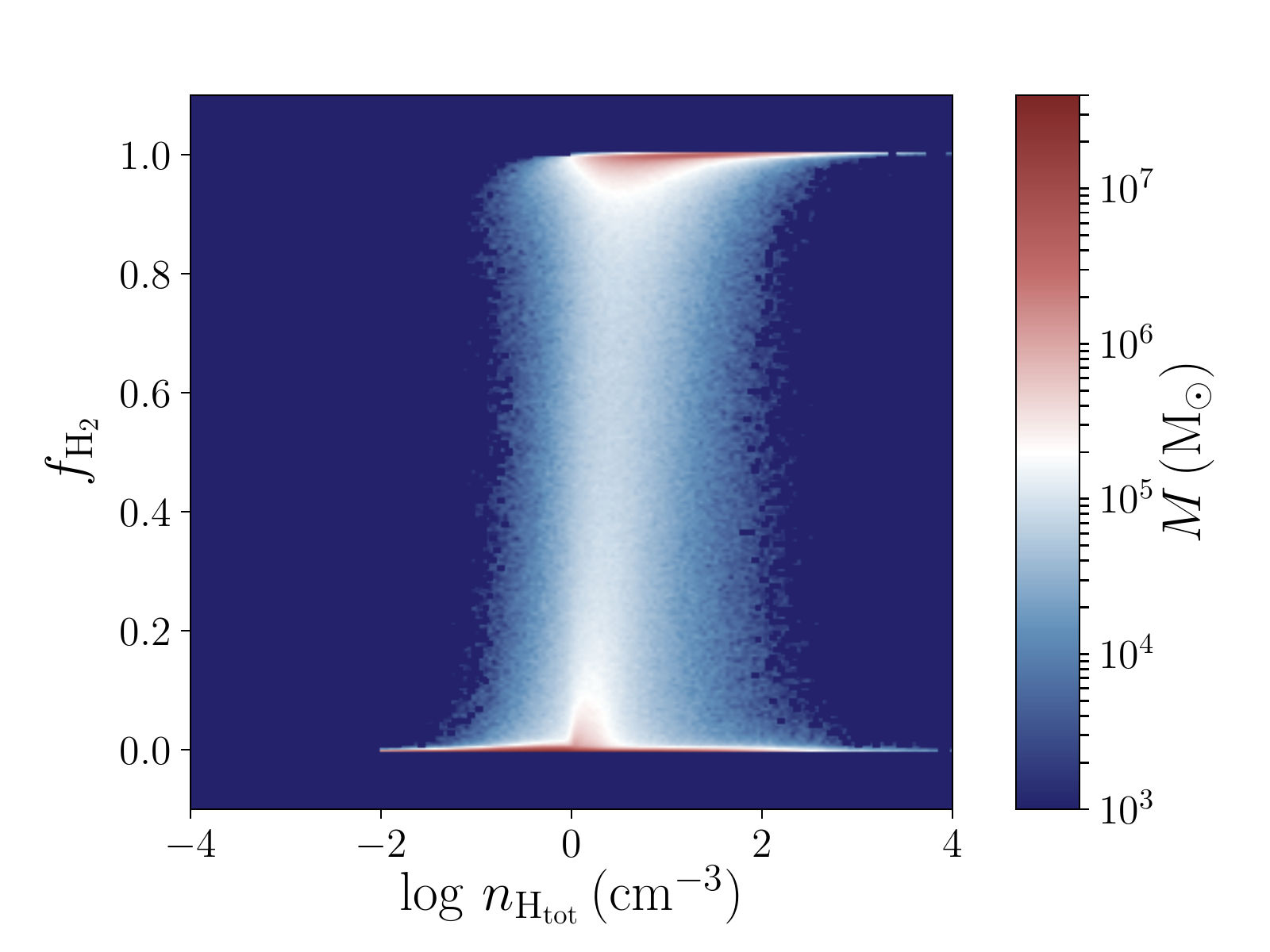}
\includegraphics[width=0.3\textwidth,trim={0cm 0cm 0cm 0.5cm},clip]{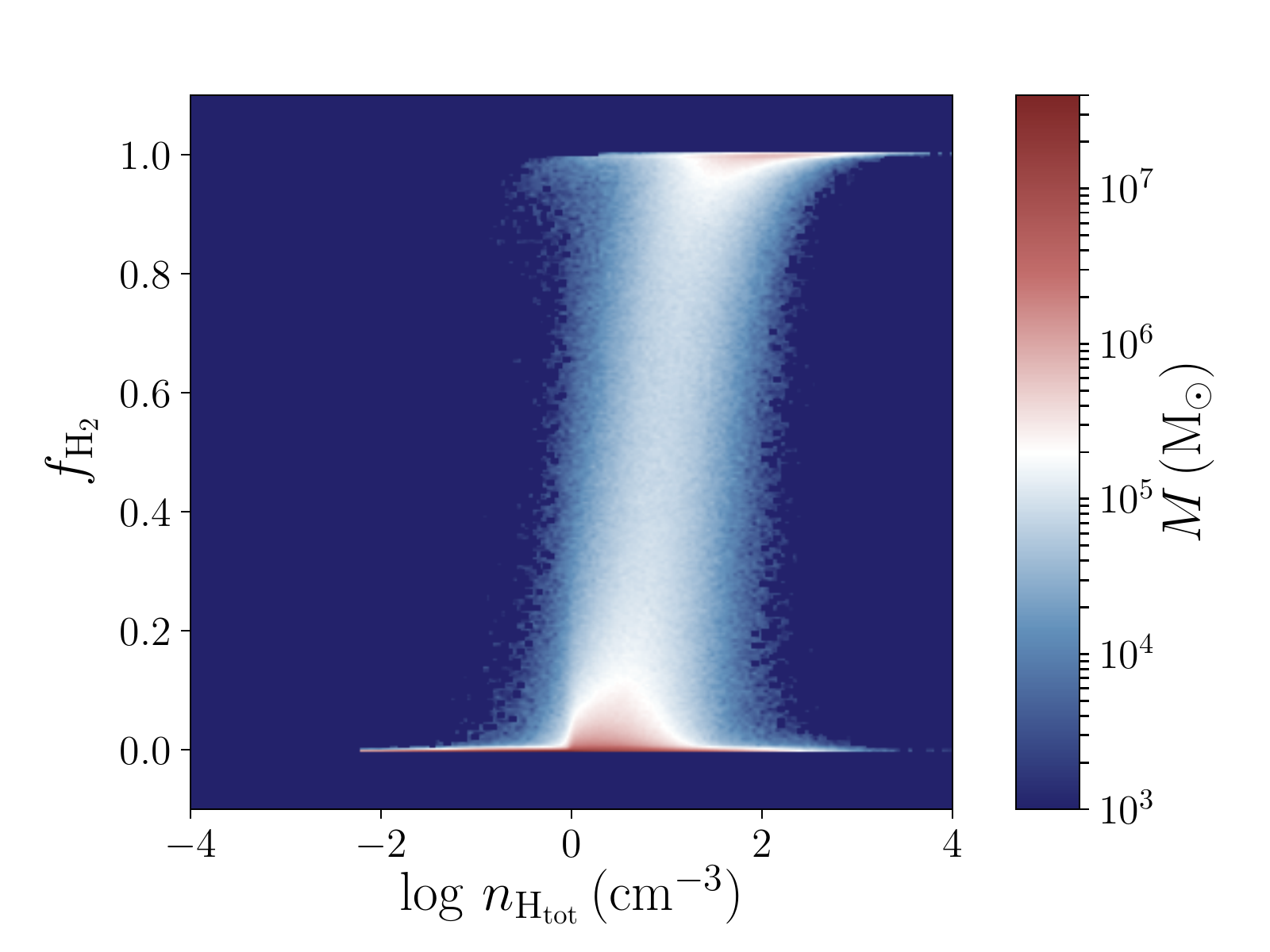}
\includegraphics[width=0.3\textwidth,trim={0cm 0cm 0cm 0.5cm},clip]{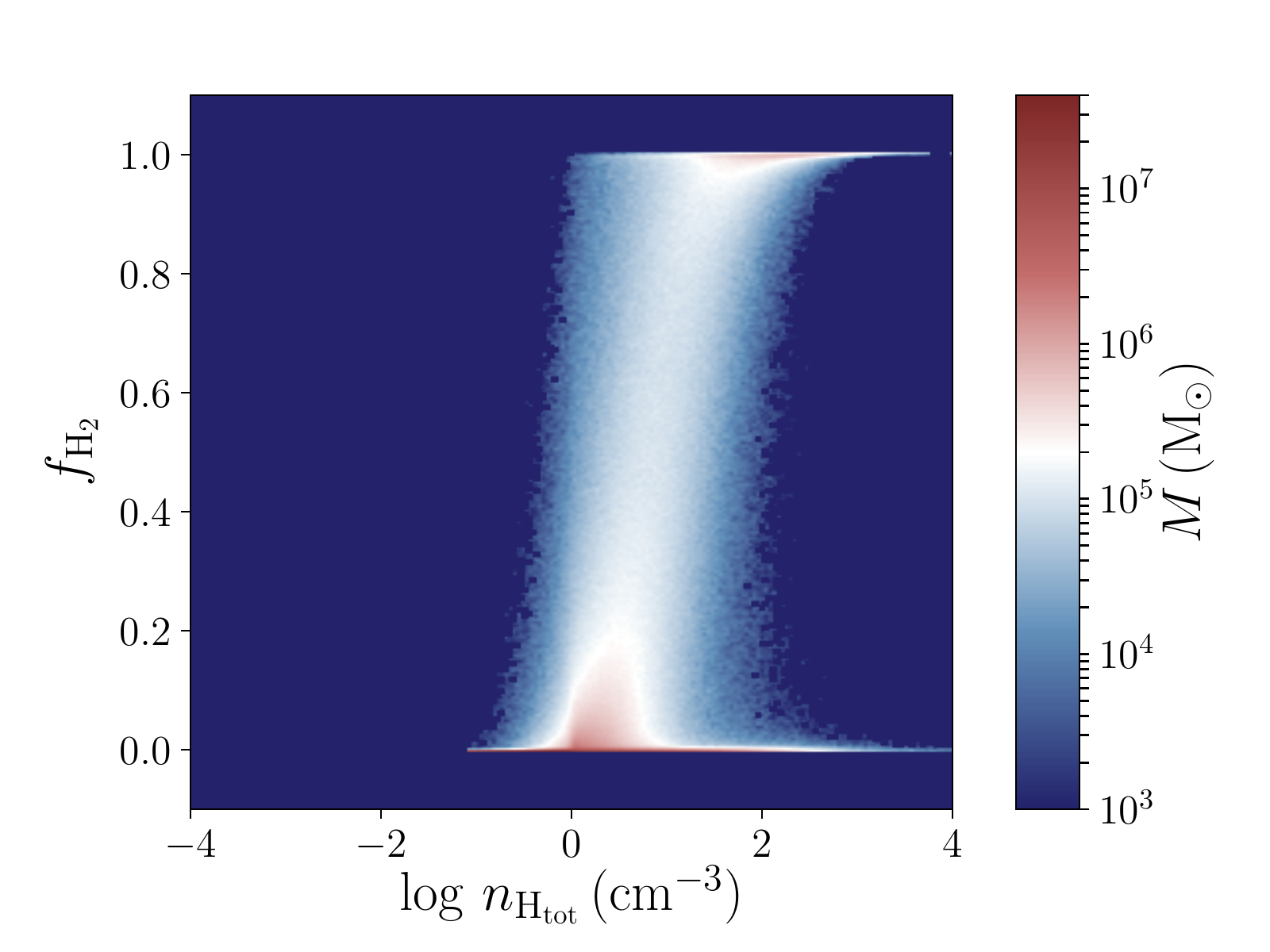}\\
\includegraphics[width=0.3\textwidth]{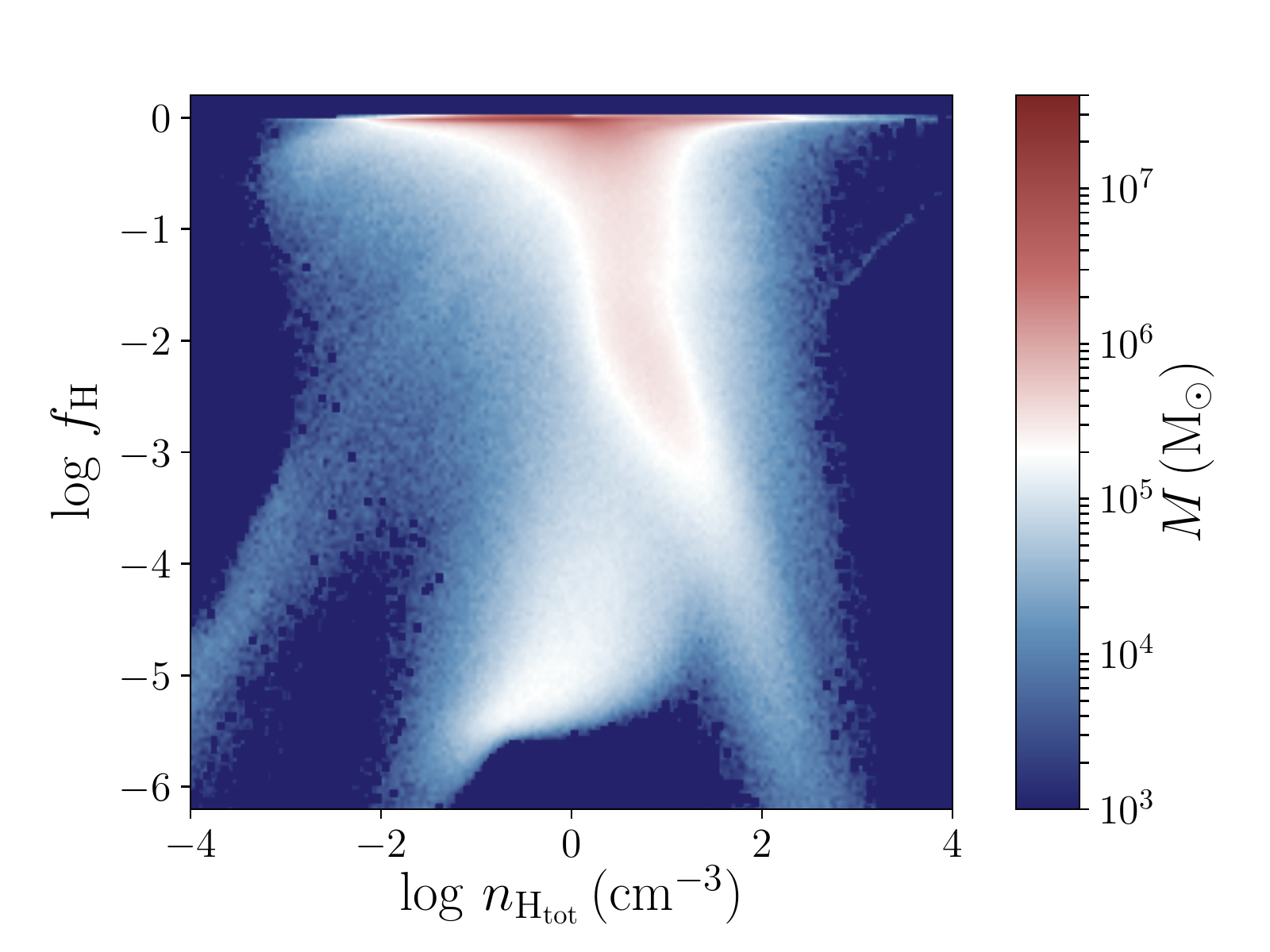}
\includegraphics[width=0.3\textwidth]{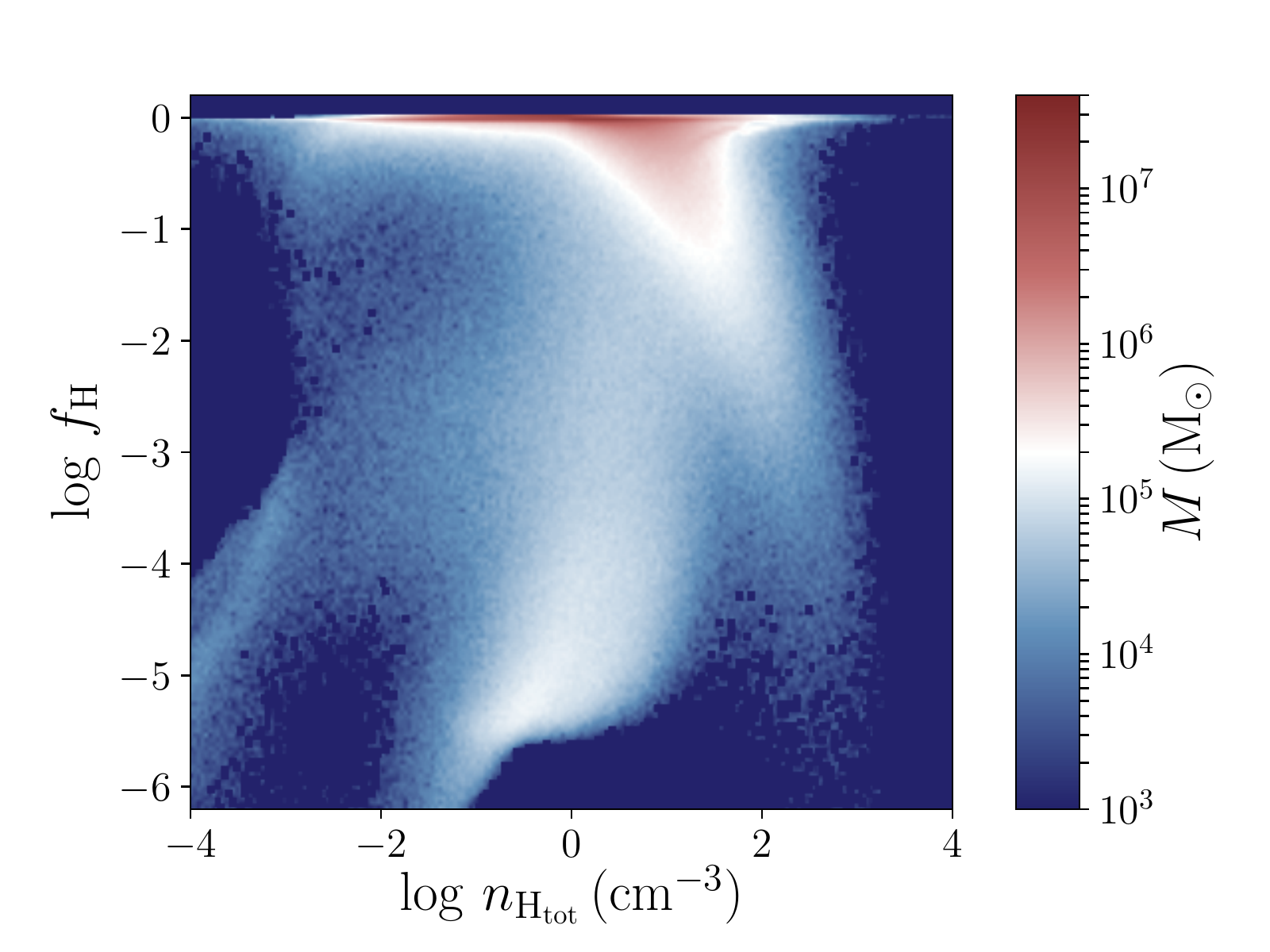}
\includegraphics[width=0.3\textwidth]{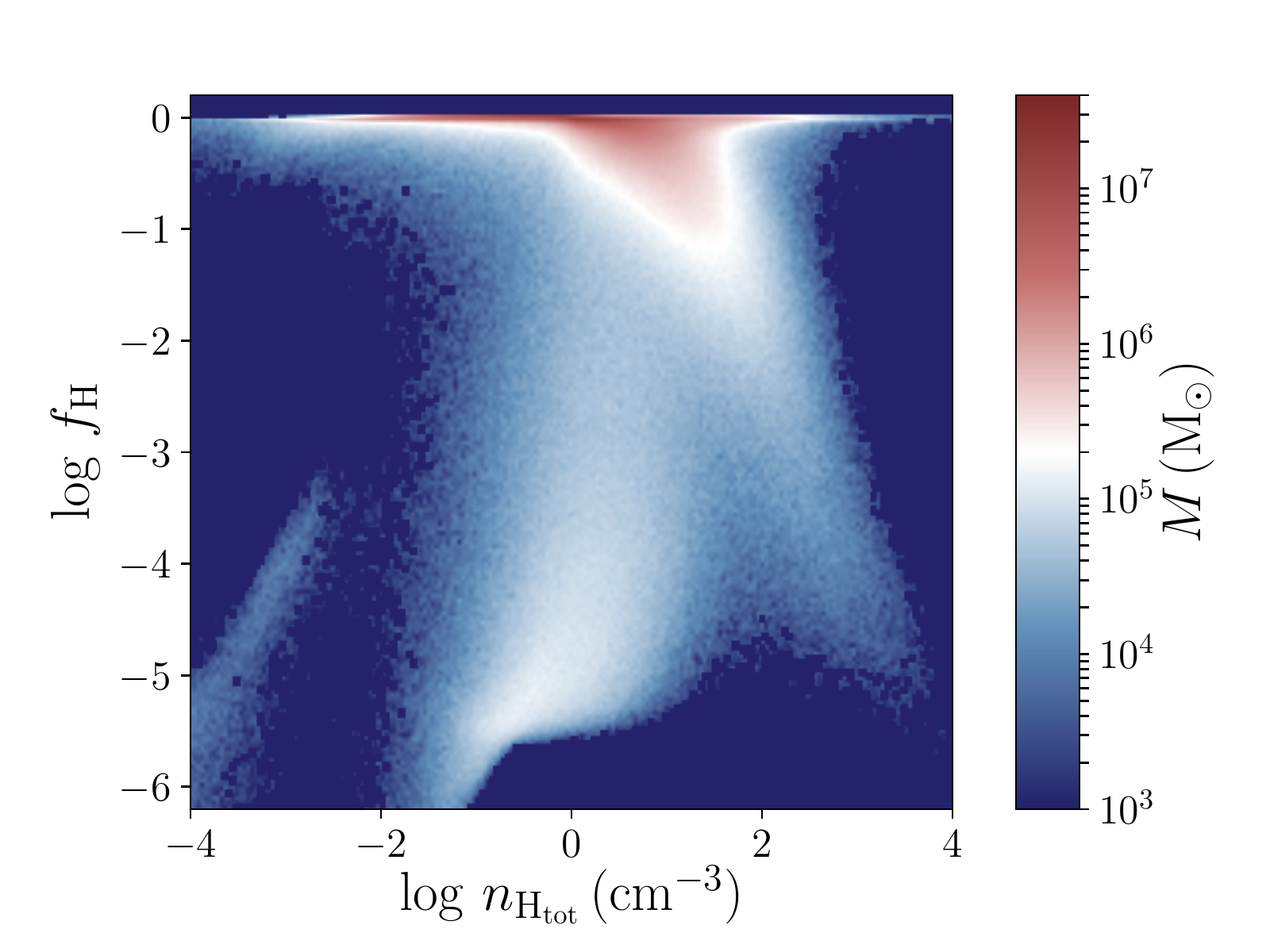}\\
\includegraphics[width=0.3\textwidth]{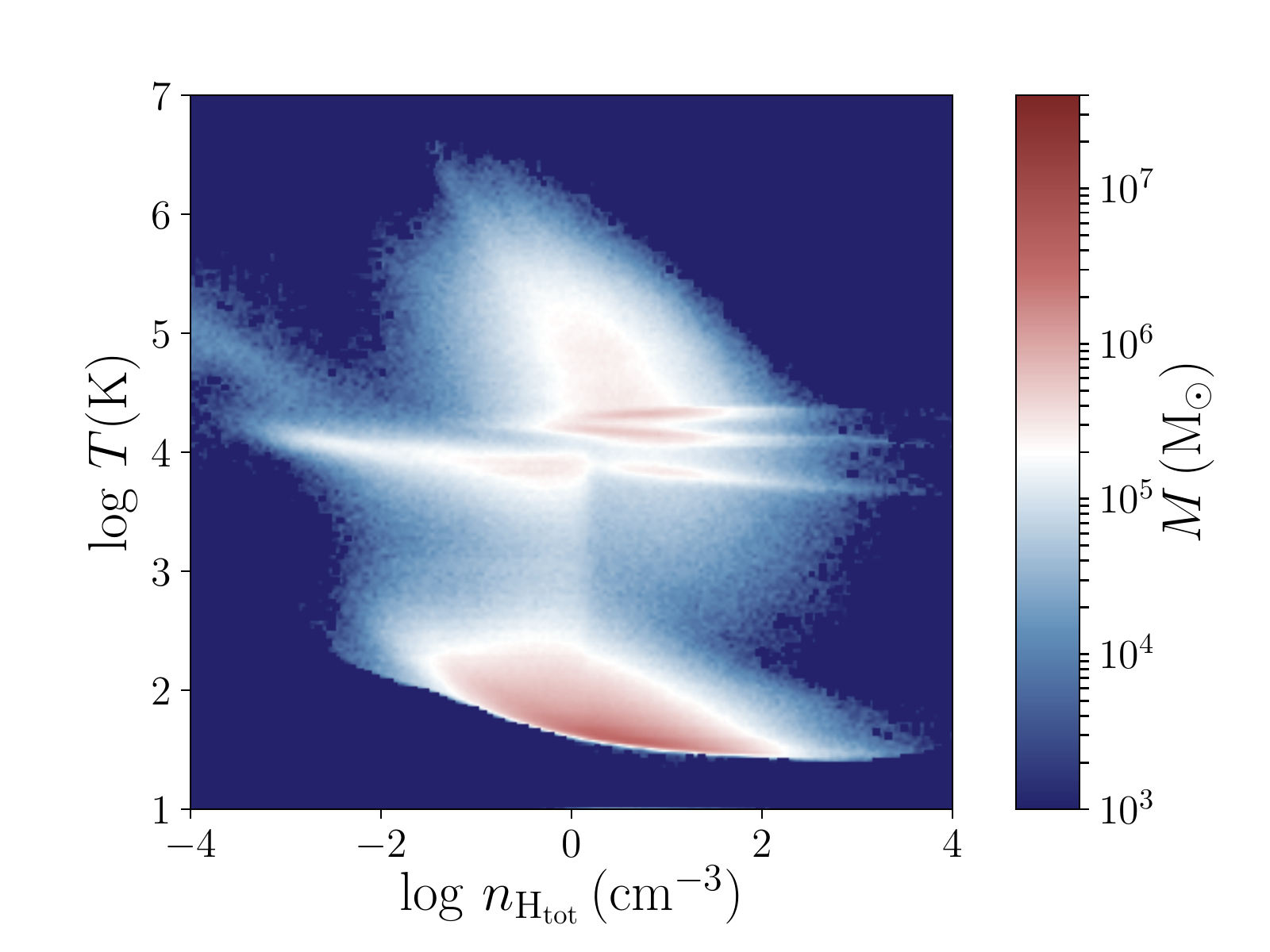}
\includegraphics[width=0.3\textwidth]{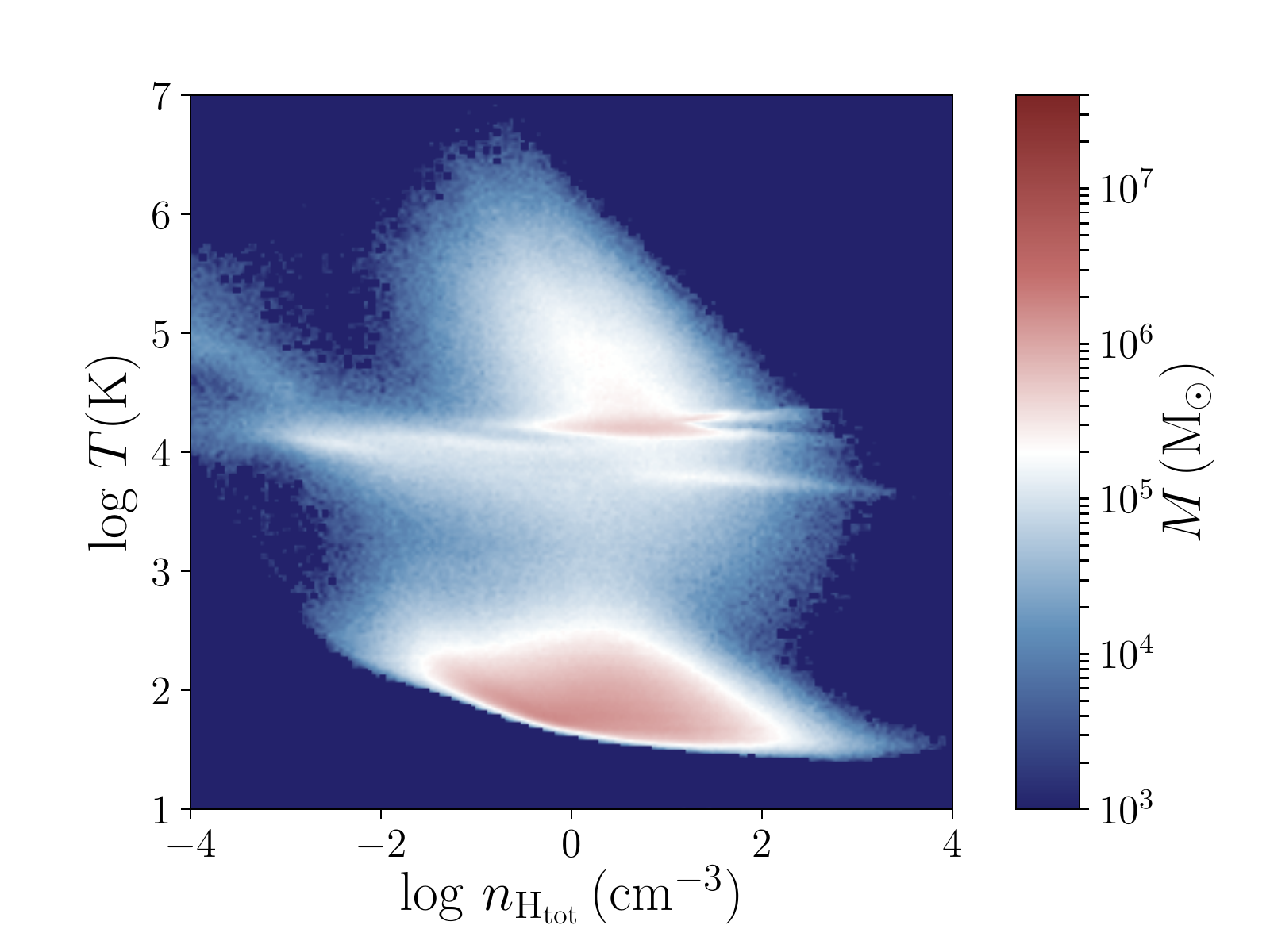}
\includegraphics[width=0.3\textwidth]{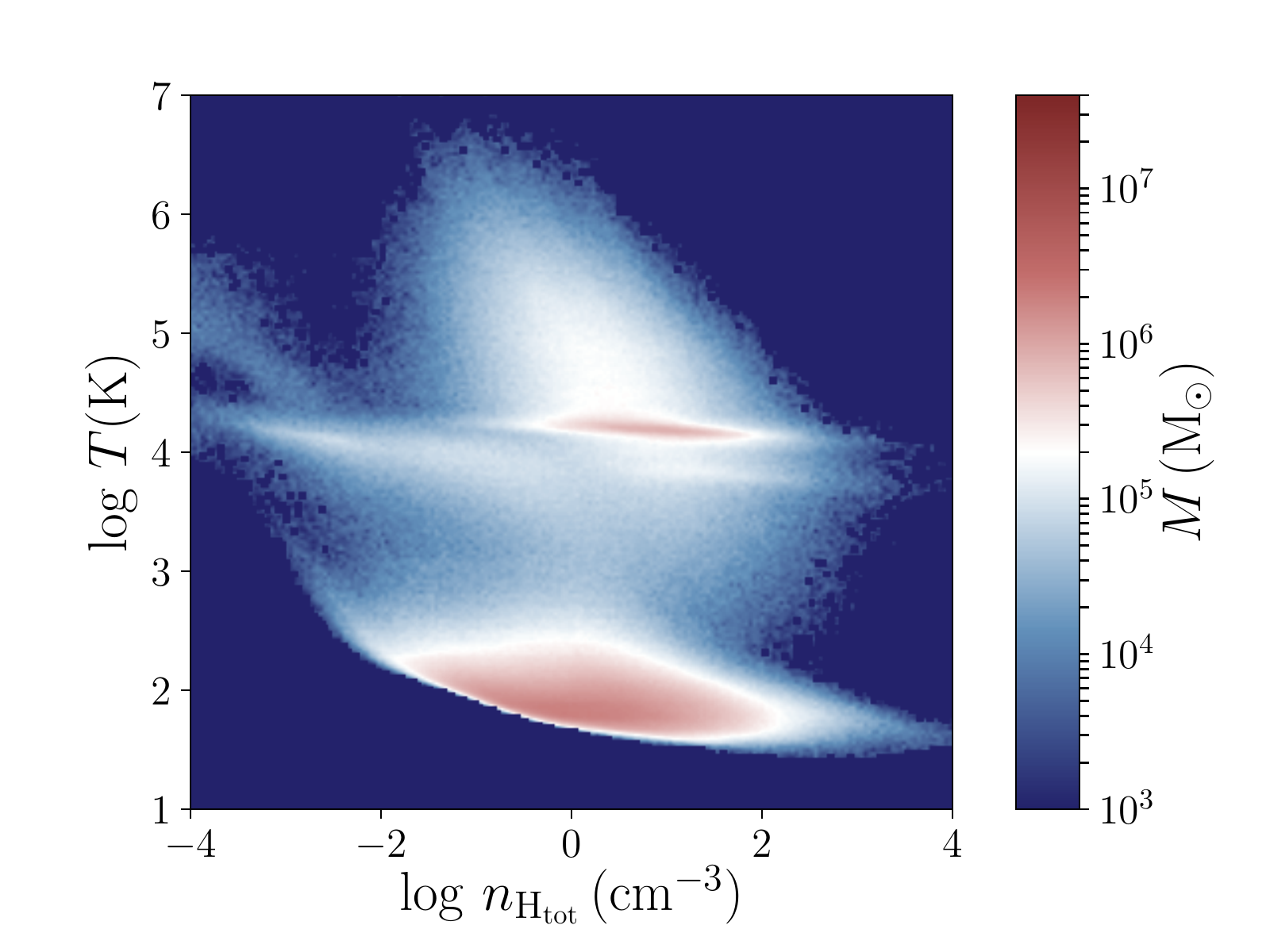}\\
\caption{Gas properties for the \grads{} (left-hand column), \grt{} (middle column), and \gradt{} (right-hand column) runs averaged over 100~Myr around $t=350$~Myr. We report molecular hydrogen (top row) and neutral hydrogen (middle row) mass fractions as a function of $n_{\rm H_{tot}}$. 
The bottom panels, instead, correspond to the gas distribution in the density--temperature plane. All the runs show similar features, with small differences in the temperature for warm gas around $10^4$~K and in the abundance of molecular gas. The strongest difference is in the abundance of atomic hydrogen in \grads{}, where the underestimation of the stellar radiation at intermediate densities, when $\tau=1$ from Eq.~\eqref{eq:tau} is never reached, results in a massive tail extending down to $f_{\rm H}\sim 10^{-3}$.}
\label{fig:compplots}
\end{figure*}

For the molecular gas distribution (Fig.~\ref{fig:compmapsH2}), the differences are more evident.\grads{} generally forms more H$_2$ compared to the other runs, whereas \grt{} and \gradt{}, instead, exhibit a more peaked distribution, with H$_2$ formation more vigorous in the high density regions of the galaxy spiral arms and suppressed in the lower density regions.

After a first qualitative comparison, we now quantitatively assess the differences amongst the models in Fig.~\ref{fig:compplots}.

In the top row, we compare the H$_2$ mass fraction as a function of the total hydrogen density for the three runs. \grads{} forms significantly more molecular gas mass at intermediate densities ($n_{\rm H_{tot}}\sim 1-10\rm\, cm^{-3}$), compared to \grt{} and \gradt, which show instead a very similar behaviour. Anyway, the transition to fully molecular gas is the same for all these three runs and corresponds to $n_{\rm H_{tot}}\sim 10\rm\, cm^{-3}$.
In the middle row, the differences are more evident. While \gradt{} and \grt{} show very similar features  except for a slightly larger amount of atomic hydrogen in \gradt{} at $n_{\rm H_{tot}} = 1-3\rm\, cm^{-3}$, \grads{} exhibits a massive tail extending down to $f_{\rm H} \sim 10^{-3}$, consistent with the more efficient conversion of H in H$_2$.
Finally, in the bottom row, we observe very small differences, with \grads{} showing slightly larger amounts of cold gas (below $T\sim 100$~K) and warm gas (around $T=10^{4-5}$~K). \grt{} and \gradt{}, instead, are in even better agreement, with only a little more warm gas in the range $1{\rm\, cm^{-3}} < n_{\rm H_{tot}} < 100\rm\, cm^{-3}$.

\begin{figure*}
\centering
\includegraphics[width=0.3\textwidth,trim={2cm 0cm 3cm 0cm},clip]{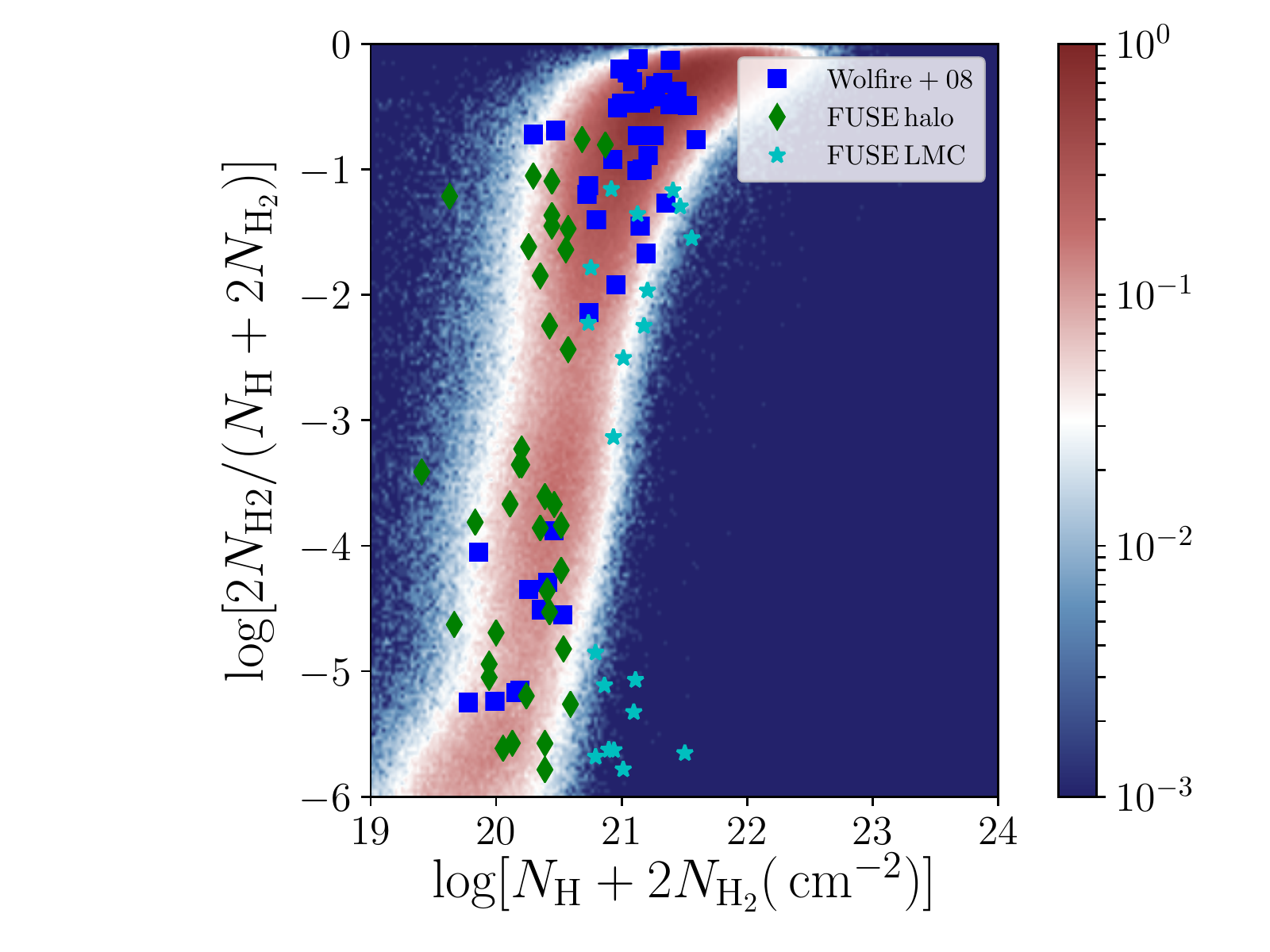}
\includegraphics[width=0.3\textwidth,trim={2cm 0cm 3cm 0cm},clip]{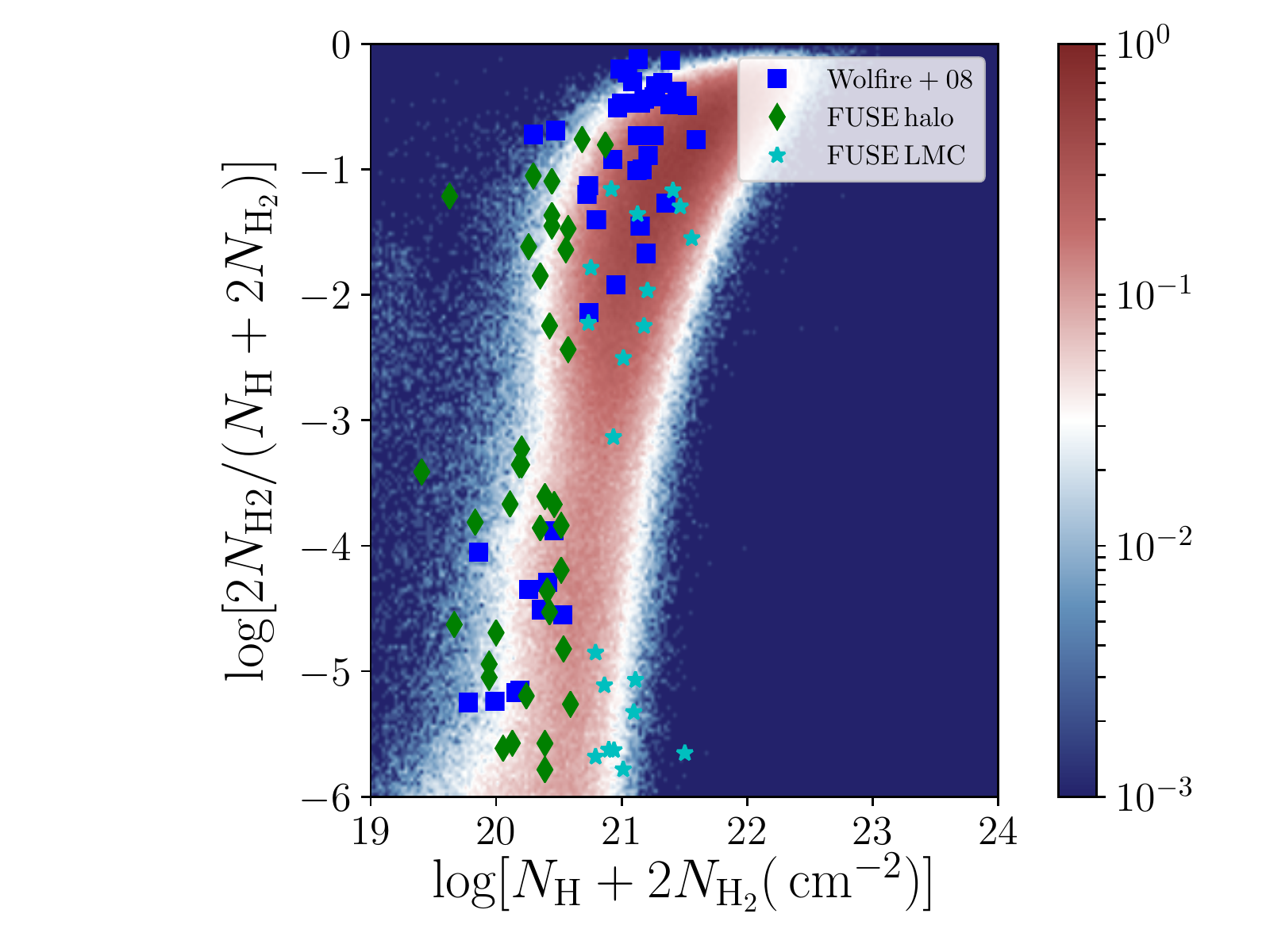}
\includegraphics[width=0.352\textwidth,trim={2cm 0cm 1cm 0cm},clip]{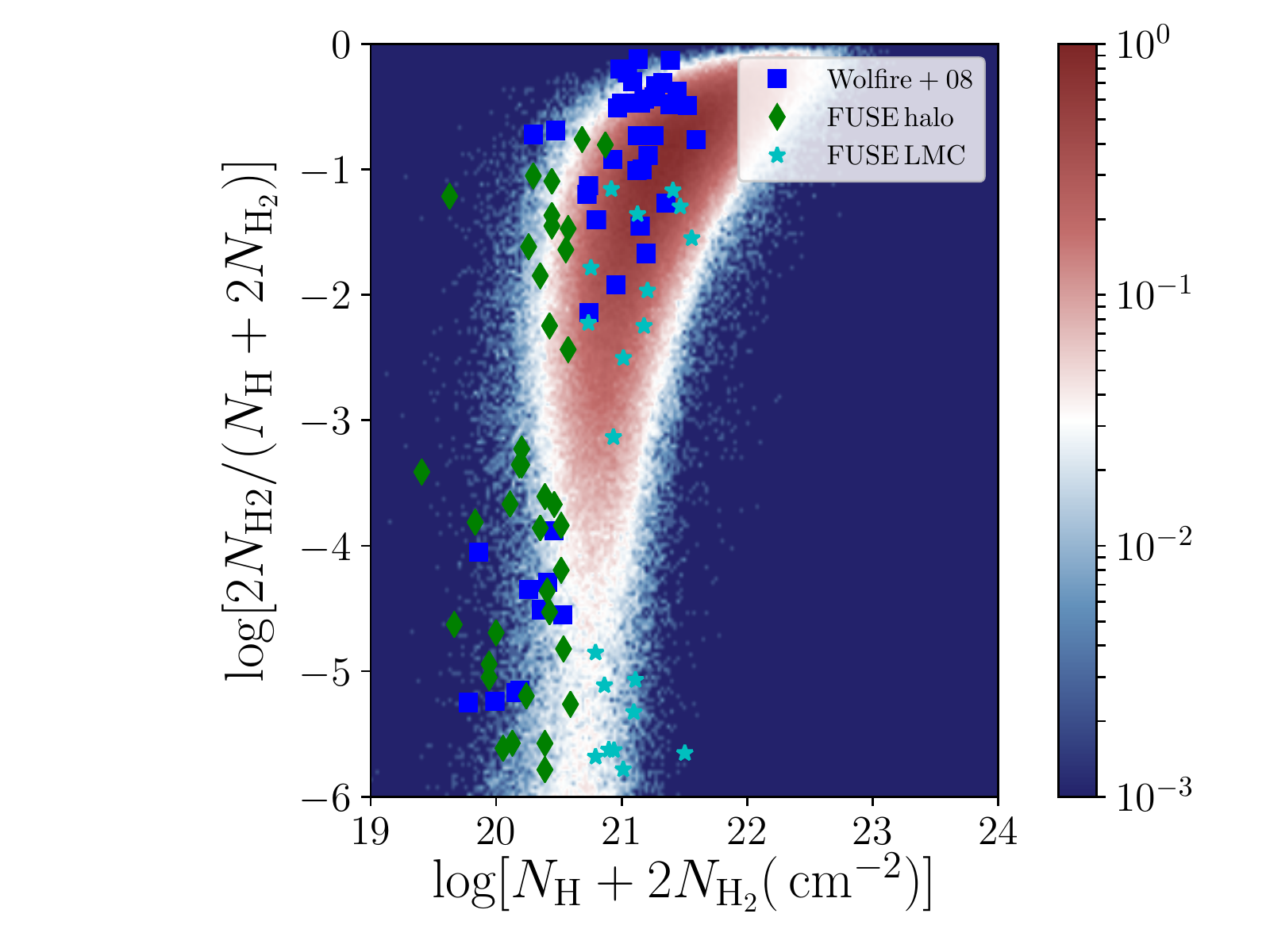}
\caption{H$_2$ column density fraction in our \grads{} (left-hand panel), \grt{} (central panel), and \gradt{} (right-hand panel) runs at $t=350\rm\, Myr$, compared with local observations of molecular clouds in the Milky Way disc and halo, and in the LMC. The observational data very well bracket our simulation results, overlapping with them in most of the available column density range. There are small differences among the three runs, especially at low column densities, where \grads{} shows a slightly higher H$_2$ column density fraction consistent with the enhanced formation of H$_2$ already observed in Fig.~\ref{fig:compplots}. \gradt{}, instead, exhibits a denser concentration at high H$_2$ fractions, consistent with the stronger shielding at high gas densities.}
\label{fig:colH2}
\end{figure*}

We now compare the H$_2$ column density with local observations. Since observations at $Z\sim 0.5 Z_\odot$ are not available, we report data for the Milky Way disc and halo (at solar metallicity) and the Large Magellanic Cloud (LMC, $0.3 Z_\odot$), which should bracket our simulation results. We compute the column density of both atomic and molecular hydrogen, respectively $N_{\rm H}$ and $N_{\rm H_2}$, in a cylinder of 10~kpc radius and 1~kpc height along 25 randomly distributed line of sights. We report in Fig.~\ref{fig:colH2} the logarithmic H$_2$ column density fraction as a function of the total logarithmic column density. The results are binned in 200 log-spaced bins along both axes, where the colour coding represents the point density in each bin. The cyan stars correspond to the LMC data by \citet{tumlinson02}, the blue squares to the Milky Way disc data by \citet{wolfire08}, and the green diamonds to the Milky Way halo from the FUSE survey. The simulation results agree very well with observations, overlapping both LMC and Milky Way data for a large fraction of the available data range. This is consistent with our expectations, considering that the initial metallicity was $Z=0.5 Z_\odot$ and SNe enriched the galaxy during the 400 Myr of evolution, moving it closer to the solar metallicity data. Nevertheless, the lower metallicity data for the LMC are still in reasonable agreement with our data, lying at the edge of the distribution. 
In this case, the different local radiation model considered clearly affects the results. In particular, \grads{} exhibits a generally higher H$_2$ fraction at all column densities, because of the underestimation of the radiation flux at intermediate gas densities, as already discussed. Compared to \grt{}, the distribution in \grads{} also extends to larger column densities. A similar behaviour is also observed in \gradt{}. For column densities between $10^{20}$ and $10^{21}$ cm$^{-2}$, instead, \gradt{} shows a similar trend to \grt{}, but with a slightly sparser distribution. A critical feature of both \grt{} and \gradt{} is that they seem to never reach H$_2$ fractions higher than 80--90 per cent, contrary to \grads{}, where the suppressed radiation boosts the formation of H$_2$, shifting the data upwards.

In  summary, in \grads{}  the assumption of a maximum search radius allowed us to accurately account for the radiation close to the stellar sources, but at the cost of neglecting part of the stellar flux for gas at intermediate or low densities, when $\tau=1$ could not be reached. This inevitably resulted in a slight overestimation of the molecular gas fraction at lower densities compared to the other models. In \gradt{} , instead, the attenuation at both the source and the absorption point enabled us to consider all the stellar sources in the galaxy, but with the drawback of overestimating the attenuation when the star--cell distance was shorter than the sum of the two absorption lengths. This weakened the effect of radiation in the very high-density regions, whereas at intermediate or low density the approximation was much more accurate. Despite these limitations, the differences among \grads, \grt, and \gradt{} remained small, and only affected the results at a few per cent level in the total molecular gas and stellar mass formed. A crucial role in suppressing the differences is played by the clumping factor, which globally shifts the transition to the fully molecular regime at lower densities, where the differences in the radiative model were stronger. In conclusion, both sub-grid models considered here represent valid alternatives to expensive RT calculations: with the employed reduced speed of light of 0.001c, the overhead of the \grt{} simulation was a factor of 2--3 compared to \gradt{} and roughly a factor of 5 compared to \grads{}.

\subsection{SF in the galaxy and the SF law}
\begin{figure*}
\centering
\includegraphics[width=1\textwidth,trim={0cm 0.3cm 0cm 1cm},clip]{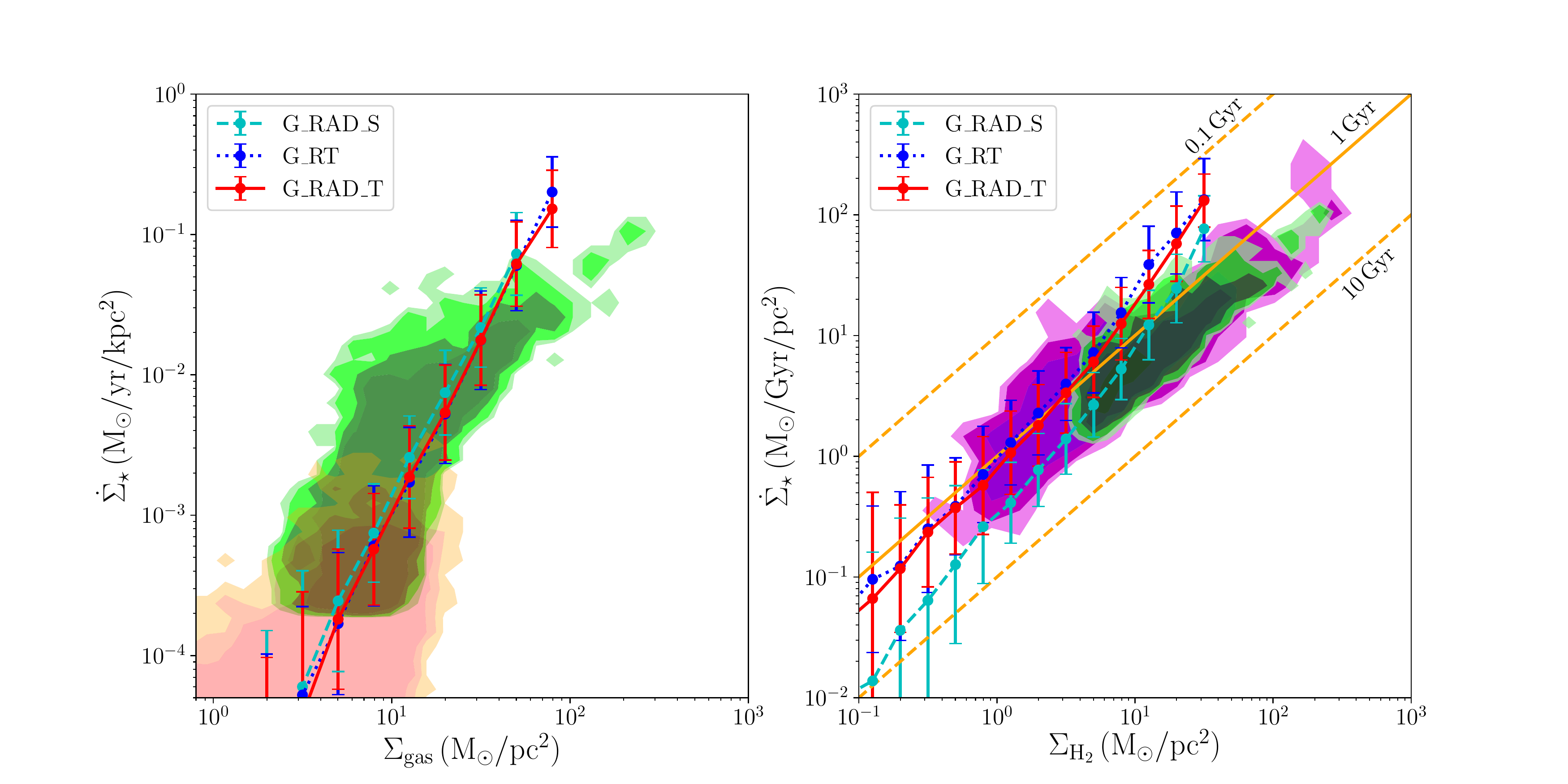}
\caption{SF law for the \grads, \grt, and \gradt{} runs in the last 100~Myr of the simulations, compared with the observational data as in Fig.~\ref{fig:noUVks}. For the simulated data, we show here the average SFR $\mu$, with its standard deviation $\sigma$ reported in the vertical error bar. \grads{} is plotted in cyan, \grt{} in blue, and \gradt{} in red, respectively. The agreement in the left-hand panel is good, with all the runs perfectly lying within the \citet{bigiel10} contours. 
In the right-hand panel, instead, the differences are a bit larger, but almost everywhere within the simulations error bars. The slope of the relation is consistent with the observational data, with a small deviation in \grads{} for very low H$_2$ surface densities, consistent with the higher H$_2$ abundance at intermediate $n_{\rm H_{tot}}$ in Fig.~\ref{fig:compplots} (top panels). The normalisation is also in good agreement with observations over more than one order of magnitude, and slightly drifting upwards only above 10 $\msun/$pc$^2$, corresponding to the central 2~kpc of the galaxy.}
\label{fig:compks}
\end{figure*}

\begin{figure*}
\centering
\includegraphics[width=0.66\textwidth,trim={0cm 1cm 0cm 1cm},clip]{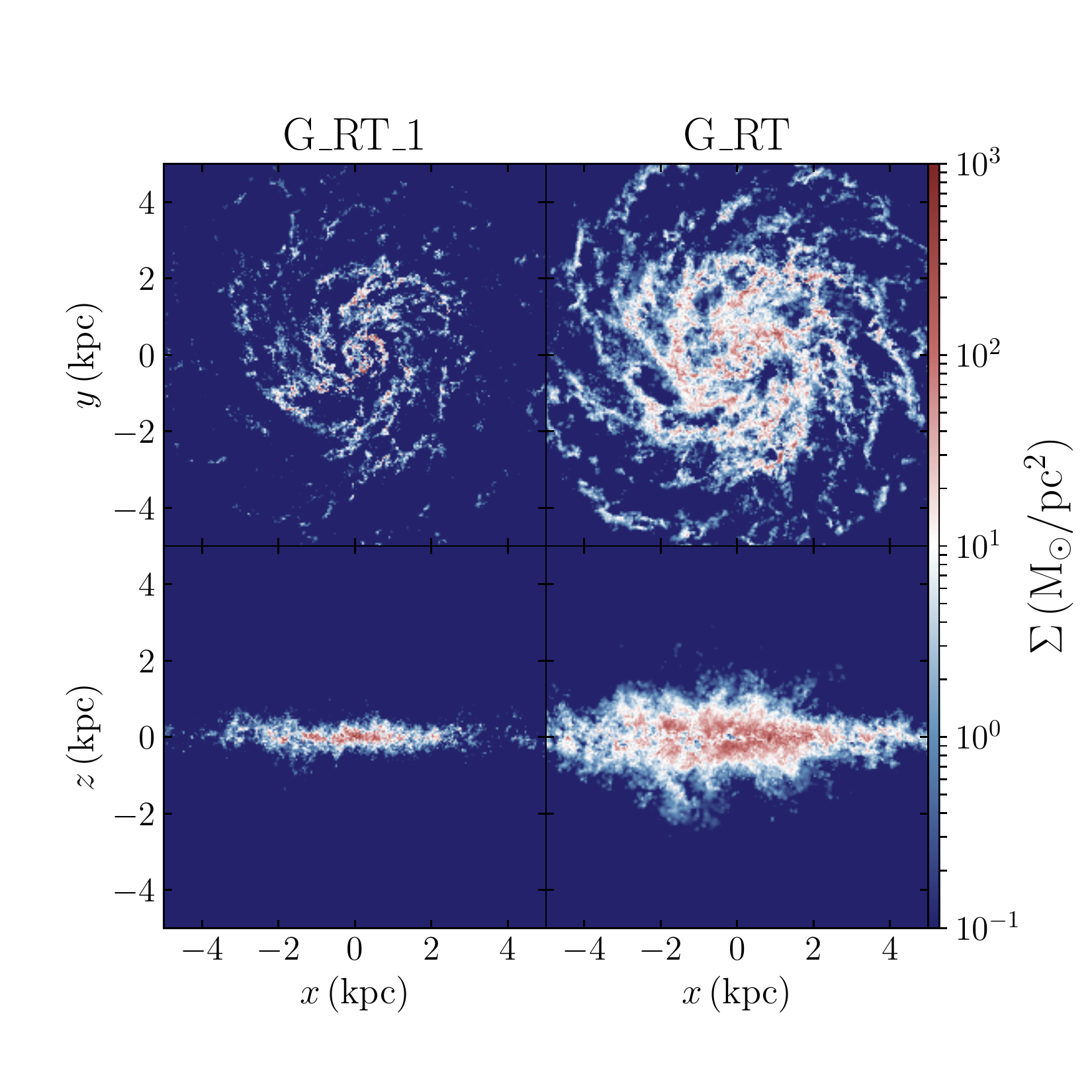}
\caption{Projected gas density maps of the H$_2$ component for the \grt\_1 (left-hand panels) and \grt{} (right-hand panels) runs at $t=350$~Myr, face-on (top panels) and edge-on (bottom panels). The clumping factor in the fiducial run enhances the amount of H$_2$ forming in the galaxy, producing a thicker molecular disc (bottom panels), and also a more spread distribution across the entire galaxy, although the highest concentration still resides in the dense regions of the spiral arms.}
\label{fig:compmapscf}
\end{figure*}
In Fig.~\ref{fig:compsfr}, we show the SFR (left-hand panel) and the formed stellar mass (right-hand panel) for the \grads, \grt, and \gradt{} runs in the last 300~Myr of the simulations. We explicitly excluded the first 100~Myr ($\sim$ three dynamical times of the galaxy) from the comparison in order to avoid introducing a bias due to the initial relaxation phase, which was slightly different in the different runs. All the runs show a similar SFR, although the SF bursts are offset in time in the different cases. This suggests that, despite the good agreement in the chemical properties amongst the three simulations, the different models for stellar radiation significantly affect the details of the evolution, changing the actual SF history of the galaxy on time-scales of about 20~to~50~Myr. In particular, the SFR in \grt{} exhibits a second burst around $t=300$~Myr not observed in the other runs, and \grads{} exhibits the lowest SFR amongst the three runs, which is likely related to a slightly stronger ionisation close to the stellar sources compared to the other models. This difference in \grads{} is also reflected in the total stellar mass formed (right-hand panel), with \grads{} producing the lowest stellar mass compared to \gradt{} and \grt, for which the agreement is much better. Nevertheless, the difference in the total stellar mass amongst the runs is smaller than 10 per cent. Moreover, if we also take into account the initial stellar mass of the galaxy, $M_\star = 2.4\times 10^9\, \msun$, the difference drops to less than 1 per cent.
\begin{figure*}
\begin{flushleft}
\large \hspace{0.24\textwidth} G\_RT\_1 \hspace{0.335\textwidth} G\_RT 
\end{flushleft}
\centering
\includegraphics[width=0.4\textwidth,trim={0cm 0cm 0cm 0.5cm},clip]{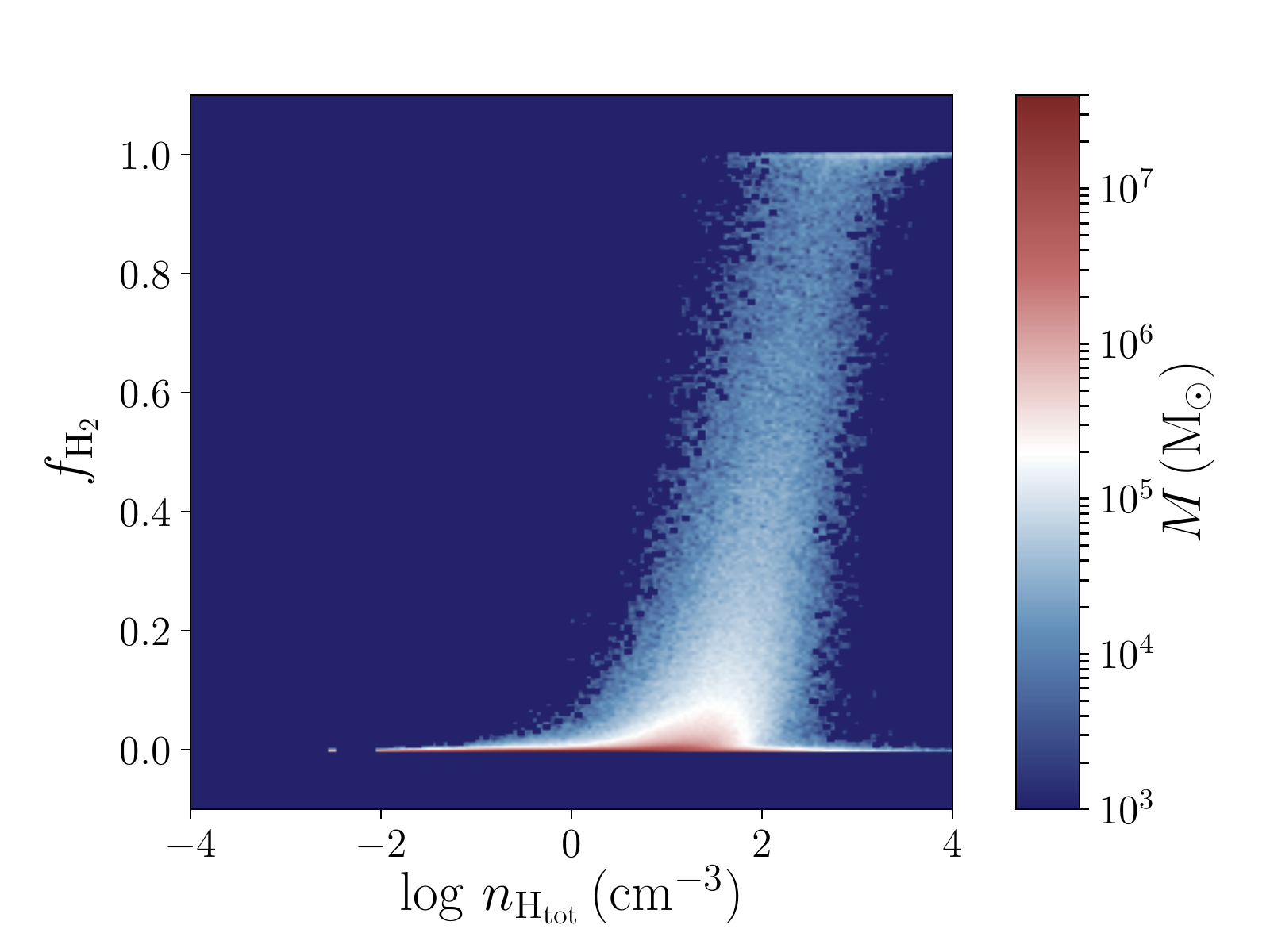}
\includegraphics[width=0.4\textwidth,trim={0cm 0cm 0cm 0.5cm},clip]{Pictures/GRTC/H2lin.pdf}\\
\includegraphics[width=0.4\textwidth]{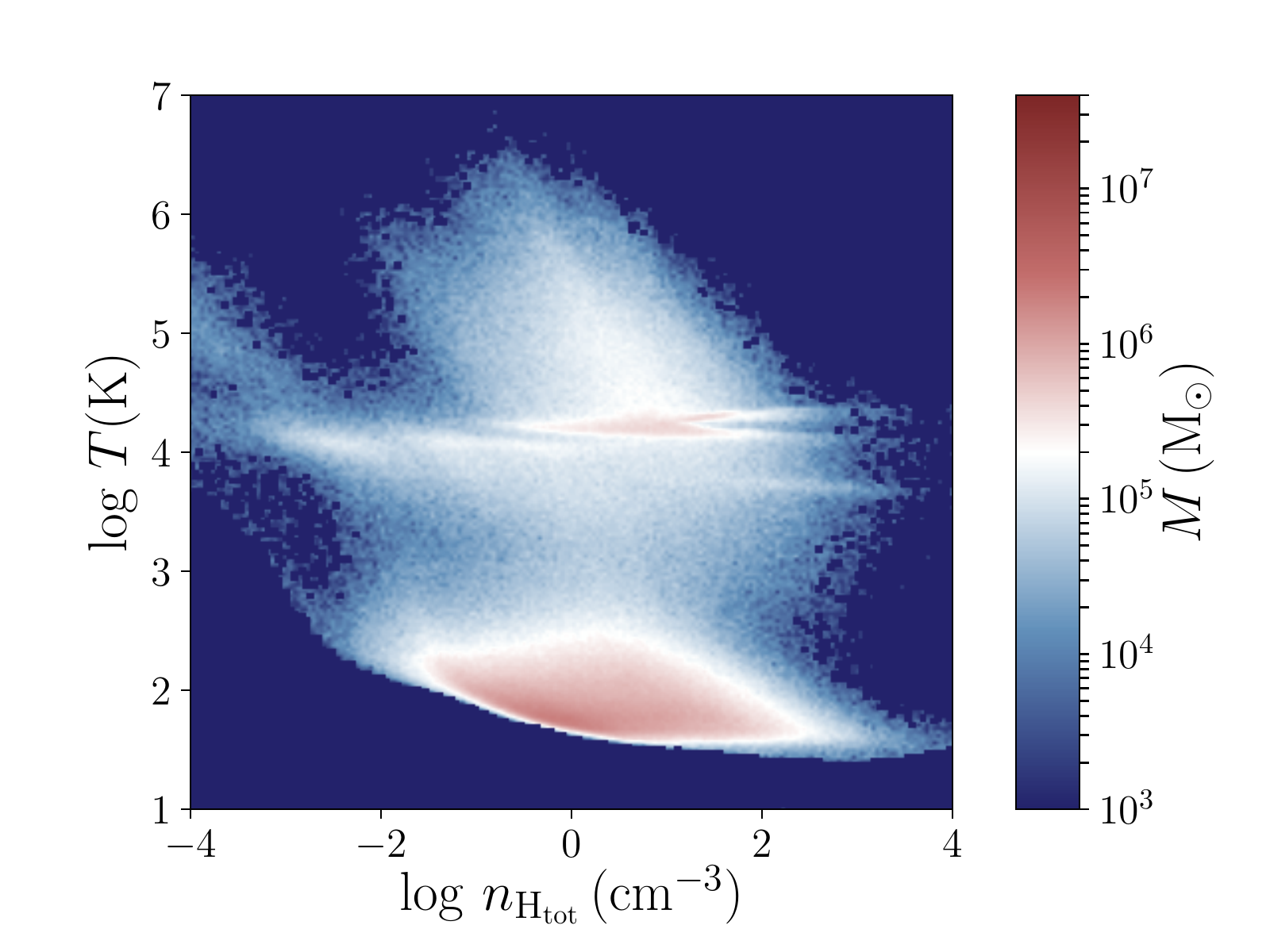}
\includegraphics[width=0.4\textwidth]{Pictures/GRTC/temp.pdf}\\
\caption{Gas properties for the \grt\_1 (left-hand column) and \grt{} (right-hand column) runs averaged over 100~Myr around $t=350$~Myr. We show the molecular hydrogen mass fraction in the top row and the gas temperature in the bottom row, as a function of the total hydrogen density. The critical density to become fully molecular shifts from $\sim 5\,\rm cm^{-3}$ (top-left panel) to $\sim 100\,\rm cm^{-3}$ (top-right panel), due to the effect of $C_\rho$ which boosts H$_2$ formation also at lower hydrogen densities. On the other hand, the temperature distribution of the gas is almost unaffected, with only a moderately larger amount of cold gas below 100 K around $n_{\rm H_{tot}} = 1\rm\, cm^{-3}$.}
\label{fig:compplotscf}
\end{figure*}

\begin{figure*}
\centering
\includegraphics[width=\textwidth,trim={0cm 0.3cm 0cm 1cm},clip]{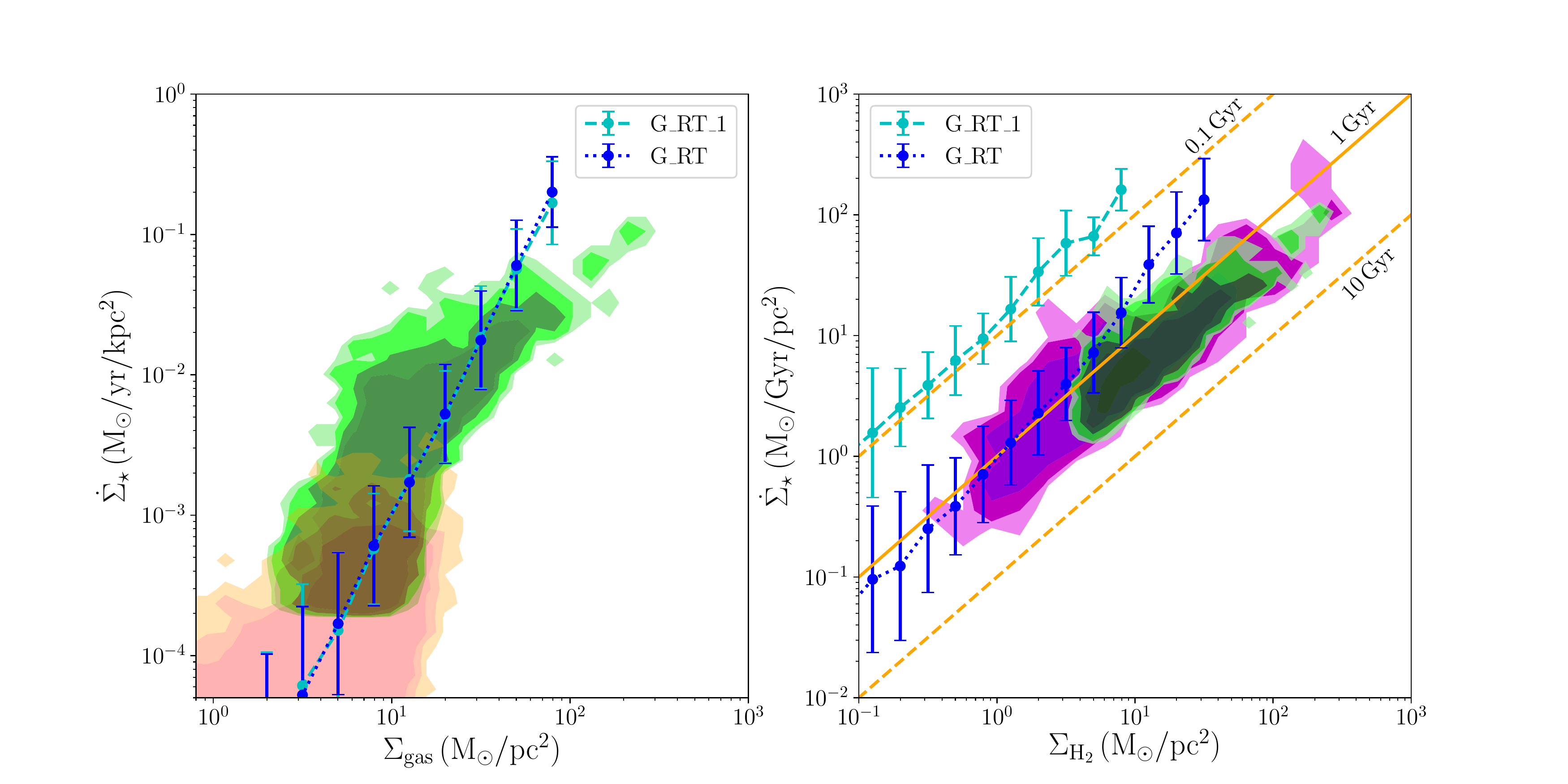}
\caption{Same as Fig.~\ref{fig:compks}, but comparing the \grt\_1 and \grt{} runs. While the KS relation in the left-hand panel does not show any variation relate to the clumping factor, the H$_2$ SF law very well matches the slope of the observed relation, but the normalisation varies by roughly an order of magnitude between the two runs. This suggests that our resolution is not high enough to properly model the very high density gas where most of H$_2$ forms, and a clumping factor is still necessary to properly match the results.}
\label{fig:comp1ks}
\end{figure*}

In Fig.~\ref{fig:compks}, we show the SF law for the same three runs. In this case, instead of the line contours, we binned our data points along the gas density axis in 0.2 decade-wide bins and reported the average SFR in each bin, using the standard deviation in the bins as the y-axis error bar. In order to exclude poorly sampled bins, we removed all the bins with less than five data points. We find very good agreement with observations by \citet{bigiel10} for the KS relation in total gas (H+H$_2$) up to $\Sigma_\star = 10^{-2}\rm\, \msun/yr/kpc^2$, the only exception being the data-points in the central 2~kpc of the galaxy which lie slightly above the orange shaded area. However, we notice that the observational data we compared with are for local galaxies, where the highest surface densities have small statistics. If we compare our results with other datasets from higher redshift galaxies \citep{orr17}, our data points are perfectly consistent in the entire range we sample. Both slope and normalisation of the H$_2$ SF law are now in agreement with the observational points. However, in \grads, the relation is slightly shifted to slightly higher H$_2$ surface densities, and we also observe a slight deviation at low H$_2$ surface densities, due to the higher abundance of H$_2$ compared to the other runs, in particular at intermediate densities (see Fig.~\ref{fig:compplots}, top-left panel).

\subsection{The role of the clumping factor}
\label{sec:clumping}
We now compare the \grt\_1 and the \grt{} runs, where we use the same model for stellar radiation, but with a different clumping factor. 
We show, in Fig.~\ref{fig:compmapscf}, the projected H$_2$ density maps for \grt{} (right-hand panels), compared with those of \grt\_1 (left-hand panels), face-on and edge-on. Expectedly, much more H$_2$ forms in \grt{}, because of the enhanced formation rate, but the spiral arm pattern is very similar in the two runs, suggesting that the effect on the global dynamics is almost negligible. Nonetheless, most of H$_2$ still resides in the dense regions of the galaxy spiral arms, where the gas is expected to be colder and turbulent. In the edge-on view, \grt{} exhibits a thicker molecular disc, where H$_2$ is abundant up to $\sim 500$~pc from the disc plane, whereas in \grt\_1 it does not extend to more than 100--200~pc. 
In Fig.~\ref{fig:compplotscf}, we compare the gas phase diagrams for the same runs. In particular, the transition from the neutral to the fully molecular phase (top panels) shifts from $\sim 5$ (right-hand panel) to $\sim 100\rm\, cm^{-3}$ (left-hand panel), because of the suppressed H$_2$ formation rate when C$_\rho = 1$. This is consistent with previous results \citep[e.g.][]{gnedin09,christensen12}, where the gas becomes fully molecular above $100\rm\, cm^{-3}$. In our fiducial runs, instead, the transition occurs at lower densities because of the clumping factor. However, we must keep in mind that the density measured in the simulation is the mean density of the clouds for which we assume a log-normal PDF and most of the molecular gas actually forms in the high-density tail of the PDF, above $100\rm\, cm^{-3}$. In the bottom panels, instead, we compare the temperature distribution of the gas. \grt{} exhibits a moderately larger amount of cold gas below 100 K, consistent with the boosted H$_2$ formation which provides a slightly more effective cooling. Small differences are also visible in the warm gas at intermediate densities, where the distribution is more spread in the \grt{} run, but they can be compatible with stochastic fluctuations of a few thousand particles and are not relevant.

Fig.~\ref{fig:comp1ks}, instead, shows the SF law in the total (H+H$_2$) and molecular gas, following the same approach used in the previous section. As expected, the KS relation in total gas is almost unaffected by the clumping factor, only depending on the net amount of neutral hydrogen. However, the molecular counterpart is significantly affected, moving by roughly an order of magnitude between the two runs. This result suggests that the resolution achieved is not sufficient to properly resolve the very high densities where the gas becomes fully molecular all over the galaxy, and a sub-grid clumping factor is necessary to account for it. Remarkably, our clumping factor model compensates for the offset without any calibration, shifting the simulation results over the observed relation.

\begin{figure}
\centering
\includegraphics[width=0.45\textwidth]{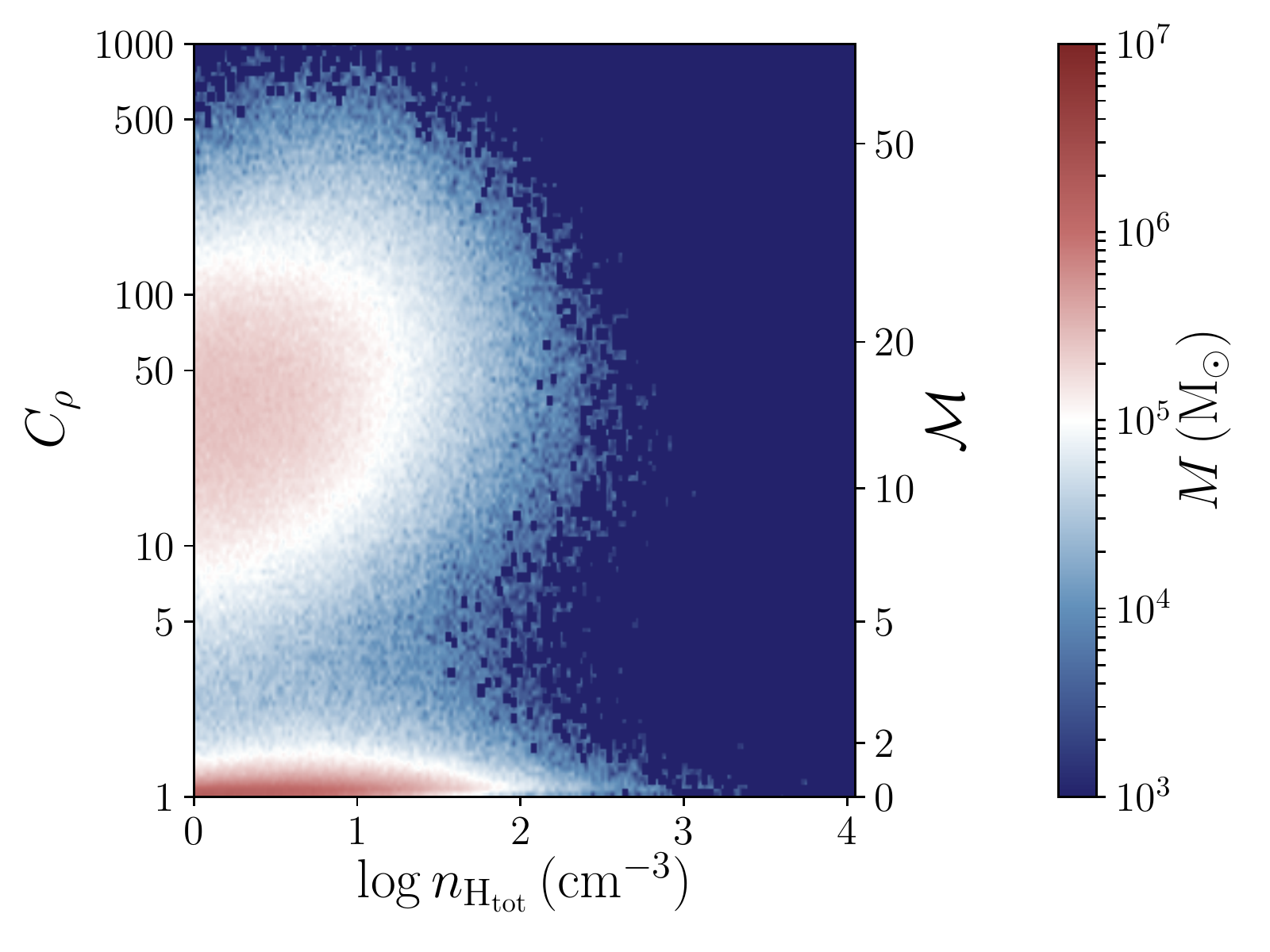}
\caption{Clumping factor distribution for \grt{} at $t=350$~Myr, binning the gas particle distribution with 0.02 decade-wide bins on both axes. For completeness, we also show the corresponding values for $\mach$, obtained reversing eq.~\eqref{eq:cfac}. Though a large fraction of gas clusters at low $C_\rho$, consistent with a pressure-dominated regime, a moderate amount of gas below $n_{\rm H_{tot}}=100\rm\, cm^{-3}$ exhibits a turbulent-dominated state, with $\mach$ up to 50, where the H$_2$ formation rate is strongly enhanced.}
\label{fig:cf}
\end{figure}

We find that the clumping factor can vary significantly in the simulation, depending on the gas properties, from 1 to up to a maximum of about $10^3$, but the distribution is far from being uniform (see Fig.~\ref{fig:cf}). Comparing the clumping factor distribution at different times we also found that the exact moment chosen could vary the distribution by a few per cent, if we look for example at a few tens of Myr after a starburst or a more quiescent phase. In particular, roughly $38\pm 3$ per cent of the gas is typically below $C_\rho=10$ and only $\sim 10\pm 2$ per cent exceeds $C_\rho=100$. The median of the distribution is $18\pm 3$ and the average is $\sim 35\pm 5$. 
Although values above 100 can appear unphysical, we emphasise that they come from gas with $25\lesssim \mach \lesssim 100$ (assuming $b=0.4$), extreme but still plausible. 
For sake of simplicity, we do not discuss in detail the runs \grads\_1 and \gradt\_1. as the same differences observed between \grt{} and \grt\_1 are reflected into the other cases.

\section{Caveats of the study}
Our results showed that the relation between SF and H$_2$ abundance is sensitive to the physical modelling in the simulations, in particular to the stellar radiative feedback. Though not explored in this study, it is possible that the relation would also be affected by different SF and SN feedback models. \begin{itemize}
\item {\it The SF model}: we used here a physically motivated parameter-free prescription which does not require a priori calibrations, unlike the standard one used in galaxy simulations, where density and (sometimes) temperature cuts are employed instead. In addition, the SF efficiency is here self-consistently computed from the gas properties, and not chosen by hand to reproduce the KS relation. However, a main limitation of the model is that it holds only for turbulent clouds, which means that the star-forming gas must satisfy $\mach \geq 2$. In spite of this, we stress that this condition does not have any impact on the results and, moreover, stars are expected to physically form in turbulent molecular clouds where $\mach >> 1$, as found in the nearby starburst NGC253 \citep{leroy15}, and not in the sub-sonic ISM, hence the employed cut has a clear physical motivation. On top of that, we also impose a density cut to prevent SF in regions where the gas is expected to be warm and not star-forming, and this could appear as an arbitrary choice. However, the motivation for this choice resides in the strong dependence of the SF efficiency on the gas density, with values as low as $10^{-4}-10^{-3}$ around the SF density threshold, as can be seen from Fig.~\ref{fig:sfe_rhoT}. For instance, at our density threshold the expected SFR is of the order of $10^{-8}-10^{-7}\rm\, \msun/yr$, corresponding to less than 1 stellar particle during the entire evolution of the galaxy. By using a higher SF density threshold, we would have introduced an additional free parameter which would have simply made the SF more bursty, but with almost no difference in the stellar spatial distribution and in the resulting KS relation.
At high densities, instead, the high SF efficiency could have played a major role in reproducing the SF-H$_2$ relation. Actually, it is possible that the high density gas was turned into stars very rapidly, and the H$_2$ associated with the gas particle instantaneously removed, reducing the total amount of H$_2$ available. This effect could have probably been attenuated using the standard SF prescription, where the SF efficiency is kept constant also for high density gas, but it is not obvious whether the final result would have been significantly different. To test this idea, we evolved our target galaxy for 200 Myr using the same model of our \gradt{} run, but with the standard SF prescription, assuming $\rho_{\rm thr}=100\rm\, cm^{-3}$ and a constant SF efficiency $\varepsilon = 0.03$. The SF--H$_2$ law did not differ significantly from that of our fiducial model.
\item {\it The SN feedback model}: in this study, we used the empirical delayed-cooling model, which was shown to better reproduce the KS relation \citep{rosdahl17}, at the expense of the ISM temperature. In particular, the gas heated up by SNe would remain warm/hot even at densities of a few cm$^{-3}$, where radiative cooling should be important, and this could affect the abundance of the chemical elements. However, although the H$_2$ abundance for the intermediate density gas kept hot by SNe could have been underestimated in our simulations, it is also true that the gas should become fully molecular around $n_{\rm tot} \sim 100\rm \, cm^{-3}$, where the amount of warm gas is subdominant compared to the cold counterpart (see for instance Fig.~\ref{fig:compplots}). In conclusion, further detailed investigations about the effect of different SN feedback models are needed to assess the main advantages and limitations of each model and they could help determining the model which better reproduces observations. Nevertheless, this is beyond the scope of the present study.
\end{itemize}

Finally, other numerical aspects could also play a role in how well we can reproduce the molecular KS relation, like the hydrodynamics method and the artificial pressure floor (not employed here). Though these effects have not been addressed here, we refer to \citet{capelo17}, where the same galaxy is evolved using a different hydrodynamics scheme \citep[SPH; with \textsc{gasoline2};][]{wadsley17}, a pressure floor, and with different chemistry networks to assess the role of non-equilibrium metal cooling at low temperature.

\section{Conclusions}
\label{sec:conclusions}
We introduced  a model to accurately follow the chemical evolution of molecular hydrogen and performed a suite of  simulations of an isolated galaxy at $z=3$ to test its validity.
To self-consistently follow H$_2$, we coupled the chemistry package \textsc{krome} with the mesh-less code \textsc{gizmo}. We used model 1a in \citet{bovino16}, including all the relevant processes regulating H$_2$ formation and dissociation, together with a new physically motivated SF prescription based on theoretical models of turbulent GMCs; delayed-cooling SN feedback; gas, dust, and H$_2$ shielding; extragalactic UV background and local UV stellar radiation. The latter is modelled with sub-grid models (\grads{} and \gradt{}) and compared to RT calculations (\grt).  We also included a cell-by-cell estimate of a clumping factor, the enhancement in the formation of H$_2$ in the very high density cores of the GMCs, unresolved in our simulations. 

We showed that the external UV background has the only effect of destroying H$_2$ at very low density, and the most important contribution to H$_2$ dissociation comes from stellar radiation inside the galaxy. The two sub-grid models for local UV stellar radiation perform similarly well against the RT simulation, and have a much higher computational efficiency. 

The KS relation in the total gas matches well observations across several orders of magnitudes, with only a small increase of the SFR at higher densities.  As for the H$_2$ counterpart, the slope is generically in  good agreement with the observational data, but the normalisation is recovered correctly only when including a clumping factor to account for unresolved dense cores in GMCs. 

In summary, we showed that, with a detailed chemical network and a physically motivated SF prescription (with a consistent sub-grid clumping factor), we have been able to naturally reproduce the observed relation between the H$_2$ and SFR surface densities for a large range of $\Sigma_{\rm H_2}$. In particular, we confirm that this relation can naturally arise when an opportune model is used, and an a priori dependence of SF on the H$_2$ abundance, as assumed in several studies \citep[e.g.][]{gnedin09,christensen12,pallottini17,hopkins17} is not necessary. We have also found that the relation seems to extend down to lower H$_2$ densities maintaining the same slope, a regime not explored yet with the available observations.

\section*{Acknowledgements}
We thank the anonymous referee for the useful suggestions which helped improve the study.
We thank Philip F. Hopkins for providing the radiative transfer implementation in \textsc{gizmo} and for useful discussions. We thank Sijing Shen for having provided the metal cooling table described in \citet{shen10,shen13}.
We acknowledge support from the European Research Council (Project No. 267117, `DARK', AL, JS; Project no. 614199, `BLACK', AL, MV). 
SB thanks for funding through the DFG priority program `The Physics of the Interstellar Medium' (projects BO 4113/1-2).
PRC acknowledges research funding by the Deutsche Forschungsgemeinschaft (DFG) Sonderforschungsbereiche (SFB) 963 and support by the Tomalla Foundation.
This work was granted access to the High Performance Computing resources of CINES under the allocation x2016046955 made by GENCI, and it has made use of the Horizon Cluster, funded by Institut d'Astrophysique de Paris, for the analysis of the simulation results. 
\bibliographystyle{mnras}
\bibliography{./Biblio}

\appendix
\section{Star formation efficiency}
\label{app:sfe}
\begin{figure}
\centering
\includegraphics[width=0.48\textwidth,trim={0cm 0.7cm 0cm 1cm},clip]{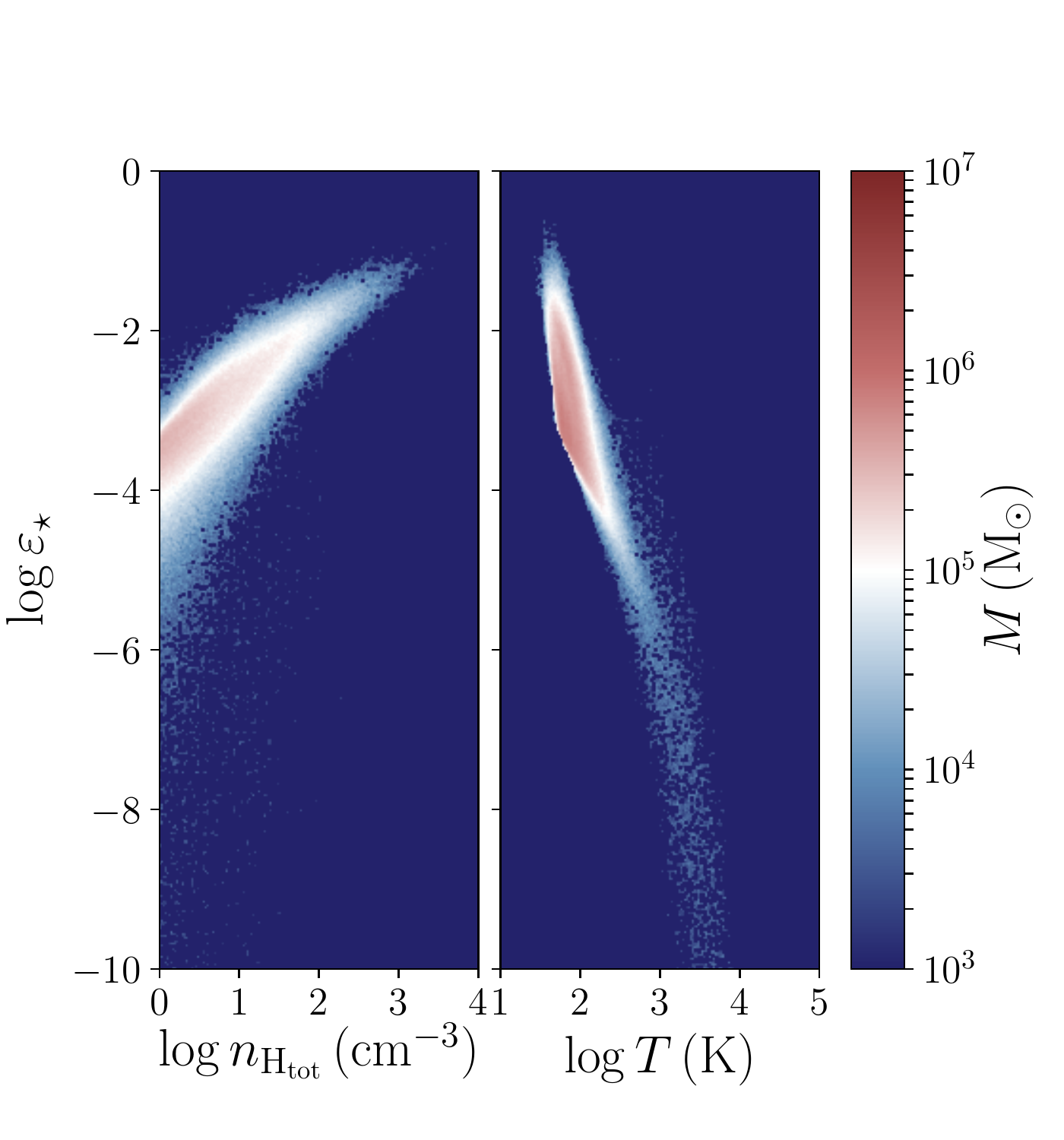}
\caption{SF efficiency versus total hydrogen density (left-hand panel) and gas temperature (right-hand panel) for our \gradt{} run at $t=350$~Myr. The SF efficiency rapidly increases with increasing density and is strongly suppressed in the warm gas, consistent with the fact that SF mainly occurs in dense, cold, and turbulent clouds.}
\label{fig:sfe_rhoT}
\end{figure}

In this study, we have implemented a new prescription for SF based on the theoretical model by \citet{padoan11} and calibrated against simulations and observations by \citet{federrath12}. Unlike the standard model used in numerical simulations, where the SF efficiency $\varepsilon$ is constant and must be tuned to get the right KS normalization, this model self-consistently gives a SF efficiency dependent on the gravo-turbulent state of the gas. According to \citet{federrath12}, the SF efficiency can vary across several orders of magnitude and, in principle, can also exceed 100 per cent 
 when the GMC is strongly bound ($\alpha_{\rm vir}<<1$) and very turbulent ($\mach>>10$), with compressive modes dominating over the solenoidal modes.\footnote{This is due to the definition of the SF efficiency per free-fall time, which is normalized with the total mass and the free-fall time at the mean density of the GMC.}
To evaluate the range spanned by the SF efficiency in our simulations and the dependence on the gas properties, we plotted in Fig.~\ref{fig:sfe_rhoT} the SF efficiency from our \gradt{} run at $t=350$~Myr, as a function of the total hydrogen density (left-hand panel) and gas temperature (right-hand panel), binning all the quantities in 0.02 decade-wide bins. The SF efficiency around the SF density threshold $n_{\rm H_{tot},thr}=1\rm\, cm^{-3}$ is much lower than the typically used value of 0.01 (most of the gas has $\varepsilon\lesssim 10^{-3}$) and rapidly increases with the gas density up to more than 10 per cent\footnote{This only occurs for a few particles at very high densities, which cannot be seen in the plot.} above $n_{\rm H_{tot}}=10^3\rm\, cm^{-3}$. This suggests that $\rho_{\rm SF,thr}$ is not a crucial parameter of the model, because of the rapid decay of $\varepsilon$ at very low densities. A moderate value of $n_{\rm H_{tot}}$ as low as $1\rm\, cm^{-3}$ is, however, a reasonable choice to avoid spending time computing the SFR in regions where stellar particles would never form. From the SF efficiency distribution it seems that we never approach 100 per cent efficiency, even when gas is fully molecular (cf. Fig.~\ref{fig:compplotscf}). However, it could simply be a selection effect due to the fact that at very high SF efficiencies the conversion from gas into stars is almost instantaneous, yielding to the disappearance of the particle from the SF efficiency plot shown here. As for the right-hand panel, instead, the dependence on the temperature is extremely steep, with all the gas particles with $\varepsilon>10^{-5}$ being colder than $T=10^3$~K and the hotter ones showing almost negligible $\varepsilon$. The model therefore provides a natural temperature threshold for SF, without the need to impose a strict criterion as usually done with the standard prescription. Moreover, the lack of a sharp vertical cut-off in the temperature--SF efficiency plot reveals that the Mach number criterion does not play a fundamental role, being most of the cold (and highly star-forming) gas in a turbulent-dominated state.

\begin{figure}
\centering
\includegraphics[width=0.48\textwidth,trim={1cm 0.5cm 1cm 1cm},clip]{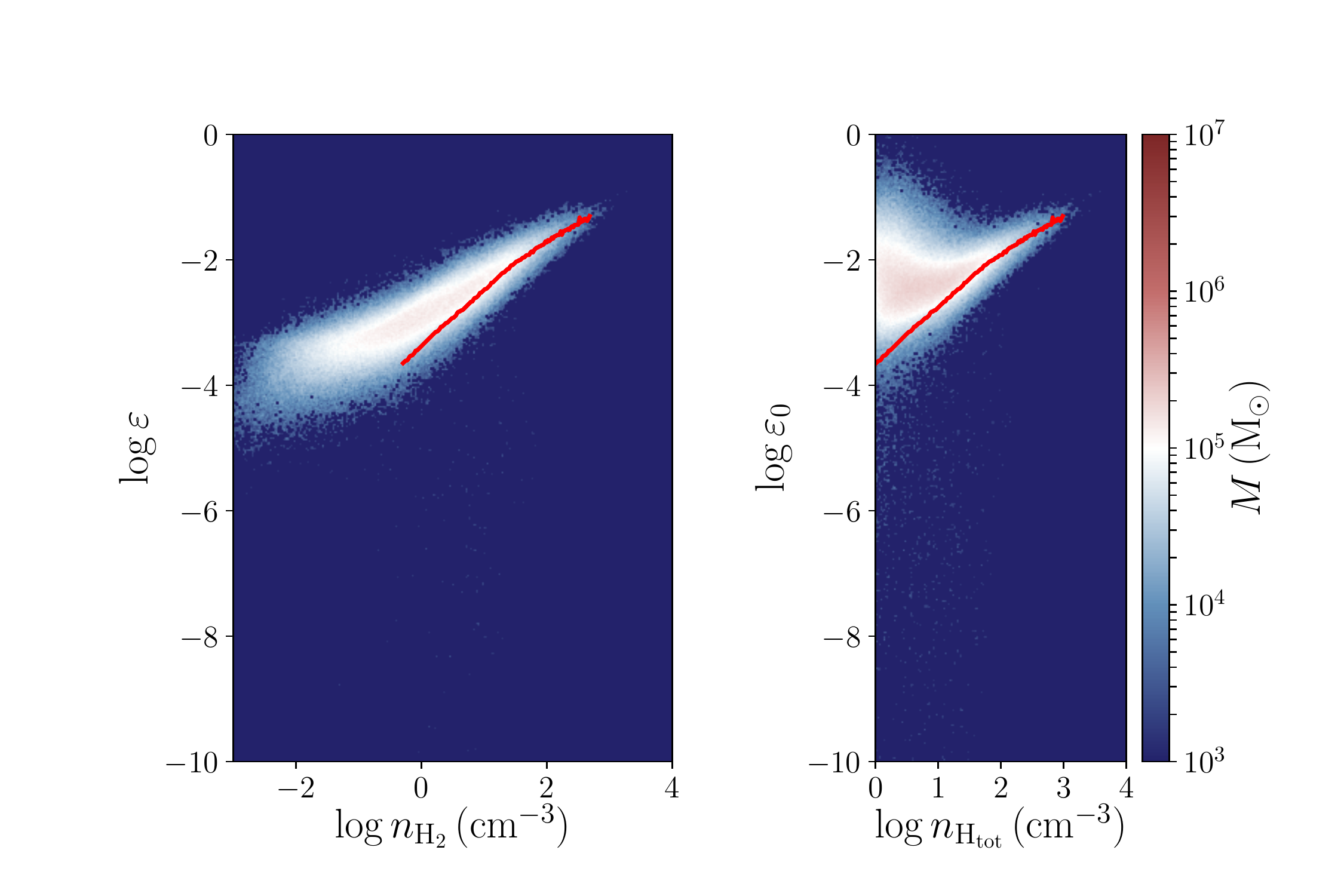}
\caption{SF efficiency versus H$_2$ density (left-hand panel) and renormalized SF efficiency versus $n_{\rm H_{tot}}$ (right-hand panel) for our \gradt{} run at $t=350$~Myr. The red curve is the median of the SF efficiency from the left-hand panel in Fig.~\ref{fig:sfe_rhoT}, reported as a function of total hydrogen density in both panels. In the left panel, it is rescaled by a factor of 2 to account for the number of hydrogen nuclei in the H$_2$ molecule and be consistent with the x-axis of the plot. Though the SF efficiency shows a strong dependence on the H$_2$ abundance, there is no apparent causal connection between the H$_2$ and SF. When renormalized to the H$_2$ abundance, the increase is still present, but a large scatter appears for low densities where SF proceeds in the H-dominated regime.}
\label{fig:sfe_H2}
\end{figure}

In Fig.~\ref{fig:sfe_H2},  we show the SF efficiency as a function of H$_2$ density (left-hand panel) and the renormalized SF efficiency $\varepsilon_0 = \varepsilon/f_{\rm H_2}$ (right-hand panel), using the same binning as in Fig.~\ref{fig:sfe_rhoT}. We overplot, in both panels, the median of the SF efficiency distribution from the left-hand panel of Fig.~\ref{fig:sfe_rhoT}, to highlight the change in the distribution when we introduce a dependence on the H$_2$ abundance. The SF efficiency, mildly increasing up to $n_{\rm H_2}\sim 0.1 \rm\, cm^{-3}$, with an average value around $10^{-4}$, shows a steeper slope at higher densities, where most of the SF occurs. $\varepsilon_0$ has a very large scatter below $n_{\rm H_{tot}}\sim 30-40\rm\,cm^{-3}$, where stars form in the atomic hydrogen regime with a very weak dependence on the abundance of H$_2$, whereas a clear overlap with the median curve (in red) is visible at higher densities. This is expected, since we are in the regime where gas becomes fully molecular ($f_{\rm H_2}\approx 1$) and the normalization factor has a negligible effect. Even at high $n_{\rm H_2}$, $\varepsilon_0$ never reaches a 100 per cent efficiency in our simulations, but it self-regulates to always be below few tens per cent.

\begin{figure}
\centering
\includegraphics[width=0.45\textwidth]{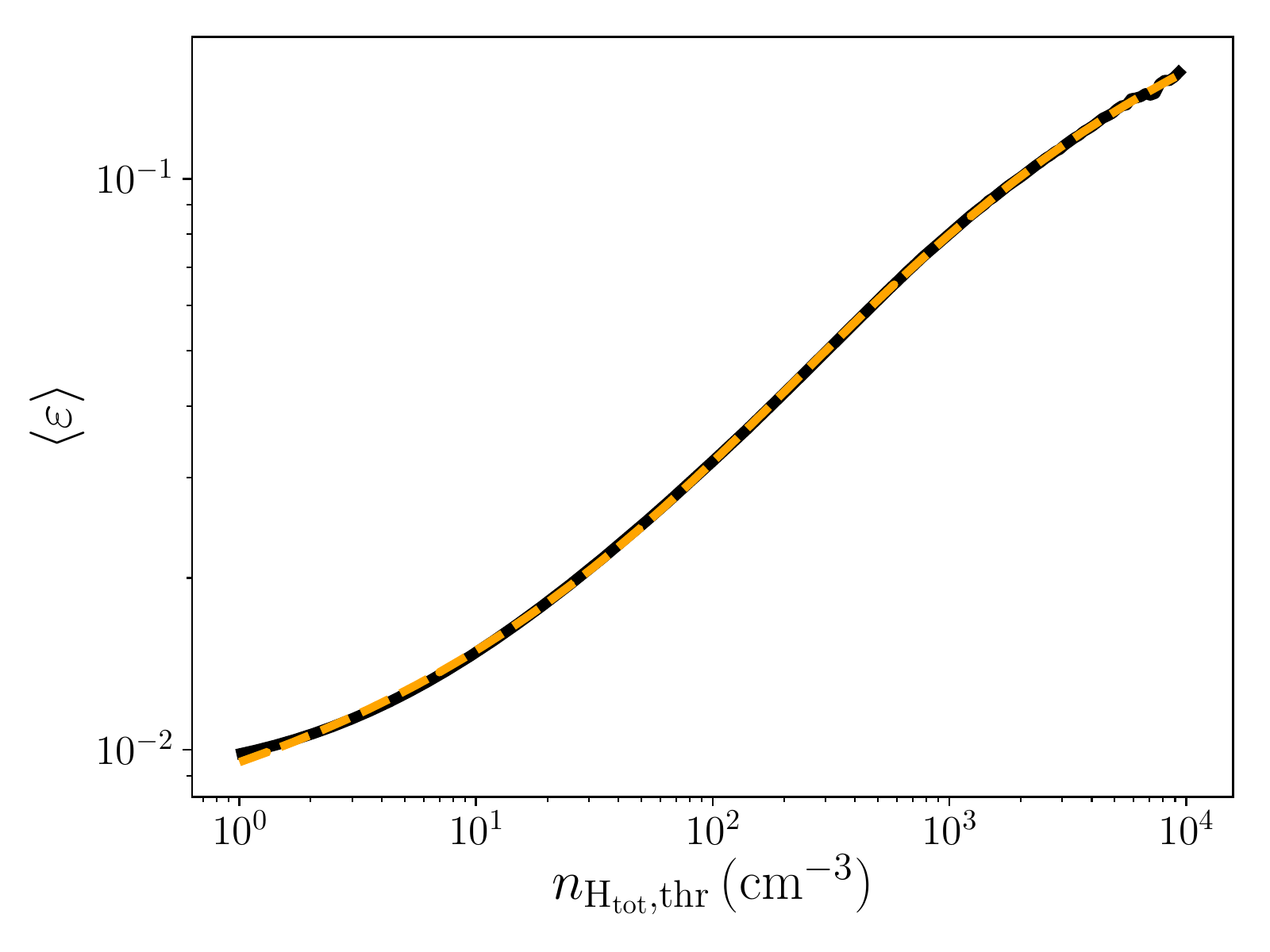}
\caption{Average SF efficiency as a function of the density threshold $n_{\rm H_{tot},thr}$. The black solid line is the result from the averaging procedure applied to the bins in the left-hand panel of Fig.~\ref{fig:sfe_rhoT} and the orange dashed one is the best fit with the analytic function defined in Eq.~\eqref{eq:bestfit}.}
\label{fig:avg_sfe}
\end{figure}

As a benchmark to compare the SF prescription implemented in this study with the standard one, where stars form in gas with $n_{\rm H_{tot}}>n_{\rm H_{tot},thr}$ with a constant $\varepsilon$, we compute the average SF efficiency $\langle\varepsilon\rangle $ for gas above a total hydrogen density $n_{\rm H_{tot},thr}$ by stacking the snapshots of the last 100~Myr of the simulation, as
\begin{equation}
\langle\varepsilon\rangle = \frac{\sum_{i=i_0}^{N_i}\left[\sum_{\varepsilon_j=0}^{N_j} \varepsilon_{i;j} M_{i;j}\right]/t_{\rm ff}(n_i)}{\sum_{i=i_0}^{N_i}\left[\sum_{j=0}^{N_j}M_{i;j}\right]/t_{\rm ff}(n_i)},
\label{eq:bestfit}
\end{equation}
where $i$ runs on the $N_i$ density bins, $j$ runs on the $N_j$ SF efficiency bins, $i_0$ corresponds to the $n_{\rm H_{tot},thr}$ bin, $M_{(i;j)}$ is the gas mass in the $(i;j)$-th bin, $\varepsilon_j$ is the SF efficiency in the $j$-th bin, and $t_{\rm ff}(n_i)$ is the free-fall time in the $i$-th bin.
Fig.~\ref{fig:avg_sfe} shows the resulting $\langle\varepsilon\rangle$, together with the best fit, expressed by the analytic expression
\begin{equation}
\langle\varepsilon\rangle = \alpha\exp\left\{\beta \arctan\left[\phi\ln(\xi n_{\rm H_{tot},thr})\right]\right\}.
\end{equation}
We report the best-fitting parameters in Table~\ref{tab:fitpars}.
The fitting function used shows a perfect agreement across four orders of magnitude with the simulation data and varies between the usual SF efficiency value of 0.01 and $\gsim 0.14$. While it is common practice to increase the SF efficiency when a high SF density threshold is used, the only way to tune it is by rescaling it to match the local KS relation. Here we provide an alternative way to do it, by means of an analytic fitting function, aimed at reproducing, on average, the same SFR but using a simpler SF prescription with constant SF efficiency.

\begin{table}
\centering
\caption{Parameter values for the best-fitting function to $\langle\varepsilon\rangle$ defined in Eq.~\eqref{eq:bestfit}.}
\label{tab:fitpars}
\begin{tabular*}{\columnwidth}{@{\extracolsep{\stretch{1}}}*{2}{c}@{}}
\hline\hline
Parameter & Value\\
\hline
$\alpha$ & 0.04415 \\
$\beta$ & 1.642\\
$\phi$ & 0.2507\\
$\xi$ & 0.004509\\
\hline\hline
\end{tabular*}
\end{table}

\section{The low-resolution case}
\label{app:lowres}
\begin{figure*}
\centering
\includegraphics[width=\textwidth]{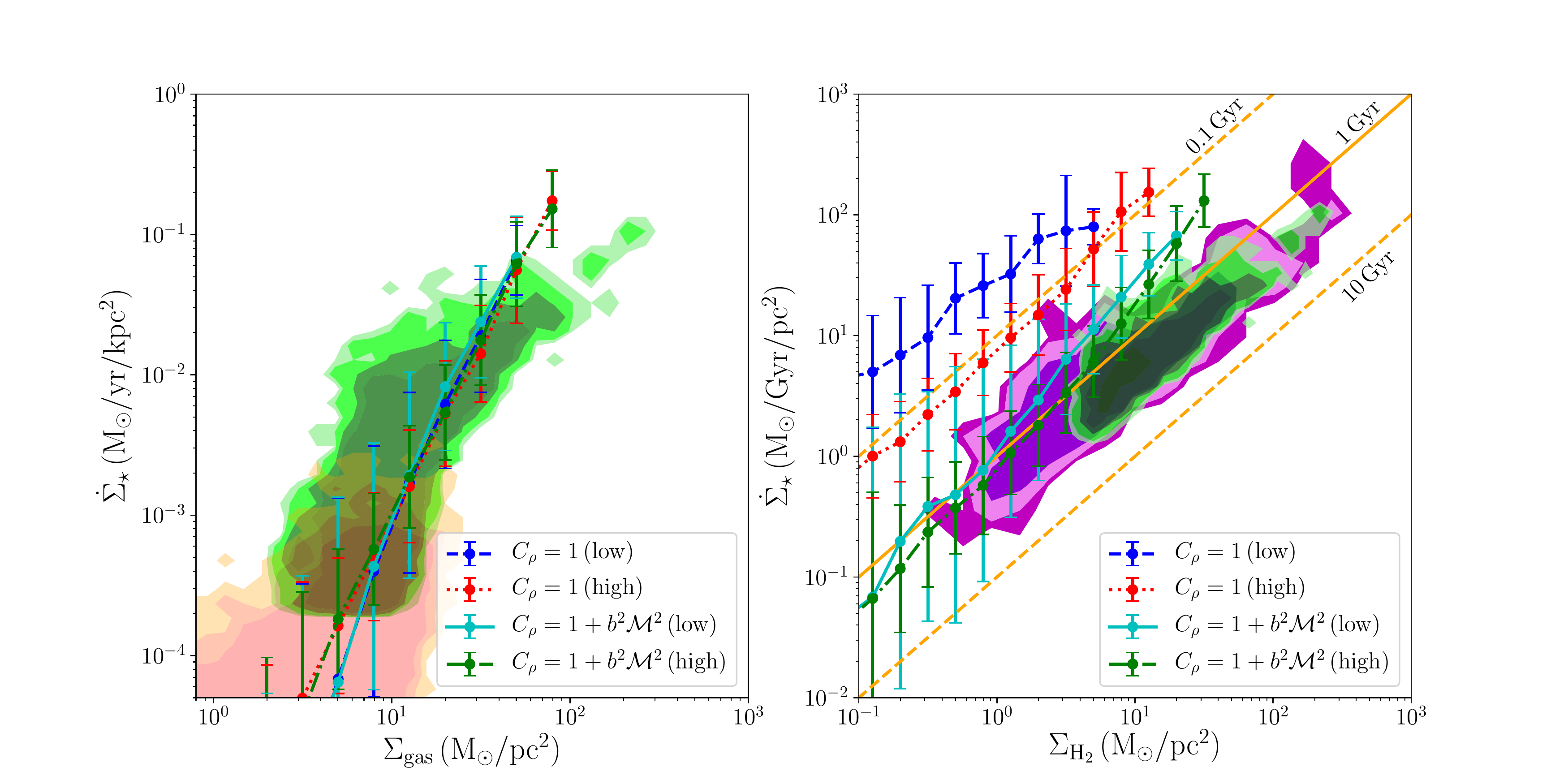}
\caption{SF law for the low- and high-resolution runs in the last 100~Myr of the simulations, with and without the clumping factor. The high-resolution cases correspond to the \gradt\_1 and \gradt{} runs described in the previous sections. We show the relation for the total gas (H+H$_2$) in the top panel and that for H$_2$ only in the bottom panel, following the same approach of Fig.~\ref{fig:compks}. All runs agree on the total gas KS relation, independently of the clumping factor and the resolution. The H$_2$ SF law is instead significantly offset from the observational data, for the runs without clumping factor, and this is more evident in the low-resolution run. When the clumping factor is used, the agreement improves, bringing both low- and high-resolution runs within the respective error bars.}
\label{fig:ks_LR}
\end{figure*}
We report here the results of two low-resolution runs analogous to the high-resolution ones described in the main sections of the paper, to assess how sensitive our clumping factor model is to the numerical resolution. These simulations are run using model (b) for stellar radiation, which showed a slightly better agreement with our \grt{} run.

The galaxy is here modelled using 100k gas particles, 500k DM particles, and 100k stellar particles in the disc and 48k stellar particles in the bulge, corresponding to a mass resolution of $23910\, \msun$ for gas, $16600$ and $15940\,\msun$ for stellar disc and bulge, respectively, and $5.9\times 10^5\,\msun$ for DM.
The softening length is rescaled here to 10~pc for stars and 6~pc (minimum) for gas.

For the sake of simplicity, we do not report here the full analysis done for the high-resolution runs, but we limit the discussion to the KS relation only. We show, in Fig.~\ref{fig:ks_LR}, the results of the two low-resolution runs, with and without our clumping factor model, obtained by binning along the gas density axis as done for Fig.~\ref{fig:compks}, and we compare them with those of the corresponding high-resolution runs \gradt\_1 and \gradt{}. For the KS relation in the total gas (H+H$_2$), the results are in very good agreement, with small differences very well within the error bars. For the relation in the molecular gas, while both the low-resolution (blue dashed) and the high-resolution (red dotted) runs without clumping factor are offset from the observed relation, the offset is more pronounced in the low-resolution case. When we use the clumping factor model (cyan solid and green dot-dashed lines), instead, the difference almost vanishes, becoming much smaller than the measured scatter, and, moreover, both curves become consistent with a depletion time of $\sim 1$~ Gyr, as found in observations.

This suggests that, although the agreement slightly improves increasing the resolution, the model is still valid in the low-resolution case. However, to ensure the validity of the gas log-normal PDF, we must be able to resolve single GMCs, at least the more common ones. If this is not the case, both the clumping factor and the SF prescription become invalid and should not be used. We therefore suggest a minimum gas mass resolution of a few $10^4\,\msun$, according to the observations by \citet{romanduval10} of 580 molecular clouds in the University of Massachusetts--Stony Brook Survey, where only $\sim 15$ per cent is more massive than $10^5\,\msun$, and most of the GMCs ($\sim 42$ per cent of the total) are in the range $10^4-10^5\,\msun$.

\label{lastpage}
\end{document}